\documentclass{jpp_gio}

\usepackage{graphicx}
\usepackage{natbib}

\ifCUPmtlplainloaded \else
  \checkfont{eurm10}
  \iffontfound
    \IfFileExists{upmath.sty}
      {\typeout{^^JFound AMS Euler Roman fonts on the system,
                   using the 'upmath' package.^^J}%
       \usepackage{upmath}}
      {\typeout{^^JFound AMS Euler Roman fonts on the system, but you
                   dont seem to have the}%
       \typeout{'upmath' package installed. JPP.cls can take advantage
                 of these fonts, if you use 'upmath' package.^^J}%
      }
  \else
  \fi
\fi

\ifCUPmtlplainloaded \else
  \checkfont{msam10}
  \iffontfound
    \IfFileExists{amssymb.sty}
      {\typeout{^^JFound AMS Symbol fonts on the system, using the
                'amssymb' package.^^J}%
       \usepackage{amssymb}%
       \let\le=\leqslant  
       \let\ge=\geqslant  
      }{}
  \fi
\fi

\usepackage{amsmath}
\usepackage{epsf}

\ifCUPmtlplainloaded \else
  \IfFileExists{amsbsy.sty}
    {\typeout{^^JFound the 'amsbsy' package on the system, using it.^^J}%
     \usepackage{amsbsy}}
    {\providecommand\boldsymbol[1]{\mbox{\boldmath $##1$}}}
\fi

\providecommand\bnabla{\boldsymbol{\nabla}}
\providecommand\bcdot{\boldsymbol{\cdot}}

\newsavebox{\astrutbox}
\sbox{\astrutbox}{\rule[-5pt]{0pt}{20pt}}

\newcommand\p{\ensuremath{\partial}}

\newcommand\bfzero{\mathbf{0}}

\newcommand\bfA{\mathbf{A}}

\newcommand\bfI{\mathbf{I}}

\newcommand\bfM{\mathbf{M}}

\newcommand\bfT{\mathbf{T}}
\newcommand\bfU{\mathbf{U}}

\newcommand\sfX{\mathsf{X}}

\newcommand\bmA{\boldsymbol{A}}
\newcommand\bmB{\boldsymbol{B}}

\newcommand\bmE{\boldsymbol{E}}
\newcommand\bmF{\boldsymbol{F}}

\newcommand\bmJ{\boldsymbol{J}}

\newcommand\bmR{\boldsymbol{R}}
\newcommand\bmS{\boldsymbol{S}}

\newcommand\bma{\boldsymbol{a}}

\newcommand\bmd{\boldsymbol{d}}
\newcommand\bme{\boldsymbol{e}}

\newcommand\bmg{\boldsymbol{g}}

\newcommand\bmk{\boldsymbol{k}}

\newcommand\bmn{\boldsymbol{n}}

\newcommand\bmu{\boldsymbol{u}}
\newcommand\bmv{\boldsymbol{v}}

\newcommand\bmx{\boldsymbol{x}}

\newcommand\bmz{\boldsymbol{z}}

\newcommand\bmeta{\boldsymbol{\eta}}
\newcommand\bxi{\boldsymbol{\xi}}
\newcommand\bomega{\boldsymbol{\omega}}

\newcommand\rmD{\mathrm{D}}
\newcommand\rmH{\mathrm{H}}
\newcommand\rmL{\mathrm{L}}
\newcommand\rmM{\mathrm{M}}
\newcommand\rmR{\mathrm{R}}

\newcommand\rma{\mathrm{a}}
\newcommand\rmc{\mathrm{c}}
\newcommand\rmd{\mathrm{d}}
\newcommand\rme{\mathrm{e}}
\newcommand\rmg{\mathrm{g}}
\newcommand\rmh{\mathrm{h}}
\newcommand\rmi{\mathrm{i}}
\newcommand\rmk{\mathrm{k}}
\newcommand\rmm{\mathrm{m}}

\newcommand\rmo{\mathrm{o}}
\newcommand\rmp{\mathrm{p}}
\newcommand\rms{\mathrm{s}}
\newcommand\rmv{\mathrm{v}}

\newcommand\real{\mathrm{Re}}

\newcommand\cst{\mathrm{constant}}

\newcommand\half{\tfrac{1}{2}}

\newcommand\quarter{\tfrac{1}{4}}

\newcommand\f{\frac}

\DeclareMathOperator\divword{div}

\DeclareMathOperator\sech{sech}

\newcommand\btimes{\boldsymbol{\times}}
\newcommand\grad{\bnabla}
\renewcommand\div{\bnabla\bcdot}
\newcommand\curl{\bnabla\times}
\newcommand\delsq{\nabla^2}

\renewcommand\le{\leqslant}
\renewcommand\ge{\geqslant}

\title[Astrophysical fluid dynamics]{LECTURE NOTES\\Astrophysical fluid dynamics}

\author[Gordon I. Ogilvie]%
{Gordon I. Ogilvie%
\thanks{e-mail address for correspondence: \texttt{gio10@cam.ac.uk}}}
\affiliation{Department of Applied Mathematics and Theoretical Physics, University of Cambridge, Centre for Mathematical Sciences, Wilberforce Road, Cambridge CB3 0WA, UK}

\pubyear{}
\volume{}
\pagerange{}
\begin{document}

\maketitle

\begin{abstract}
  These lecture notes and example problems are based on a course given
  at the University of Cambridge in Part~III of the Mathematical
  Tripos.

  Fluid dynamics is involved in a very wide range of astrophysical
  phenomena, such as the formation and internal dynamics of stars and
  giant planets, the workings of jets and accretion discs around stars
  and black holes, and the dynamics of the expanding Universe.
  Effects that can be important in astrophysical fluids include
  compressibility, self-gravitation and the dynamical influence of the
  magnetic field that is `frozen in' to a highly conducting plasma.

  The basic models introduced and applied in this course are Newtonian
  gas dynamics and magnetohydrodynamics (MHD) for an ideal
  compressible fluid.  The mathematical structure of the governing
  equations and the associated conservation laws are explored in some
  detail because of their importance for both analytical and numerical
  methods of solution, as well as for physical interpretation.  Linear
  and nonlinear waves, including shocks and other discontinuities, are
  discussed.  The spherical blast wave resulting from a supernova, and
  involving a strong shock, is a classic problem that can be solved
  analytically.  Steady solutions with spherical or axial symmetry
  reveal the physics of winds and jets from stars and discs.  The
  linearized equations determine the oscillation modes of
  astrophysical bodies, as well as determining their stability and
  their response to tidal forcing.
\end{abstract}


\tableofcontents

\section{Introduction}

\subsection{Areas of application}

Astrophysical fluid dynamics (AFD) is a theory relevant to the
description of the interiors of stars and planets, exterior phenomena
such as discs, winds and jets, and also the interstellar medium, the
intergalactic medium and cosmology itself.  A fluid description is not
applicable (i) in regions that are solidified, such as the rocky or
icy cores of giant planets (under certain conditions) and the crusts
of neutron stars, and (ii) in very tenuous regions where the medium is
not sufficiently collisional (see Section~\ref{s:validity}).

\newpage

Important areas of application include:
\begin{itemize}
\item Instabilities in astrophysical fluids
\item Convection
\item Differential rotation and meridional flows in stars
\item Stellar oscillations driven by convection, instabilities or tidal forcing
\item Astrophysical dynamos
\item Magnetospheres of stars, planets and black holes
\item Interacting binary stars and Roche-lobe overflow
\item Tidal disruption and stellar collisions
\item Supernovae
\item Planetary nebulae
\item Jets and winds from stars and discs
\item Star formation and the physics of the interstellar medium
\item Astrophysical discs, including protoplanetary discs, accretion
  discs in interacting binary stars and galactic nuclei, planetary
  rings, etc.
\item Other accretion flows (Bondi, Bondi--Hoyle, etc.)
\item Processes related to planet formation and planet--disc
  interactions
\item Planetary atmospheric dynamics
\item Galaxy clusters and the physics of the intergalactic medium
\item Cosmology and structure formation
\end{itemize}

\subsection{Theoretical varieties}

There are various flavours of AFD in common use. The basic model
involves a compressible, inviscid fluid and is Newtonian (i.e.\
non-relativistic).  This is known as hydrodynamics (HD) or gas
dynamics (to distinguish it from incompressible hydrodynamics).  The
thermal physics of the fluid may be treated in different ways, either
by assuming it to be isothermal or adiabatic, or by including
radiative processes in varying levels of detail.

Magnetohydrodynamics (MHD) generalizes this theory by including the
dynamical effects of a magnetic field. Often the fluid is assumed to
be perfectly electrically conducting (ideal MHD).  One can also
include the dynamical (rather than thermal) effects of radiation,
resulting in a theory of radiation (magneto)hydrodynamics. Dissipative
effects such as viscosity and resistivity can be included. All these
theories can also be formulated in a relativistic framework.

\begin{itemize}
\item HD: hydrodynamics
\item MHD: magnetohydrodynamics
\item RHD: radiation hydrodynamics
\item RMHD: radiation magnetohydrodynamics
\item GRHD: general relativistic hydrodynamics
\item GRRMHD: general relativistic radiation magnetohydrodynamics, etc.
\end{itemize}

\subsection{Characteristic features}

AFD typically differs from `laboratory' or `engineering' fluid
dynamics in the relative importance of certain effects.
Compressibility and gravitation are often important in AFD, while
magnetic fields, radiation forces and relativistic phenomena are
important in some applications. Effects that are often unimportant in
AFD include viscosity, surface tension and the presence of solid
boundaries.

\section{Ideal gas dynamics}

\subsection{Fluid variables}

A fluid is characterized by a \emph{velocity field} $\bmu(\bmx,t)$ and
two independent thermodynamic properties.  Most useful are the
dynamical variables: the \emph{pressure} $p(\bmx,t)$ and the
\emph{mass density} $\rho(\bmx,t)$.  Other properties, e.g.\
temperature $T$, can be regarded as functions of $p$ and $\rho$.  The
\emph{specific volume} (volume per unit mass) is $v=1/\rho$.

We neglect the possible complications of variable chemical composition
associated with chemical and nuclear reactions, ionization and
recombination.

\subsection{Eulerian and Lagrangian viewpoints}

In the \emph{Eulerian viewpoint} we consider how fluid properties vary
in time at a point that is fixed in space, i.e.\ attached to the
(usually inertial) coordinate system.  The \emph{Eulerian
  time-derivative} is simply the partial differential operator
\begin{equation}
  \f{\p}{\p t}.
\end{equation}

In the \emph{Lagrangian viewpoint} we consider how fluid properties
vary in time at a point that moves with the fluid at velocity
$\bmu(\bmx,t)$.  The \emph{Lagrangian time-derivative} is then
\begin{equation}
  \f{\rmD}{\rmD t}=\f{\p}{\p t}+\bmu\bcdot\grad.
\end{equation}

\subsection{Material points and structures}
\label{s:material}

A \emph{material point} is an idealized \emph{fluid element}, a point
that moves with the bulk velocity $\bmu(\bmx,t)$ of the fluid.  (Note
that the true particles of which the fluid is composed have in
addition a random thermal motion.)  \emph{Material curves, surfaces
  and volumes} are geometrical structures composed of fluid elements;
they move with the fluid flow and are distorted by it.

\begin{figure}
  \centerline{\epsfbox{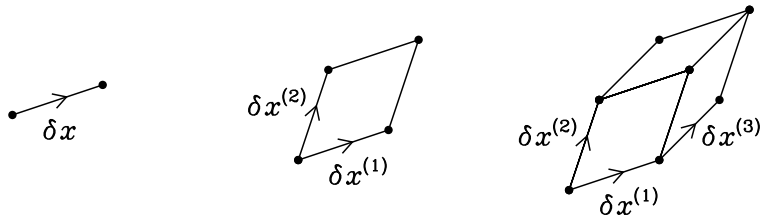}}
  \caption{Examples of material line, surface and volume elements.}
  \label{f:material_elements}
\end{figure}

An infinitesimal material line element $\delta\bmx$
(Figure~\ref{f:material_elements}) evolves according to
\begin{equation}
  \f{\rmD\,\delta\bmx}{\rmD t}=\delta\bmu=\delta\bmx\bcdot\grad\bmu.
\label{ddxdt}
\end{equation}
It changes its length and/or orientation in the presence of a velocity
gradient.  (Since $\delta\bmx$ is only a time-dependent vector rather
than a vector field, the time-derivative could be written as an
ordinary derivative $\rmd/\rmd t$.  The notation $\rmD/\rmD t$ is used
here to remind us that $\delta\bmx$ is a material structure that moves
with the fluid.)

Infinitesimal material surface and volume elements can be defined from
two or three material line elements according to the vector product
and the triple scalar product (Figure~\ref{f:material_elements})
\begin{equation}
  \delta\bmS=\delta\bmx^{(1)}\btimes\delta\bmx^{(2)},\qquad
  \delta V=\delta\bmx^{(1)}\bcdot\delta\bmx^{(2)}\btimes\delta\bmx^{(3)}.
\end{equation}
They therefore evolve according to
\begin{equation}
  \f{\rmD\,\delta\bmS}{\rmD t}=(\div\bmu)\,\delta\bmS-(\grad\bmu)\bcdot\delta\bmS,\qquad
  \f{\rmD\,\delta V}{\rmD t}=(\div\bmu)\,\delta V,
\end{equation}
as follows from the above equations (\textbf{exercise}).  The second
result is easier to understand: the volume element increases when the
flow is divergent.  These equations are most easily derived using
Cartesian tensor notation.  In this notation the equation for
$\delta\bmS$ reads
\begin{equation}
  \f{\rmD\,\delta S_i}{\rmD t}=\f{\p u_j}{\p x_j}\,\delta S_i-\f{\p u_j}{\p x_i}\,\delta S_j.
\end{equation}

\subsection{Equation of mass conservation}

The equation of mass conservation,
\begin{equation}
  \f{\p\rho}{\p t}+\div(\rho\bmu)=0,
\end{equation}
has the typical form of a conservation law: $\rho$ is the mass density
(mass per unit volume) and $\rho\bmu$ is the mass flux density (mass
flux per unit area).  An alternative form of the same equation is
\begin{equation}
  \f{\rmD\rho}{\rmD t}=-\rho\div\bmu.
\end{equation}
If $\delta m=\rho\,\delta V$ is a material mass element, it can be
seen that mass is conserved in the form
\begin{equation}
  \f{\rmD\,\delta m}{\rmD t}=0.
\end{equation}

\subsection{Equation of motion}

The equation of motion,
\begin{equation}
  \rho\f{\rmD\bmu}{\rmD t}=-\rho\grad\Phi-\grad p,
\end{equation}
derives from Newton's second law per unit volume with gravitational
and pressure forces.  $\Phi(\bmx,t)$ is the gravitational potential
and $\bmg=-\grad\Phi$ is the gravitational field.  The force due to
pressure acting on a volume $V$ with bounding surface $S$ is
\begin{equation}
  -\int_Sp\,\rmd\bmS=\int_V(-\grad p)\,\rmd V.
\end{equation}
Viscous forces are neglected in ideal gas dynamics.

\subsection{Poisson's equation}

The gravitational potential is related to the mass density by
Poisson's equation,
\begin{equation}
  \delsq\Phi=4\pi G\rho,
\end{equation}
where $G$ is Newton's constant.  The solution
\begin{equation}
  \Phi(\bmx,t)=\Phi_\text{int}+\Phi_\text{ext}=-G\int_V\f{\rho(\bmx',t)}{|\bmx'-\bmx|}\,\rmd^3\bmx'-G\int_{\hat V}\f{\rho(\bmx',t)}{|\bmx'-\bmx|}\,\rmd^3\bmx'
\end{equation}
generally involves contributions from both the fluid region $V$ under
consideration and the exterior region $\hat V$.

A \emph{non-self-gravitating} fluid is one of negligible mass for
which $\Phi_\text{int}$ can be neglected.  More generally, the
\emph{Cowling approximation}\footnote{Thomas George Cowling
  (1906--1990), British.} consists of treating $\Phi$ as being
specified in advance, so that Poisson's equation is not coupled to the
other equations.

\subsection{Thermal energy equation}

In the absence of non-adiabatic heating (e.g.\ by viscous dissipation
or nuclear reactions) and cooling (e.g.\ by radiation or conduction),
\begin{equation}
  \f{\rmD s}{\rmD t}=0,
\end{equation}
where $s$ is the \textit{specific entropy} (entropy per unit mass).
Fluid elements undergo reversible thermodynamic changes and preserve
their entropy.

This condition is violated in shocks (see Section~\ref{s:shock}).

The thermal variables $(T,s)$ can be related to the dynamical
variables $(p,\rho)$ via an \emph{equation of state} and standard
thermodynamic identities.  The most important case is that of an
\emph{ideal gas} together with \emph{black-body radiation},
\begin{equation}
  p=p_\text{g}+p_\text{r}=\f{k\rho T}{\mu m_\rmH}+
  \f{4\sigma T^4}{3c},
\end{equation}
where $k$ is Boltzmann's constant, $m_\rmH$ is the mass of the
hydrogen atom, $\sigma$ is Stefan's constant and $c$ is the speed of
light.  $\mu$ is the mean molecular weight (the average mass of the
particles in units of $m_\rmH$), equal to $2.0$ for molecular
hydrogen, $1.0$ for atomic hydrogen, $0.5$ for fully ionized hydrogen
and about $0.6$ for ionized matter of typical cosmic abundances.
Radiation pressure is usually negligible except in the centres of
high-mass stars and in the immediate environments of neutron stars and
black holes.  The pressure of an ideal gas is often written in the
form $\mathcal{R}\rho T/\mu$, where $\mathcal{R}=k/m_\rmH$ is a
version of the universal gas constant.

We define the first \emph{adiabatic exponent}
\begin{equation}
  \Gamma_1=\left(\f{\p\ln p}{\p\ln\rho}\right)_s,
\end{equation}
which is related to the ratio of specific heat capacities
\begin{equation}
  \gamma=\f{c_\rmp}{c_\rmv}=\f{T\displaystyle{\left(\f{\p s}{\p T}\right)_p}}{T\displaystyle{\left(\f{\p s}{\p T}\right)_v}}
\end{equation}
by (\textbf{exercise})
\begin{equation}
  \Gamma_1=\chi_\rho\gamma,
\end{equation}
where
\begin{equation}
  \chi_\rho=\left(\f{\p\ln p}{\p\ln\rho}\right)_T
\end{equation}
can be found from the equation of state.  We can then rewrite the
thermal energy equation as
\begin{equation}
  \f{\rmD p}{\rmD t}=\f{\Gamma_1p}{\rho}\f{\rmD\rho}{\rmD t}=-\Gamma_1p\div\bmu.
\end{equation}

For an ideal gas with negligible radiation pressure, $\chi_\rho=1$ and
so $\Gamma_1=\gamma$.  Adopting this very common assumption, we write
\begin{equation}
  \f{\rmD p}{\rmD t}=-\gamma p\div\bmu.
\end{equation}

\subsection{Simplified models}

A \emph{perfect gas} may be defined as an ideal gas with constant
$c_\text{v}$, $c_\rmp$, $\gamma$ and $\mu$.  Equipartition of energy
for a classical gas with $n$ degrees of freedom per particle gives
$\gamma=1+2/n$.  For a classical monatomic gas with $n=3$
translational degrees of freedom, $\gamma=5/3$.  This is relevant for
fully ionized matter.  For a classical diatomic gas with two
additional rotational degrees of freedom, $n=5$ and $\gamma=7/5$.
This is relevant for molecular hydrogen.  In reality $\Gamma_1$ is
variable when the gas undergoes ionization or when the gas and
radiation pressures are comparable.  The specific \textit{internal
  energy} (or \textit{thermal energy}) of a perfect gas is
\begin{equation}
  e=\f{p}{(\gamma-1)\rho}\qquad\left[=\f{n}{\mu m_\rmH}\half kT\right].
\end{equation}
(Note that each particle has an internal energy of $\half kT$ per
degree of freedom, and the number of particles per unit mass is $1/\mu
m_\rmH$.)

A \emph{barotropic fluid} is an idealized situation in which the
relation $p(\rho)$ is known in advance.  We can then dispense with the
thermal energy equation.  e.g.\ if the gas is strictly isothermal and
perfect, then $p=c_\text{s}^2\rho$ with $c_\text{s}=\cst$ being the
isothermal sound speed.  Alternatively, if the gas is strictly
homentropic and perfect, then $p=K\rho^\gamma$ with $K=\cst$.

An \emph{incompressible fluid} is an idealized situation in which
$\rmD\rho/\rmD t=0$, implying $\div\bmu=0$.  This can be achieved
formally by taking the limit $\gamma\to\infty$.  The approximation of
incompressibility eliminates acoustic phenomena from the dynamics.

The ideal gas law itself is not valid at very high densities or where
quantum degeneracy is important.

\subsection{Microphysical basis}

It is useful to understand the way in which the fluid-dynamical
equations are derived from microphysical considerations.  The simplest
model involves identical neutral particles of mass $m$ of negligible
size with no internal degrees of freedom.

\subsubsection{The Boltzmann equation}

Between collisions, particles follow Hamiltonian trajectories in their
six-dimensional $(\bmx,\bmv)$ phase space:
\begin{equation}
  \dot x_i=v_i,\qquad
  \dot v_i=a_i=-\f{\p\Phi}{\p x_i}.
\end{equation}
The \textit{distribution function} $f(\bmx,\bmv,t)$ specifies the
number density of particles in phase space.  The velocity moments of
$f$ define the number density $n(\bmx,t)$ in real space, the bulk
velocity $\bmu(\bmx,t)$ and the velocity dispersion $c(\bmx,t)$
according to
\begin{equation}
  \int f\,\rmd^3\bmv=n,\qquad
  \int \bmv f\,\rmd^3\bmv=n\bmu,\qquad
  \int|\bmv-\bmu|^2f\,\rmd^3\bmv=3nc^2.
\end{equation}
Equivalently,
\begin{equation}
  \int v^2f\,\rmd^3\bmv=n(u^2+3c^2).
\end{equation}
The relation between velocity dispersion and temperature is $kT=mc^2$.

In the absence of collisions, $f$ is conserved following the
Hamiltonian flow in phase space.  This is because particles are
conserved and the flow in phase space is incompressible (Liouville's
theorem).  More generally, $f$ evolves according to \emph{Boltzmann's
  equation},
\begin{equation}
  \f{\p f}{\p t}+v_j\f{\p f}{\p x_j}+a_j\f{\p f}{\p v_j}=\left(\f{\p f}{\p t}\right)_{\rm c}.
\end{equation}
The collision term on the right-hand side is a complicated integral
operator but has three simple properties corresponding to the
conservation of mass, momentum and energy in collisions:
\begin{equation}
  \int m\left(\f{\p f}{\p t}\right)_{\rm c}\,\rmd^3\bmv=0,\qquad
  \int m\bmv\left(\f{\p f}{\p t}\right)_{\rm c}\,\rmd^3\bmv=\bfzero,\qquad
  \int\half mv^2\left(\f{\p f}{\p t}\right)_{\rm c}\,\rmd^3\bmv=0.
\end{equation}
The collision term is local in $\bmx$ (not even involving derivatives)
although it does involve integrals over $\bmv$.  The equation $(\p
f/\p t)_{\rm c}=0$ has the general solution
\begin{equation}
  f=f_\text{M}=(2\pi c^2)^{-3/2}n\,\exp\left(-\f{|\bmv-\bmu|^2}{2c^2}\right),
\end{equation}
with parameters $n$, $\bmu$ and $c$ that may depend on $\bmx$.  This
is the \emph{Maxwellian distribution}.

\subsubsection{Derivation of fluid equations}

A crude but illuminating model of the collision operator is the
\textit{BGK approximation}
\begin{equation}
  \left(\f{\p f}{\p t}\right)_{\rm c}\approx-\f{1}{\tau}(f-f_\text{M})
\end{equation}
where $f\rmm$ is a Maxwellian distribution with the same $n$,
$\bmu$ and $c$ as $f$, and $\tau$ is the \textit{relaxation time}.
This can be identified approximately with the mean free flight time of
particles between collisions.  In other words the collisions attempt
to restore a Maxwellian distribution on a characteristic time-scale
$\tau$.  They do this by randomizing the particle velocities in a way
consistent with the conservation of momentum and energy.

If the characteristic time-scale of the fluid flow is much greater
than $\tau$, then the collision term dominates the Boltzmann equation
and $f$ must be very close to $f_\text{M}$.  This is the
\textit{hydrodynamic limit}.

The velocity moments of $f_\rmM$ can be determined from standard
Gaussian integrals, in particular (\textbf{exercise})
\begin{equation}
  \int f_\rmM\,\rmd^3\bmv=n,\qquad
  \int v_if_\rmM\,\rmd^3\bmv=nu_i,
\end{equation}
\begin{equation}
  \int v_iv_jf_\rmM\,\rmd^3\bmv=n(u_iu_j+c^2\delta_{ij}),\qquad
  \int v^2v_if_\rmM\,\rmd^3\bmv=n(u^2+5c^2)u_i.
\end{equation}

We obtain equations for mass, momentum and energy by taking moments of
the Boltzmann equation weighted by $(m,mv_i,\half mv^2)$.  In each
case the collision term integrates to zero because of its conservative
properties, and the $\p/\p v_j$ term can be integrated by parts.  We
replace $f$ with $f_\rmM$ when evaluating the left-hand sides and
note that $mn=\rho$:
\begin{equation}
  \f{\p\rho}{\p t}+\f{\p}{\p x_i}(\rho u_i)=0,
\end{equation}
\begin{equation}
  \f{\p}{\p t}(\rho u_i)+\f{\p}{\p x_j}\left[\rho(u_iu_j+c^2\delta_{ij})\right]-\rho a_i=0,
\end{equation}
\begin{equation}
  \f{\p}{\p t}\left(\half\rho u^2+\tfrac{3}{2}\rho c^2\right)+\f{\p}{\p x_i}\left[(\half\rho u^2+\tfrac{5}{2}\rho c^2)u_i\right]-\rho u_ia_i=0.
\end{equation}
These are equivalent to the equations of ideal gas dynamics in
conservative form (see Section~\ref{s:conservation}) for a monatomic
ideal gas ($\gamma=5/3$).  The specific internal energy is
$e=\tfrac{3}{2}c^2=\tfrac{3}{2}kT/m$.

This approach can be generalized to deal with molecules with internal
degrees of freedom and also to plasmas or partially ionized gases
where there are various species of particle with different charges and
masses.  The equations of MHD can be derived by including the
electromagnetic forces in Boltzmann's equation.

\subsubsection{Validity of a fluid approach}
\label{s:validity}

The essential idea here is that deviations from the Maxwellian
distribution are small when collisions are frequent compared to the
characteristic time-scale of the flow.  In higher-order approximations
these deviations can be estimated, leading to the equations of
\textit{dissipative gas dynamics} including \textit{transport effects}
(viscosity and heat conduction).

The fluid approach breaks down if the mean flight time $\tau$ is not
much less than the characteristic time-scale of the flow, or if the
mean free path $\lambda\approx c\tau$ between collisions is not much
less than the characteristic length-scale of the flow.  $\lambda$ can
be very long (measured in AU or pc) in very tenuous gases such as the
interstellar medium, but may still be smaller than the size of the
system.

Some typical order-of-magnitude estimates:

Solar-type star: centre $\rho\sim10^2\,\text{g}\,\text{cm}^{-3}$,
$T\sim10^7\,\text{K}$; photosphere $\rho\sim10^{-7}\,\text{g}\,\text{cm}^{-3}$, $T\sim10^4\,\text{K}$; corona
$\rho\sim10^{-15}\,\text{g}\,\text{cm}^{-3}$, $T\sim10^6\,\text{K}$.

Interstellar medium: molecular clouds $n\sim10^3\,\text{cm}^{-3}$,
$T\sim10\,\text{K}$; cold medium (neutral)
$n\sim10-100\,\text{cm}^{-3}$, $T\sim10^2\,K$; warm medium
(neutral/ionized) $n\sim0.1-1\,\text{cm}^{-3}$, $T\sim10^4\,K$; hot
medium (ionized) $n\sim10^{-3}-10^{-2}\,\text{cm}^{-3}$,
$T\sim10^6\,K$.

The Coulomb cross-section for `collisions' (i.e.\ large-angle
scatterings) between charged particles (electrons or ions) is
$\sigma\approx1\times10^{-4}(T/\text{K})^{-2}\,\text{cm}^2$.  The mean
free path is $\lambda=1/(n\sigma)$.

\medskip
\noindent Related examples (see Appendix~\ref{s:examples}):
\ref{e:collisions}, \ref{e:vorticity}, \ref{e:expansion},
\ref{e:ellipsoidal}.

\section{Ideal magnetohydrodynamics}

\subsection{Elementary derivation of the MHD equations}

\emph{Magnetohydrodynamics} (MHD) is the dynamics of an electrically
conducting fluid (a fully or partially ionized gas or a liquid metal)
containing a magnetic field.  It is a fusion of fluid dynamics and
electromagnetism.

\subsubsection{Galilean electromagnetism}

The equations of Newtonian gas dynamics are invariant under
  the \emph{Galilean transformation} to a frame of reference moving
  with uniform velocity $\bmv$,
\begin{equation}
  \bmx'=\bmx-\bmv t,\qquad
  t'=t.
\end{equation}
Under this change of frame, the fluid velocity transforms according to
\begin{equation}
  \bmu'=\bmu-\bmv,
\end{equation}
while scalar variables such as $p$, $\rho$ and $\Phi$ are invariant.
The Lagrangian time-derivative $\rmD/\rmD t$ is also invariant,
because the partial derivatives transform according to
\begin{equation}
  \grad'=\grad,\qquad
  \f{\p}{\p t'}=\f{\p}{\p t}+\bmv\bcdot\grad.
\end{equation}

In Maxwell's electromagnetic theory the electric and magnetic fields
$\bmE$ and $\bmB$ are governed by the equations
\begin{equation}
  \f{\p\bmB}{\p t}=-\curl\bmE,\qquad
  \div\bmB=0,\qquad
  \curl\bmB=\mu_0\left(\bmJ+\epsilon_0\f{\p\bmE}{\p t}\right),\qquad
  \div\bmE=\f{\rho_\rme}{\epsilon_0},
\end{equation}
where $\mu_0$ and $\epsilon_0$ are the vacuum permeability and
permittivity, $\bmJ$ is the electric current density and $\rho_\rme$
is the electric charge density.  (In these notes we use rationalized
(e.g.\ SI) units for electromagnetism.  In astrophysics it is also
common to use Gaussian units, which are discussed in
Appendix~\ref{s:gaussian}.)

It is well known that Maxwell's equations are invariant under the
Lorentz transformation of special relativity, with
$c=(\mu_0\epsilon_0)^{-1/2}$ being the speed of light.  These
equations cannot be consistently coupled with those of Newtonian gas
dynamics, which are invariant under the Galilean transformation.  To
derive a consistent Newtonian theory of MHD, valid for situations in
which the fluid motions are slow compared to the speed of light, we
must use Maxwell's equations without the displacement current
$\epsilon_0\,\p\bmE/\p t$,
\begin{equation}
  \f{\p\bmB}{\p t}=-\curl\bmE,\qquad
  \div\bmB=0,\qquad
  \curl\bmB=\mu_0\bmJ.
\end{equation}  
(We will not require the fourth Maxwell equation, involving
$\div\bmE$, because the charge density will be found to be
unimportant.)  It is easily verified (\textbf{exercise}) that these
\emph{pre-Maxwell equations}\footnote{It was by introducing the
  displacement current that Maxwell identified electromagnetic waves,
  so it is appropriate that a highly subluminal approximation should
  neglect this term.} are indeed invariant under the Galilean
transformation, provided that the fields transform according to
\begin{equation}
  \bmE'=\bmE+\bmv\btimes\bmB,\qquad
  \bmB'=\bmB,\qquad
  \bmJ'=\bmJ.
\end{equation}
These relations correspond to the limit of the Lorentz transformation
for electromagnetic fields\footnote{This was called the \emph{magnetic
    limit} of Galilean electromagnetism by \citet{LeBellac73}.} when $|\bmv|\ll c$ and $|\bmE|\ll c|\bmB|$.

Under the pre-Maxwell theory, the equation of charge conservation
takes the simplified form $\div\bmJ=0$; this is analogous to the use
of $\div\bmu=0$ as the equation of mass conservation in the
incompressible (highly subsonic) limit of gas dynamics.  The equation
of energy conservation takes the simplified form
\begin{equation}
  \f{\p}{\p t}\left(\f{B^2}{2\mu_0}\right)+\div\left(\f{\bmE\btimes\bmB}{\mu_0}\right)=0,
\end{equation}
in which the energy density, $B^2/2\mu_0$, is purely magnetic (because
$|\bmE|\ll c|\bmB|$), while the energy flux density has the usual form
of the \emph{Poynting vector} $\bmE\btimes\bmB/\mu_0$.  We will verify
the self-consistency of the approximations made in Newtonian MHD in
Section~\ref{s:consistency}.

\subsubsection{Induction equation}

In the \emph{ideal MHD approximation} we regard the fluid as a perfect
electrical conductor.  The electric field in the rest frame of the
fluid therefore vanishes, implying that
\begin{equation}
  \bmE=-\bmu\btimes\bmB
\label{e}
\end{equation}
in a frame in which the fluid velocity is $\bmu(\bmx,t)$.  This
condition can be regarded as the limit of a constitutive relation such
as Ohm's law, in which the effects of resistivity (i.e.\ finite
conductivity) are neglected.

From Maxwell's equations, we then obtain the ideal \emph{induction
  equation},
\begin{equation}
  \f{\p\bmB}{\p t}=\curl(\bmu\btimes\bmB).
\end{equation}
This is an evolutionary equation for $\bmB$ alone, $\bmE$ and $\bmJ$
having been eliminated.  The divergence of the induction equation,
\begin{equation}
  \f{\p}{\p t}(\div\bmB)=0,
\end{equation}
ensures that the solenoidal character of $\bmB$ is preserved.

\subsubsection{The Lorentz force}

A fluid carrying a current density $\bmJ$ in a magnetic field $\bmB$
experiences a bulk \emph{Lorentz force}
\begin{equation}
  \bmF_\rmm=\bmJ\btimes\bmB=\f{1}{\mu_0}(\curl\bmB)\btimes\bmB
\end{equation}
per unit volume.  This can be understood as the sum of the Lorentz
forces on individual particles of charge $q$ and velocity $\bmv$,
\begin{equation}
  \sum q\bmv\btimes\bmB=\left(\sum q\bmv\right)\btimes\bmB.
\end{equation}
(The electrostatic force can be shown to be negligible in the limit
relevant to Newtonian MHD; see Section~\ref{s:consistency}.)

In Cartesian coordinates
\begin{equation}
\begin{split}
  (\mu_0\bmF_\rmm)_i&=\epsilon_{ijk}\left(\epsilon_{jlm}\f{\p B_m}{\p x_l}\right)B_k\\
  &=\left(\f{\p B_i}{\p x_k}-\f{\p B_k}{\p x_i}\right)B_k\\
  &=B_k\f{\p B_i}{\p x_k}-\f{\p}{\p x_i}\left(\f{B^2}{2}\right).
\end{split}
\end{equation}
Thus
\begin{equation}
  \bmF_\rmm=\f{1}{\mu_0}\bmB\bcdot\grad\bmB-
  \grad\left(\f{B^2}{2\mu_0}\right).
\end{equation}
The first term can be interpreted as a \emph{curvature force} due to a
\emph{magnetic tension} $T_\rmm=B^2/\mu_0$ per unit area in the field
lines; if the field is of constant magnitude then this term is equal
to $T_\rmm$ times the curvature of the field lines, and is directed
towards the centre of curvature.  The second term is the gradient of
an isotropic \emph{magnetic pressure}
\begin{equation}
  p_\rmm=\f{B^2}{2\mu_0},
\end{equation}
which is also equal to the energy density of the magnetic field.

The magnetic tension gives rise to \emph{Alfv\'en
  waves}\footnote{Hannes Olof G\"osta Alfv\'en (1908--1995), Swedish.
  Nobel Prize in Physics (1970) `for fundamental work and discoveries
  in magnetohydro-dynamics with fruitful applications in different
  parts of plasma physics'.} (see later), which travel parallel to the
magnetic field with characteristic speed
\begin{equation}
  v_{\rm a}=\left(\f{T_\rmm}{\rho}\right)^{1/2}
  =\f{B}{(\mu_0\rho)^{1/2}},
\end{equation}
the \emph{Alfv\'en speed}.  This is often considered as a vector
\emph{Alfv\'en velocity},
\begin{equation}
  \bmv_{\rm a}=\f{\bmB}{(\mu_0\rho)^{1/2}}.
\end{equation}
The magnetic pressure also affects the propagation of sound waves,
which become \emph{magnetoacoustic waves} (or \emph{magnetosonic
  waves}; see later).

The combination
\begin{equation}
  \Pi=p+\f{B^2}{2\mu_0}
\end{equation}
is often referred to as the \emph{total pressure}, while the ratio
\begin{equation}
  \beta=\f{p}{B^2/2\mu_0}
\end{equation}
is known as the \emph{plasma beta}.

\subsubsection{Self-consistency of approximations}
\label{s:consistency}

Three effects neglected in a Newtonian theory of MHD are (i) the
displacement current in Maxwell's equations (compared to the electric
current), (ii) the bulk electrostatic force on the fluid (compared to
the magnetic Lorentz force) and (iii) the electrostatic energy
(compared to the magnetic energy).  We can verify the self-consistency
of these approximations by using order-of-magnitude estimates or
scaling relations.  If the fluid flow has a characteristic
length-scale $L$, time-scale $T$, velocity $U\sim L/T$ and magnetic
field $B$, then the electric field can be estimated from
equation~(\ref{e}) as $E\sim UB$.  The electric current density and
charge density can be estimated from Maxwell's equations as
$J\sim\mu_0^{-1}B/L$ and $\rho_\rme\sim\epsilon_0E/L$.  Hence the
ratios of the three neglected effects to the terms that are retained
in Newtonian MHD can be estimated as follows:
\begin{equation}
  \f{\epsilon_0|\p\bmE/\p t|}{|\bmJ|}\sim\f{\epsilon_0UB/T}{\mu_0^{-1}B/L}\sim\f{U^2}{c^2},
\end{equation}
\begin{equation}
  \f{|\rho_\rme\bmE|}{|\bmJ\btimes\bmB|}\sim\f{\epsilon_0E^2/L}{\mu_0^{-1}B^2/L}\sim\f{U^2}{c^2},
\end{equation}
\begin{equation}
  \f{\epsilon_0|\bmE|^2/2}{|\bmB|^2/2\mu_0}\sim\f{U^2}{c^2}.
\end{equation}
Therefore Newtonian MHD corresponds to a consistent approximation of
relativistic MHD for highly subluminal flows that is correct to the
leading order in the small parameter $U^2/c^2$.

\subsubsection{Summary of the MHD equations}

The full set of ideal MHD equations is
\begin{equation}
  \f{\p\rho}{\p t}+\div(\rho\bmu)=0,
\end{equation}
\begin{equation}
  \rho\f{\rmD\bmu}{\rmD t}=-\rho\grad\Phi-\grad p+
  \f{1}{\mu_0}(\curl\bmB)\btimes\bmB,
\end{equation}
\begin{equation}
  \f{\rmD s}{\rmD t}=0,
\end{equation}
\begin{equation}
  \f{\p\bmB}{\p t}=\curl(\bmu\btimes\bmB),
\end{equation}
\begin{equation}
  \div\bmB=0,
\end{equation}
together with the equation of state, Poisson's equation, etc., as
required.  Most of these equations can be written in at least one
other way that may be useful in different circumstances.

These equations display the essential \emph{nonlinearity} of MHD.
When the velocity field is prescribed, an artifice known as the
\emph{kinematic approximation}, the induction equation is a relatively
straightforward linear evolutionary equation for the magnetic field.
However, a sufficiently strong magnetic field will modify the velocity
field through its dynamical effect, the Lorentz force.  This nonlinear
coupling leads to a rich variety of behaviour.  Of course, the purely
hydrodynamic nonlinearity of the $\bmu\bcdot\grad\bmu$ term, which is
responsible for much of the complexity of fluid dynamics, is still
present.

\subsection{Physical interpretation of MHD}

There are two aspects to MHD: the advection of $\bmB$ by
$\bmu$ (induction equation) and the dynamical back-reaction of $\bmB$
on $\bmu$ (Lorentz force).

\subsubsection{Kinematics of the magnetic field}

The ideal induction equation
\begin{equation}
  \f{\p\bmB}{\p t}=\curl(\bmu\btimes\bmB)
\end{equation}
has a beautiful geometrical interpretation: the magnetic
field lines are `frozen in' to the fluid and can be identified with
material curves.  This is sometimes known as \emph{Alfv\'en's theorem}.

One way to show this result is to use the identity
\begin{equation}
  \curl(\bmu\btimes\bmB)=\bmB\bcdot\grad\bmu-\bmB(\div\bmu)-\bmu\bcdot\grad\bmB+\bmu(\div\bmB)
\end{equation}
to write the induction equation in the form
\begin{equation}
  \f{\rmD\bmB}{\rmD t}=\bmB\bcdot\grad\bmu-\bmB(\div\bmu),
\end{equation}
and use the equation of mass conservation,
\begin{equation}
  \f{\rmD\rho}{\rmD t}=-\rho\div\bmu,
\end{equation}
to obtain
\begin{equation}
  \f{\rmD}{\rmD t}\left(\f{\bmB}{\rho}\right)=
  \left(\f{\bmB}{\rho}\right)\bcdot\grad\bmu.
\end{equation}
This is exactly the same equation satisfied by a material line element
$\delta\bmx$ (equation~\ref{ddxdt}).  Therefore a magnetic field line
(an integral curve of $\bmB/\rho$) is advected and distorted by the
fluid in the same way as a material curve.

A complementary property is that the magnetic flux
$\delta\Phi=\bmB\bcdot\delta\bmS$ through a material surface element
is conserved:
\begin{equation}
\begin{split}
  \f{\rmD\,\delta\Phi}{\rmD t}&=\f{\rmD\bmB}{\rmD t}\bcdot\delta\bmS+\bmB\bcdot\f{\rmD\,\delta\bmS}{\rmD t}\\
  &=\left(B_j\f{\p u_i}{\p x_j}-B_i\f{\p u_j}{\p x_j}\right)\delta S_i+B_i\left(\f{\p u_j}{\p x_j}\,\delta S_i-\f{\p u_j}{\p x_i}\,\delta S_j\right)\\
  &=0.
\end{split}
\end{equation}
By extension, we have conservation of the magnetic flux passing
through any material surface.

Precisely the same equation as the ideal induction equation,
\begin{equation}
  \f{\p\bomega}{\p t}=\curl(\bmu\btimes\bomega),
\end{equation}
is satisfied by the \emph{vorticity} $\bomega=\curl\bmu$ in
homentropic or barotropic ideal fluid dynamics in the absence of a
magnetic field, in which case the vortex lines are `frozen in' to the
fluid (see Example~\ref{e:vorticity}).  The conserved quantity that is
analogous to the magnetic flux through a material surface is the flux
of vorticity through that surface, which, by Stokes's theorem, is
equivalent to the circulation $\oint\bmu\bcdot\rmd\bmx$ around the
bounding curve.  However, the fact that $\bomega$ and $\bmu$ are
directly related by the curl operation, whereas in MHD $\bmB$ and
$\bmu$ are indirectly related through the equation of motion and the
Lorentz force, means that the analogy between vorticity dynamics and
MHD is limited in scope.

\medskip
\noindent Related examples: \ref{e:resistive}, \ref{e:euler}.

\subsubsection{The Lorentz force}

The Lorentz force per unit volume,
\begin{equation}
  \bmF_\text{m}=\f{1}{\mu_0}\bmB\bcdot\grad\bmB-\grad\left(\f{B^2}{2\mu_0}\right),
\end{equation}
can also be written as the divergence of the \textit{Maxwell stress tensor}:
\begin{equation}
  \bmF_\text{m}=\div\bfM,\qquad
  \bfM=\f{1}{\mu_0}\left(\bmB\bmB-\f{B^2}{2}\bfI\right),
\end{equation}
where $\bfI$ is the identity tensor.  (The electric part of the
electromagnetic stress tensor is negligible in the limit relevant for
Newtonian MHD, for the same reason that the electrostatic energy is
negligible.)  In Cartesian coordinates
\begin{equation}
  (\bmF_\text{m})_i=\f{\p M_{ji}}{\p x_j},\qquad
  M_{ij}=\f{1}{\mu_0}\left(B_iB_j-\f{B^2}{2}\delta_{ij}\right).
\end{equation}
If the magnetic field is locally aligned with the $x$-axis, then
\begin{equation}
  \bfM=\begin{bmatrix}T_\text{m}&0&0\\0&0&0\\0&0&0\end{bmatrix}-\begin{bmatrix}p_\text{m}&0&0\\0&p_\text{m}&0\\0&0&p_\text{m}\end{bmatrix},
\end{equation}
showing the magnetic tension and pressure.

Combining the ideas of magnetic tension and a frozen-in field leads to
the picture of field lines as elastic strings embedded in the fluid.
Indeed there is a close analogy between MHD and the dynamics of dilute
solutions of long-chain polymer molecules.  The magnetic field imparts
elasticity to the fluid.

\subsubsection{Differential rotation and torsional Alfv\'en waves}

We first consider the kinematic behaviour of a magnetic field in the
presence of a prescribed velocity field involving \textit{differential
rotation}.  In cylindrical polar coordinates $(r,\phi,z)$, let
\begin{equation}
  \bmu=r\Omega(r,z)\,\bme_\phi.
\end{equation}
Consider an axisymmetric magnetic field, which we separate into
\textit{poloidal} (meridional: $r$ and $z$) and \textit{toroidal}
(azimuthal: $\phi$) parts:
\begin{equation}
  \bmB=\bmB_\rmp(r,z,t)+B_\phi(r,z,t)\,\bme_\phi.
\end{equation}
The ideal induction equation reduces to (\textbf{exercise})
\begin{equation}
  \f{\p\bmB_\rmp}{\p t}=0,\qquad
  \f{\p B_\phi}{\p t}=r\bmB_\rmp\bcdot\grad\Omega.
\end{equation}
Differential rotation winds the poloidal field to generate a toroidal
field.  To obtain a steady state without winding, we require the
angular velocity to be constant along each magnetic field line:
\begin{equation}
  \bmB_\rmp\bcdot\grad\Omega=0,
\end{equation}
a result known as \textit{Ferraro's law of
  isorotation}\footnote{Vincenzo Ferraro (1902--1974), British.}.

There is an energetic cost to winding the field, as work is done
against magnetic tension.  In a dynamical situation a strong magnetic
field tends to enforce isorotation along its length.

We now generalize the analysis to allow for axisymmetric
\textit{torsional oscillations}:
\begin{equation}
  \bmu=r\Omega(r,z,t)\,\bme_\phi.
\end{equation}
The azimuthal component of equation of motion is (\textbf{exercise})
\begin{equation}
  \rho r\f{\p\Omega}{\p t}=\f{1}{\mu_0r}\bmB_\rmp\bcdot\grad(rB_\phi).
\end{equation}
This combines with the induction equation to give
\begin{equation}
  \f{\p^2\Omega}{\p t^2}=\f{1}{\mu_0\rho r^2}\bmB_\rmp\bcdot\grad(r^2\bmB_\rmp\bcdot\grad\Omega).
\end{equation}
This equation describes \textit{torsional Alfv\'en waves}.  For
example, if $\bmB_\rmp=B_z\,\bme_z$ is vertical and uniform, then
\begin{equation}
  \f{\p^2\Omega}{\p t^2}=v_\text{a}^2\f{\p^2\Omega}{\p z^2}.
\end{equation}
This is not strictly an exact nonlinear analysis because we have
neglected the force balance (and indeed motion) in the meridional
plane.

\subsubsection{Force-free fields}

In regions of low density, such as the solar corona, the magnetic
field may be dynamically dominant over the effects of inertia, gravity
and gas pressure.  Under these circumstances we have (approximately) a
\emph{force-free magnetic field} such that
\begin{equation}
  (\curl\bmB)\btimes\bmB=\bfzero.
\end{equation}
Vector fields $\bmB$ satisfying this equation are known in a wider
mathematical context as \emph{Beltrami fields}.  Since $\curl\bmB$
must be parallel to $\bmB$, we may write
\begin{equation}
  \curl\bmB=\lambda\bmB,
\label{beltrami}
\end{equation}
for some scalar field $\lambda(\bmx)$.  The divergence of this equation is
\begin{equation}
  0=\bmB\bcdot\grad\lambda,
\end{equation}
so that $\lambda$ is constant along each magnetic field line.  In the
special case $\lambda=\cst$, known as a \emph{linear force-free
  magnetic field}, the curl of equation~(\ref{beltrami}) results in
the Helmholtz equation
\begin{equation}
  -\delsq\bmB=\lambda^2\bmB,
\end{equation}
which admits a wide variety of non-trivial solutions.

A subset of force-free magnetic fields consists of \emph{potential}
or \emph{current-free} magnetic fields for which
\begin{equation}
  \curl\bmB=\bfzero.
\end{equation}
In a true vacuum, the magnetic field must be potential.  However, only
an extremely low density of charge carriers (i.e.\ electrons) is
needed to make the force-free description more relevant.

\begin{figure}
  \centerline{\epsfysize9cm\epsfbox{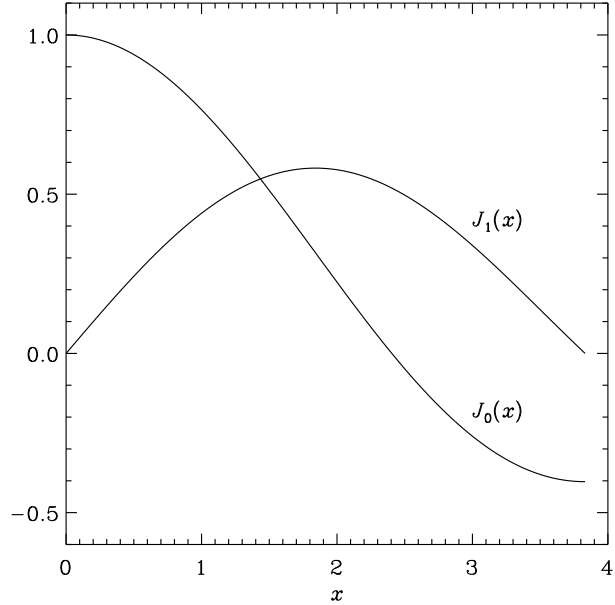}}
  \caption{The Bessel functions $J_0(x)$ and $J_1(x)$ from the origin
    to the first zero of $J_1$.}
  \label{f:bessel}
\end{figure}

An example of a force-free field in cylindrical polar coordinates
$(r,\phi,z)$ is
\begin{equation}
\begin{split}
  \bmB&=B_\phi(r)\,\bme_\phi+B_z(r)\,\bme_z,\\
  \curl\bmB&=-\f{\rmd B_z}{\rmd r}\,\bme_\phi+\f{1}{r}\f{\rmd}{\rmd r}(rB_\phi)\,\bme_z.
\end{split}
\end{equation}
Now $\curl\bmB=\lambda\bmB$ implies
\begin{equation}
  -\f{1}{r}\f{\rmd}{\rmd r}\left(r\f{\rmd B_z}{\rmd r}\right)=\lambda^2B_z,
\end{equation}
which is the $z$ component of the Helmholtz equation.  The solution
regular at $r=0$ is
\begin{equation}
  B_z=B_0J_0(\lambda r),\qquad
  B_\phi=B_0J_1(\lambda r),
\end{equation}
where $J_n$ is the Bessel function of order $n$
(Figure~\ref{f:bessel}).  [Note that $J_0(x)$ satisfies
$(xJ_0')'+xJ_0=0$ and $J_1(x)=-J_0'(x)$.] The helical nature of this
field is typical of force-free fields with $\lambda\ne0$.

When applied to a infinite cylinder (e.g.\ as a simplified model of a
magnetized astrophysical jet), the solution could be extended from the
axis to the first zero of $J_1$ and then matched to a uniform external
axial field $B_z$.  In this case the net axial current is zero.
Alternatively the solution could be extended from the axis to the
first zero of $J_0$ and matched to an external azimuthal field
$B_\phi\propto r^{-1}$ generated by the net axial current.

\subsubsection{Magnetostatic equilibrium and magnetic buoyancy}

A \emph{magnetostatic equilibrium} is a static solution
$(\bmu=\bfzero)$ of the equation of motion, i.e.\ one satisfying
\begin{equation}
  \bfzero=-\rho\grad\Phi-\grad p+
  \f{1}{\mu_0}(\curl\bmB)\btimes\bmB,
\end{equation}
together with $\div\bmB=0$.

While it is possible to find solutions in which the forces balance in
this way, inhomogeneities in the magnetic field typically result in a
lack of equilibrium.  A \textit{magnetic flux tube} is an idealized
situation in which the magnetic field is localized to the interior of
a tube and vanishes outside.  To balance the total pressure at the
interface, the gas pressure must be lower inside.  Unless the
temperatures are different, the density is lower inside.  In a
gravitational field the tube therefore experiences an upward buoyancy
force and tends to rise.

\begin{figure}
  \centerline{\epsfysize5cm\epsfbox{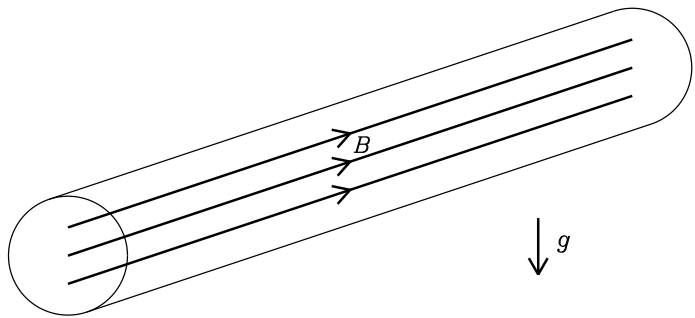}}
  \caption{A buoyant magnetic flux tube.}
\label{f:tube}
\end{figure}

\medskip
\noindent Related examples: \ref{e:prominence}, \ref{e:equilibrium},
\ref{e:fff}.

\section{Conservation laws, symmetries and hyperbolic structure}
\label{s:conservation}

\subsection{Introduction}

There are various ways in which a quantity can be said to be
`conserved' in fluid dynamics or MHD.  If a quantity has a density
(amount per unit volume) $q(\bmx,t)$ that satisfies an equation of the
conservative form
\begin{equation}
  \f{\p q}{\p t}+\div\bmF=0,
\end{equation}
then the vector field $\bmF(\bmx,t)$ can be identified as the flux
density (flux per unit area) of the quantity.  The rate of change of
the total amount of the quantity in a time-independent volume $V$,
\begin{equation}
  Q=\int_Vq\,\rmd V,
\end{equation}
is then equal to minus the flux of $\bmF$ through the bounding surface
$S$:
\begin{equation}
  \f{\rmd Q}{\rmd t}=-\int_V(\div\bmF)\,\rmd V=-\int_S\bmF\bcdot\rmd\bmS.
\end{equation}
If the boundary conditions on $S$ are such that this flux vanishes,
then $Q$ is constant; otherwise, changes in $Q$ can be accounted for
by the flux of $\bmF$ through $S$.  In this sense the quantity is said
to be conserved.  The prototype is mass, for which $q=\rho$ and
$\bmF=\rho\bmu$.

A \emph{material invariant} is a scalar field $f(\bmx,t)$ for which
\begin{equation}
  \f{\rmD f}{\rmD t}=0,
\end{equation}
which implies that $f$ is constant for each fluid element, and is
therefore conserved following the fluid motion.  A simple example is
the specific entropy in ideal fluid dynamics.  When combined with mass
conservation, this yields an equation in conservative form,
\begin{equation}
  \f{\p}{\p t}(\rho f)+\div(\rho f\bmu)=0.
\end{equation}

\subsection{Synthesis of the total energy equation}

Starting from the ideal MHD equations, we construct the total energy
equation piece by piece.

\textit{Kinetic energy}:
\begin{equation}
  \rho\f{\rmD}{\rmD t}(\half u^2)=\rho\bmu\bcdot\f{\rmD\bmu}{\rmD t}=-\rho\bmu\bcdot\grad\Phi-\bmu\bcdot\grad p+\f{1}{\mu_0}\bmu\bcdot\left[(\curl\bmB)\btimes\bmB\right].
\end{equation}
\textit{Gravitational energy} (assuming initially that the system is
non-self-gravitating and that $\Phi$ is independent of $t$):
\begin{equation}
  \rho\f{\rmD\Phi}{\rmD t}=\rho\bmu\bcdot\grad\Phi.
\end{equation}
\textit{Internal (thermal) energy} (using the fundamental
thermodynamic identity $\rmd e=T\,\rmd s-p\,\rmd v$):
\begin{equation}
  \rho\f{\rmD e}{\rmD t}=\rho T\f{\rmD s}{\rmD t}+p\f{\rmD\ln\rho}{\rmD t}=-p\div\bmu.
\end{equation}
Sum of these three:
\begin{equation}
  \rho\f{\rmD}{\rmD t}(\half u^2+\Phi+e)=-\div(p\bmu)+\f{1}{\mu_0}\bmu\bcdot\left[(\curl\bmB)\btimes\bmB\right].
\end{equation}
The last term can be rewritten as
\begin{equation}
  \f{1}{\mu_0}\bmu\bcdot\left[(\curl\bmB)\btimes\bmB\right]=\f{1}{\mu_0}(\curl\bmB)\bcdot(-\bmu\btimes\bmB)=\f{1}{\mu_0}(\curl\bmB)\bcdot\bmE.
\end{equation}
Using mass conservation:
\begin{equation}
  \f{\p}{\p t}\left[\rho(\half u^2+\Phi+e)\right]+
  \div\left[\rho\bmu(\half u^2+\Phi+e)+p\bmu\right]=
  \f{1}{\mu_0}(\curl\bmB)\bcdot\bmE.
\end{equation}
\textit{Magnetic energy}:
\begin{equation}
  \f{\p}{\p t}\left(\f{B^2}{2\mu_0}\right)=\f{1}{\mu_0}\bmB\bcdot\f{\p\bmB}{\p t}=-\f{1}{\mu_0}\bmB\bcdot\curl\bmE.
\end{equation}
\textit{Total energy}:
\begin{equation}
  \f{\p}{\p t}\left[\rho(\half u^2+\Phi+e)+
  \f{B^2}{2\mu_0}\right]+
  \div\left[\rho\bmu(\half u^2+\Phi+h)+
  \f{\bmE\btimes\bmB}{\mu_0}\right]=0,
\end{equation}
where $h=e+p/\rho$ is the \emph{specific enthalpy} and we have used
the identity
$\div(\bmE\btimes\bmB)=\bmB\bcdot\curl\bmE-\bmE\bcdot\curl\bmB$.  Note
that $(\bmE\btimes\bmB)/\mu_0$ is the Poynting vector, the
electromagnetic energy flux density.  The total energy is therefore
conserved.

For a self-gravitating system satisfying Poisson's equation, the
gravitational energy density can instead be regarded as $-g^2/8\pi G$:
\begin{equation}
  \f{\p}{\p t}\left(-\f{g^2}{8\pi G}\right)=-\f{1}{4\pi G}\grad\Phi\bcdot\f{\p\grad\Phi}{\p t}
\end{equation}
\begin{equation}
  \f{\p}{\p t}\left(-\f{g^2}{8\pi G}\right)+\div\left(\f{\Phi}{4\pi G}\f{\p\grad\Phi}{\p t}\right)=\f{\Phi}{4\pi G}\f{\p\delsq\Phi}{\p t}=\Phi\f{\p\rho}{\p t}=-\Phi\div(\rho\bmu)
\end{equation}
\begin{equation}
  \f{\p}{\p t}\left(-\f{g^2}{8\pi G}\right)+\div\left(\rho\bmu\Phi+\f{\Phi}{4\pi G}\f{\p\grad\Phi}{\p t}\right)=\rho\bmu\bcdot\grad\Phi.
\end{equation}
The total energy equation is then
\begin{equation}
  \f{\p}{\p t}\left[\rho(\half u^2+e)-\f{g^2}{8\pi G}+
  \f{B^2}{2\mu_0}\right]+
  \div\left[\rho\bmu(\half u^2+\Phi+h)+\f{\Phi}{4\pi G}\f{\p\grad\Phi}{\p t}+
  \f{\bmE\btimes\bmB}{\mu_0}\right]=0.
\end{equation}
It is important to note that some of the gravitational and magnetic
energy of an astrophysical body is stored in the exterior region, even
if the mass density vanishes there.

\subsection{Other conservation laws in ideal MHD}

In ideal fluid dynamics there are certain invariants with a
geometrical or topological interpretation.  In homentropic or
barotropic flow, for example, vorticity (or, equivalently,
circulation) and kinetic helicity are conserved, while, in
non-barotropic flow, potential vorticity is conserved (see
Example~\ref{e:vorticity}).  The Lorentz force breaks these
conservation laws because the curl of the Lorentz force per unit mass
does not vanish in general.  However, some new topological invariants
associated with the magnetic field appear.

The \emph{magnetic helicity} in a volume $V$ with bounding surface $S$
is defined as
\begin{equation}
  H_\rmm=\int_V\bmA\bcdot\bmB\,\rmd V,
\end{equation}
where $\bmA$ is the magnetic vector potential, such that
$\bmB=\curl\bmA$.  Now
\begin{equation}
  \f{\p\bmA}{\p t}=-\bmE-\grad\Phi_\rme=\bmu\btimes\bmB-\grad\Phi_\rme,
\end{equation}
where $\Phi_\rme$ is the electrostatic potential.  This can be
thought of as the `uncurl' of the induction equation.  Thus
\begin{equation}
  \f{\p}{\p t}(\bmA\bcdot\bmB)=-\bmB\bcdot\grad\Phi_\rme+\bmA\bcdot\curl(\bmu\btimes\bmB).
\end{equation}
In ideal MHD, therefore, magnetic helicity is conserved:
\begin{equation}
  \f{\p}{\p t}(\bmA\bcdot\bmB)+
  \div\left[\Phi_\rme\bmB+\bmA\btimes(\bmu\btimes\bmB)\right]=0.
\end{equation}
However, care is needed because $\bmA$ is not uniquely defined.  Under
a gauge transformation $\bmA\mapsto\bmA+\grad\chi$,
$\Phi_\rme\mapsto\Phi_{\rm e}-\p\chi/\p t$, where $\chi(\bmx,t)$ is a
scalar field, $\bmE$ and $\bmB$ are invariant, but $H_\rmm$ changes
by an amount
\begin{equation}
  \int_V\bmB\bcdot\grad\chi\,\rmd V=\int_V\div(\chi\bmB)\,\rmd V=
  \int_S\chi\bmB\bcdot\bmn\,\rmd S.
\end{equation}
Therefore $H_\rmm$ is not uniquely defined unless
$\bmB\bcdot\bmn=0$ on the surface $S$.

Magnetic helicity is a \emph{pseudoscalar} quantity: it changes sign
under a reflection of the spatial coordinates.  Indeed, it is non-zero
only when the magnetic field lacks reflectional symmetry.  It can also
be interpreted topologically in terms of the twistedness and
knottedness of the magnetic field (see Example~\ref{e:helicity}).
Since the field is `frozen in' to the fluid and deformed continuously
by it, the topological properties of the field are conserved.  The
equivalent conserved quantity in homentropic or barotropic ideal gas
dynamics (without a magnetic field) is the \emph{kinetic helicity}
\begin{equation}
  H_{\rm k}=\int_V\bmu\bcdot(\curl\bmu)\,\rmd V.
\end{equation}

The \emph{cross-helicity} in a volume $V$ is
\begin{equation}
  H_{\rm c}=\int_V\bmu\bcdot\bmB\,\rmd V.
\end{equation}
It is helpful here to write the equation of motion in ideal MHD in the
form
\begin{equation}
  \f{\p\bmu}{\p t}+(\curl\bmu)\btimes\bmu+\grad(\half u^2+\Phi+h)=T\grad s+\f{1}{\mu_0\rho}(\curl\bmB)\btimes\bmB,
\label{dudtalt}
\end{equation}
using the relation $\rmd h=T\,\rmd s+v\,\rmd p$.  Thus
\begin{equation}
  \f{\p}{\p t}(\bmu\bcdot\bmB)+
  \div\left[\bmu\btimes(\bmu\btimes\bmB)+(\half u^2+\Phi+h)\bmB\right]=
  T\bmB\bcdot\grad s,
\end{equation}
and so cross-helicity is conserved in ideal MHD in homentropic or
barotropic flow.

\emph{Bernoulli's theorem} follows from the inner product of
equation~(\ref{dudtalt}) with $\bmu$.  In steady flow
\begin{equation}
  \bmu\bcdot\grad(\half u^2+\Phi+h)=0,
\end{equation}
which implies that the \emph{Bernoulli function} $\half u^2+\Phi+h$ is
constant along streamlines, but only if $\bmu\bcdot\bmF_\text{m}=0$
(e.g.\ if $\bmu\,\|\,\bmB$), i.e.\ if the magnetic field does no work on
the flow.

\medskip
\noindent Related examples: \ref{e:helicity}, \ref{e:variational}.

\subsection{Symmetries of the equations}

The equations of ideal gas dynamics and MHD have numerous symmetries.
In the case of an isolated, self-gravitating system, these include:
\begin{itemize}
\item Translations of time and space, and rotations of space: related
  (via Noether's theorem) to the conservation of energy, momentum and
  angular momentum.
\item Reversal of time: related to the absence of dissipation.
\item Reflections of space (but note that $\bmB$ is a
  \emph{pseudovector} and behaves oppositely to~$\bmu$ under a
  reflection).
\item Galilean transformations.
\item Reversal of the sign of $\bmB$.
\item Similarity transformations (\textbf{exercise}): if space and
  time are rescaled by independent factors $\lambda$ and $\mu$, i.e.\
\begin{equation}
  \bmx\mapsto\lambda\,\bmx,\qquad
  t\mapsto\mu\,t,
\end{equation}
then
\begin{equation}
  \bmu\mapsto\lambda\mu^{-1}\,\bmu,\quad
  \rho\mapsto\mu^{-2}\,\rho,\quad
  p\mapsto\lambda^2\mu^{-4}\,p,\quad
  \Phi\mapsto\lambda^2\mu^{-2}\,\Phi,\quad
  \bmB\mapsto\lambda\mu^{-2}\,\bmB.
\end{equation}
(This symmetry requires a perfect gas so that the thermodynamic
relations are scale-free.)
\end{itemize}

In the case of a non-isolated system with an external potential
$\Phi_\mathrm{ext}$, these symmetries (other than $\bmB\mapsto-\bmB$)
apply only if $\Phi_\mathrm{ext}$ has them.  However, in the
approximation of a non-self-gravitating system, the mass can be
rescaled by any factor $\lambda$ such that
\begin{equation}
  \rho\mapsto\lambda\,\rho,\qquad
  p\mapsto\lambda\,p,\qquad
  \bmB\mapsto\lambda^{1/2}\,\bmB.
\end{equation}
(This symmetry also requires a perfect gas.)

\subsection{Hyperbolic structure}
\label{s:hyperbolic}

Analysing the so-called hyperbolic structure of the equations of AFD
is one way of understanding the wave modes of the system and the way
in which information propagates in the fluid.  It is fundamental to
the construction of some types of numerical method for solving the
equations.  We temporarily neglect the gravitational force here,
because in a Newtonian theory it involves instantaneous action at a
distance and is not associated with a finite wave speed.

In ideal gas dynamics, the equation of mass conservation, the thermal
energy equation and the equation of motion (omitting gravity) can be
written as
\begin{equation}
\begin{split}
  &\f{\p\rho}{\p t}+\bmu\bcdot\grad\rho+\rho\div\bmu=0,\\
  &\f{\p p}{\p t}+\bmu\bcdot\grad p+\gamma p\div\bmu=0,\\
  &\f{\p\bmu}{\p t}+\bmu\bcdot\grad\bmu+\f{1}{\rho}\grad p=\bfzero,
\end{split}
\end{equation}
and then combined into the form
\begin{equation}
  \f{\p\bfU}{\p t}+\bfA_i\f{\p\bfU}{\p x_i}=\bfzero,
\end{equation}
where
\begin{equation}
  \bfU=\begin{bmatrix}\rho\\p\\u_x\\u_y\\u_z\end{bmatrix}
\end{equation}
is a five-dimensional `state vector' and
$\bfA_x$, $\bfA_y$ and $\bfA_z$ are the three $5\times5$ matrices
\begin{equation}
 \begin{bmatrix}u_x&0&\rho&0&0\\0&u_x&\gamma p&0&0\\0&\f{1}{\rho}&u_x&0&0\\0&0&0&u_x&0\\0&0&0&0&u_x\end{bmatrix},\qquad
 \begin{bmatrix}u_y&0&0&\rho&0\\0&u_y&0&\gamma p&0\\0&0&u_y&0&0\\0&\f{1}{\rho}&0&u_y&0\\0&0&0&0&u_y\end{bmatrix},\qquad
 \begin{bmatrix}u_z&0&0&0&\rho\\0&u_z&0&0&\gamma p\\0&0&u_z&0&0\\0&0&0&u_z&0\\0&\f{1}{\rho}&0&0&u_z\end{bmatrix}.
\end{equation}
This works because every term in the equations involves a first
derivative with respect to either time or space.

The system of equations is said to be \emph{hyperbolic} if the
eigenvalues of $\bfA_in_i$ are real for any unit vector $\bmn$ and if
the eigenvectors span the five-dimensional space.  As will be seen in
Section~\ref{s:simple}, the eigenvalues can be identified as wave
speeds, and the eigenvectors as wave modes, with $\bmn$ being the unit
wavevector, locally normal to the wavefronts.

Taking $\bmn=\bme_x$ without loss of generality, we find
(\textbf{exercise})
\begin{equation}
  \det(\bfA_x-v\bfI)=-(v-u_x)^3\left[(v-u_x)^2-v_\text{s}^2\right],
\end{equation}
where
\begin{equation}
  v_\text{s}=\left(\f{\gamma p}{\rho}\right)^{1/2}
\end{equation}
is the \emph{adiabatic sound speed}.  The wave speeds $v$ are real and
the system is indeed hyperbolic.

Two of the wave modes are \emph{sound waves} (acoustic waves), which
have wave speeds $v=u_x\pm v_\rms$ and therefore propagate at the
sound speed relative to the moving fluid.  Their eigenvectors are
\begin{equation}
  \begin{bmatrix}\rho\\\gamma p\\\pm v_\rms\\0\\0\end{bmatrix}
\end{equation}
and involve perturbations of density, pressure and longitudinal
velocity.

The remaining three wave modes have wave speed $v=u_x$ and do not
propagate relative to the fluid.  Their eigenvectors are
\begin{equation}
  \begin{bmatrix}1\\0\\0\\0\\0\end{bmatrix},\qquad
  \begin{bmatrix}0\\0\\0\\1\\0\end{bmatrix},\qquad
  \begin{bmatrix}0\\0\\0\\0\\1\end{bmatrix}.
\end{equation}
The first is the \emph{entropy wave}, which involves only a density
perturbation but no pressure perturbation.  Since the entropy can be
considered as a function of the density and pressure, this wave
involves an entropy perturbation.  It must therefore propagate at the
fluid velocity because the entropy is a material invariant.  The other
two modes with $v=u_x$ are \emph{vortical waves}, which involve
perturbations of the transverse velocity components, and therefore of
the vorticity.  Conservation of vorticity implies that these waves
propagate with the fluid velocity.

To extend the analysis to ideal MHD, we may consider the induction
equation in the form
\begin{equation}
  \f{\p\bmB}{\p t}+\bmu\bcdot\grad\bmB-\bmB\bcdot\grad\bmu+\bmB(\div\bmu)=\bfzero,
\end{equation}
and include the Lorentz force in the equation of motion.  Every new
term involves a first derivative.  So the equation of mass
conservation, the thermal energy equation, the equation of motion and
the induction equation can be written in the combined form
\begin{equation}
  \f{\p\bfU}{\p t}+\bfA_i\f{\p\bfU}{\p x_i}=\bfzero,
\end{equation}
where
\begin{equation}
  \bfU=\begin{bmatrix}\rho\\p\\u_x\\u_y\\u_z\\B_x\\B_y\\B_z\end{bmatrix}
\end{equation}
is now an eight-dimensional `state vector' and the $\bfA_i$ are three
$8\times8$ matrices, e.g.
\begin{equation}
 \bfA_x=\begin{bmatrix}u_x&0&\rho&0&0&0&0&0\\0&u_x&\gamma p&0&0&0&0&0\\0&\f{1}{\rho}&u_x&0&0&0&\f{B_y}{\mu_0\rho}&\f{B_z}{\mu_0\rho}\\0&0&0&u_x&0&0&-\f{B_x}{\mu_0\rho}&0\\0&0&0&0&u_x&0&0&-\f{B_x}{\mu_0\rho}\\0&0&0&0&0&u_x&0&0\\0&0&B_y&-B_x&0&0&u_x&0\\0&0&B_z&0&-B_x&0&0&u_x\end{bmatrix}.
\end{equation}

We now find, after some algebra,
\begin{equation}
  \det(\bfA_x-v\bfI)=(v-u_x)^2\left[(v-u_x)^2-v_{\text{a}x}^2\right]\left[(v-u_x)^4-(v_\text{s}^2+v_\text{a}^2)(v-u_x)^2+v_\text{s}^2v_{\text{a}x}^2\right].
\end{equation}
The wave speeds $v$ are real and the system is indeed hyperbolic.  The
various MHD wave modes will be examined later
(Section~\ref{s:linear}).

In this representation, there are two modes that have $v=u_x$ and do
not propagate relative to the fluid.  One is still the entropy wave,
which is physical and involves only a density perturbation.  The other
is the `$\divword\bmB$' mode, which is unphysical and involves a
perturbation of $\div\bmB$ (i.e.\ of $B_x$, in the case
$\bmn=\bme_x$).  This must be eliminated by imposing the constraint
$\div\bmB=0$.  (In fact the equations in the form we have written them
imply that $(\div\bmB)/\rho$ is a material invariant and could be
non-zero unless the initial condition $\div\bmB=0$ is imposed.)  The
vortical waves are replaced by Alfv\'en waves with speeds $u_x\pm
v_{\rma x}$.

\subsection{Stress tensor and virial theorem}

In the absence of external forces, the equation of motion of a fluid
can usually be written in the form
\begin{equation}
  \rho\f{\rmD\bmu}{\rmD t}=\div\bfT\qquad\hbox{or}\qquad
  \rho\f{\rmD u_i}{\rmD t}=\f{\p T_{ji}}{\p x_j},
\end{equation}
where $\bfT$ is the \emph{stress tensor}, a symmetric second-rank
tensor field.  Using the equation of mass conservation, we can relate
this to the conservative form of the momentum equation,
\begin{equation}
  \f{\p}{\p t}(\rho\bmu)+\div(\rho\bmu\bmu-\bfT)=\bfzero,
\label{stress}
\end{equation}
which shows that $-\bfT$ is the momentum flux density excluding the
advective flux of momentum.

For a self-gravitating system in ideal MHD, the stress tensor is
\begin{equation}
  \bfT=-p\,\bfI-\f{1}{4\pi G}\left(\bmg\bmg-\half g^2\,\bfI\right)+\f{1}{\mu_0}\left(\bmB\bmB-\half B^2\,\bfI\right),
\end{equation}
or, in Cartesian components,
\begin{equation}
  T_{ij}=-p\,\delta_{ij}-\f{1}{4\pi G}\left(g_ig_j-\half g^2\,\delta_{ij}\right)+\f{1}{\mu_0}\left(B_iB_j-\half B^2\,\delta_{ij}\right).
\end{equation}
We have already identified the Maxwell stress tensor associated with
the magnetic field.  The idea of a gravitational stress tensor works
for a self-gravitating system in which the gravitational field
$\bmg=-\grad\Phi$ and the density $\rho$ are related through Poisson's
equation $-\div\bmg=\delsq\Phi=4\pi G\rho$.
In fact, for a general vector field $\bmv$, it can be shown that
(\textbf{exercise})
\begin{equation}
  \div(\bmv\bmv-\half v^2\,\bfI)=(\div\bmv)\bmv+\bmv\bcdot\grad\bmv-\grad(\half v^2)=(\div\bmv)\bmv+(\curl\bmv)\btimes\bmv.
\end{equation}
In the magnetic case ($\bmv=\bmB$) the first term in the final
expression vanishes, while in the gravitational case ($\bmv=\bmg$) the
second term vanishes, leaving $-4\pi G\rho\bmg$, which becomes the
force per unit volume, $\rho\bmg$, when divided by $-4\pi G$.

The \emph{virial equations} are the spatial moments of the equation of
motion, and provide integral measures of the balance of forces acting
on the fluid.  The first moments are generally the most useful.
Consider
\begin{equation}
  \rho\f{\rmD^2}{\rmD t^2}(x_ix_j)=\rho\f{\rmD}{\rmD t}(u_ix_j+x_iu_j)=2\rho u_iu_j+x_j\f{\p T_{ki}}{\p x_k}+x_i\f{\p T_{kj}}{\p x_k}.
\end{equation}
Integrate this equation over a material volume $V$ bounded by a
surface $S$ (with material invariant mass element $\rmd m=\rho\,\rmd
V$):
\begin{equation}
\begin{split}
  \f{\rmd^2}{\rmd t^2}\int_Vx_ix_j\,\rmd m&=\int_V\left(2\rho u_iu_j+x_j\f{\p T_{ki}}{\p x_k}+x_i\f{\p T_{kj}}{\p x_k}\right)\,\rmd V\\
  &=\int_V(2\rho u_iu_j-T_{ji}-T_{ij})\,\rmd V+\int_S(x_jT_{ki}+x_iT_{kj})n_k\,\rmd S,
\end{split}
\end{equation}
where we have integrated by parts using the divergence theorem.  In
the case of an isolated system with no external sources of gravity or
magnetic field, $\bmg$ decays proportional to $|\bmx|^{-2}$ at large
distance, and $\bmB$ decays faster.  Therefore $T_{ij}$ decays
proportional to $|\bmx|^{-4}$ and the surface integral can be
eliminated if we let $V$ occupy the whole of space.  We then obtain
(after division by~$2$) the \emph{tensor virial theorem}
\begin{equation}
  \f{1}{2}\f{\rmd^2I_{ij}}{\rmd t^2}=2K_{ij}-\mathcal{T}_{ij},
\end{equation}
where
\begin{equation}
  I_{ij}=\int x_ix_j\,\rmd m
\end{equation}
is related to the inertia tensor of the system,
\begin{equation}
  K_{ij}=\int\half u_iu_j\,\rmd m
\end{equation}
is a kinetic energy tensor and
\begin{equation}
  \mathcal{T}_{ij}=\int T_{ij}\,\rmd V
\end{equation}
is the integrated stress tensor.  (If the conditions above are not
satisfied, there will be an additional contribution from the surface
integral.)

The \emph{scalar virial theorem} is the trace of this expression,
which we write as
\begin{equation}
  \f{1}{2}\f{\rmd^2I}{\rmd t^2}=2K-\mathcal{T}.
\end{equation}
Note that $K$ is the total kinetic energy.  Now
\begin{equation}
  -\mathcal{T}=\int\left(3p-\f{g^2}{8\pi G}+\f{B^2}{2\mu_0}\right)\,\rmd V=3(\gamma-1)U+W+M,
\end{equation}
for a perfect gas with no external gravitational field, where
$U$, $W$ and $M$ are the total internal, gravitational and magnetic
energies.
Thus
\begin{equation}
  \f{1}{2}\f{\rmd^2I}{\rmd t^2}=2K+3(\gamma-1)U+W+M.
\end{equation}
On the right-hand side, only $W$ is negative.  For the system to be
bound (i.e.\ not fly apart) the kinetic, internal and magnetic
energies are limited by
\begin{equation}
  2K+3(\gamma-1)U+M\le|W|.
\end{equation}
In fact equality must hold, at least on average, unless the system is
collapsing or contracting.

The tensor virial theorem provides more specific information relating
to the energies associated with individual directions, and is
particularly relevant in cases where anisotropy is introduced by
rotation or a magnetic field.  It has been used in estimating the
conditions required for gravitational collapse in molecular clouds.  A
higher-order tensor virial method was used by Chandrasekhar and
Lebovitz to study the equilibrium and stability of rotating
ellipsoidal bodies \citep{Chandrasekhar69}.

\section{Linear waves in homogeneous media}
\label{s:linear}

In ideal MHD the density, pressure and magnetic field evolve
according to
\begin{equation}
\begin{split}
  \f{\p\rho}{\p t}&=-\bmu\bcdot\grad\rho-\rho\div\bmu,\\
  \f{\p p}{\p t}&=-\bmu\bcdot\grad p-\gamma p\div\bmu,\\
  \f{\p\bmB}{\p t}&=\curl(\bmu\btimes\bmB).
\end{split}
\end{equation}
Consider a magnetostatic equilibrium in which the density, pressure
and magnetic field are $\rho_0(\bmx)$, $p_0(\bmx)$ and $\bmB_0(\bmx)$.
The above equations are exactly satisfied in this basic state because
$\bmu=\bfzero$ and the time-derivatives vanish.  Now consider small
perturbations from equilibrium, such that
$\rho(\bmx,t)=\rho_0(\bmx)+\delta\rho(\bmx,t)$ with
$|\delta\rho|\ll\rho_0$, etc.  The linearized equations are
\begin{equation}
\begin{split}
  \f{\p\,\delta\rho}{\p t}&=-\delta\bmu\bcdot\grad\rho_0-\rho_0\div\delta\bmu,\\
  \f{\p\,\delta p}{\p t}&=-\delta\bmu\bcdot\grad p_0-\gamma p_0\div\delta\bmu,\\
  \f{\p\,\delta\bmB}{\p t}&=\curl(\delta\bmu\btimes\bmB_0).
\end{split}
\end{equation}
By introducing the \emph{displacement} $\bxi(\bmx,t)$ such that
$\delta\bmu=\p\bxi/\p t$, we can integrate these equations
to obtain
\begin{equation}
\begin{split}
  \delta\rho&=-\bxi\bcdot\grad\rho-\rho\div\bxi,\\
  \delta p&=-\bxi\bcdot\grad p-\gamma p\div\bxi,\\
  \delta\bmB&=\curl(\bxi\btimes\bmB)\\
  &=\bmB\bcdot\grad\bxi-\bmB(\div\bxi)-\bxi\bcdot\grad\bmB.
\end{split}
\end{equation}
We have now dropped the subscript `0' without danger of confusion.

(The above relations allow some freedom to add arbitrary functions of
$\bmx$.  At least when studying wavelike solutions in which all
variables have the same harmonic time-dependence, such additional
terms can be discarded.)

The linearized equation of motion is
\begin{equation}
  \rho\f{\p^2\bxi}{\p
  t^2}=-\rho\grad\delta\Phi-\delta\rho\grad\Phi-\grad\delta\Pi+
  \f{1}{\mu_0}(\delta\bmB\bcdot\grad\bmB+\bmB\bcdot\grad\delta\bmB),
\end{equation}
where the perturbation of total pressure is
\begin{equation}
  \delta\Pi=\delta p+\f{\bmB\bcdot\delta\bmB}{\mu_0}=-\bxi\bcdot\grad\Pi-
  \left(\gamma p+\f{B^2}{\mu_0}\right)\div\bxi+
  \f{1}{\mu_0}\bmB\bcdot(\bmB\bcdot\grad\bxi).
\end{equation}
The gravitational potential perturbation satisfies the linearized
Poisson equation
\begin{equation}
  \delsq\delta\Phi=4\pi G\,\delta\rho.
\end{equation}

We consider a basic state of uniform density, pressure and
magnetic field, in the absence of gravity.  Such a system is
homogeneous but anisotropic, because the uniform field distinguishes a
particular direction.  The problem simplifies to
\begin{equation}
  \rho\f{\p^2\bxi}{\p t^2}=-\grad\delta\Pi+
  \f{1}{\mu_0}\bmB\bcdot\grad
  \left[\bmB\bcdot\grad\bxi-\bmB(\div\bxi)\right],
\label{wave1}
\end{equation}
with
\begin{equation}
  \delta\Pi=-\left(\gamma p+\f{B^2}{\mu_0}\right)\div\bxi+
  \f{1}{\mu_0}\bmB\bcdot(\bmB\bcdot\grad\bxi).
\end{equation}
Owing to the symmetries of the basic state, plane-wave solutions exist,
of the form
\begin{equation}
  \bxi(\bmx,t)=\real\left[\tilde\bxi\,\exp(\rmi\bmk\bcdot\bmx-\rmi\omega t)\right],
\end{equation}
where $\omega$ and $\bmk$ are the frequency and wavevector, and
$\tilde\bxi$ is a constant vector representing the amplitude of the
wave.  For such solutions, equation~(\ref{wave1}) gives
\begin{equation}
  \rho\omega^2\bxi=\left[\left(\gamma p+\f{B^2}{\mu_0}\right)\bmk\bcdot\bxi-
  \f{1}{\mu_0}(\bmk\bcdot\bmB)\bmB\bcdot\bxi\right]\bmk+\f{1}{\mu_0}(\bmk\bcdot\bmB)
  \left[(\bmk\bcdot\bmB)\bxi-\bmB(\bmk\bcdot\bxi)\right],
\label{wave2}
\end{equation}
where we have changed the sign and omitted the tilde.

For transverse displacements that are orthogonal to both the
wavevector and the magnetic field,
i.e.\ $\bmk\bcdot\bxi=\bmB\bcdot\bxi=0$, this equation simplifies to
\begin{equation}
  \rho\omega^2\bxi=\f{1}{\mu_0}(\bmk\bcdot\bmB)^2\bxi.
\end{equation}
Such solutions are called \emph{Alfv\'en waves}.  Their \emph{dispersion
  relation} is
\begin{equation}
  \omega^2=(\bmk\bcdot\bmv_{\rm a})^2.
\end{equation}

Given the dispersion relation $\omega(\bmk)$ of any wave mode, the
\textit{phase and group velocities} of the wave can be identified as
\begin{equation}
  \bmv_\rmp=\f{\omega}{k}\,\hat\bmk,\qquad
  \bmv_\text{g}=\f{\p\omega}{\p\bmk}=\grad_{\!\bmk}\omega,
\end{equation}
where $\hat\bmk=\bmk/k$.  The phase velocity is that with which the
phase of the wave travels, while the group velocity is that which the
energy of the wave (or the centre of a wavepacket) is transported.

For Alfv\'en waves, therefore,
\begin{equation}
  \bmv_\rmp=\pm v_\text{a}\cos\theta\,\hat\bmk,\qquad
  \bmv_\text{g}=\pm\bmv_\text{a},
\end{equation}
where $\theta$ is the angle between $\bmk$ and $\bmB$.

To find the other solutions, we take the inner product of
equation~(\ref{wave2}) with $\bmk$ and then with $\bmB$ to obtain
first
\begin{equation}
  \rho\omega^2\bmk\bcdot\bxi=\left[\left(\gamma p+\f{B^2}{\mu_0}\right)
  \bmk\bcdot\bxi-\f{1}{\mu_0}(\bmk\bcdot\bmB)\bmB\bcdot\bxi\right]k^2
\end{equation}
and then
\begin{equation}
  \rho\omega^2\bmB\bcdot\bxi=\gamma p(\bmk\bcdot\bxi)\bmk\bcdot\bmB.
\end{equation}
These equations can be written in the form
\begin{equation}
  \begin{bmatrix}\rho\omega^2-\left(\gamma p+\f{B^2}{\mu_0}\right)k^2&
  \f{1}{\mu_0}(\bmk\bcdot\bmB)k^2\\
  -\gamma p(\bmk\bcdot\bmB)&\rho\omega^2\end{bmatrix}
  \begin{bmatrix}\bmk\bcdot\bxi\\ \bmB\bcdot\bxi\end{bmatrix}=
  \begin{bmatrix}0\\ 0\end{bmatrix}.
\end{equation}
The `trivial solution' $\bmk\bcdot\bxi=\bmB\bcdot\bxi=0$ corresponds to
the Alfv\'en wave that we have already identified.  The other
solutions satisfy
\begin{equation}
  \rho\omega^2\left[\rho\omega^2-\left(\gamma p+\f{B^2}{\mu_0}\right)k^2
  \right]+\gamma pk^2\f{1}{\mu_0}(\bmk\bcdot\bmB)^2=0,
\end{equation}
which simplifies to
\begin{equation}
  v_\rmp^4-(v_{\rm s}^2+v_{\rm a}^2)v_\rmp^2+v_{\rm s}^2v_{\rm a}^2\cos^2\theta=0.
\end{equation}
The two solutions
\begin{equation}
  v_\rmp^2=\half(v_{\rm s}^2+v_{\rm a}^2)\pm
  \left[\quarter(v_{\rm s}^2+v_{\rm a}^2)^2-
  v_{\rm s}^2v_{\rm a}^2\cos^2\theta\right]^{1/2}
\end{equation}
are called \emph{fast and slow magnetoacoustic} (or
\emph{magnetosonic}) waves, respectively.

In the special case $\theta=0$ ($\bmk\|\bmB$), we have
\begin{equation}
  v_\rmp^2=v_{\rm s}^2\quad\text{or}\quad v_{\rm a}^2,
\end{equation}
together with $v_\rmp^2=v_{\rm a}^2$ for the Alfv\'en wave.  Note that the
fast wave could be either $v_\rmp^2=v_{\rm s}^2$ or $v_\rmp^2=v_{\rm a}^2$,
whichever is greater.

In the special case $\theta=\pi/2$ ($\bmk\perp\bmB$), we have
\begin{equation}
  v_\rmp^2=v_{\rm s}^2+v_{\rm a}^2\quad\text{or}\quad0,
\end{equation}
together with $v_\rmp^2=0$ for the Alfv\'en wave.

The effects of the magnetic field on wave propagation can be
understood as resulting from the two aspects of the Lorentz force.
The magnetic tension gives rise to Alfv\'en waves, which are similar
to waves on an elastic string, and are trivial in the absence of the
magnetic field.  In addition, the magnetic pressure affects the
response of the fluid to compression, and therefore modifies the
propagation of acoustic waves.

\begin{figure}
  \centerline{\epsfysize6cm\epsfbox{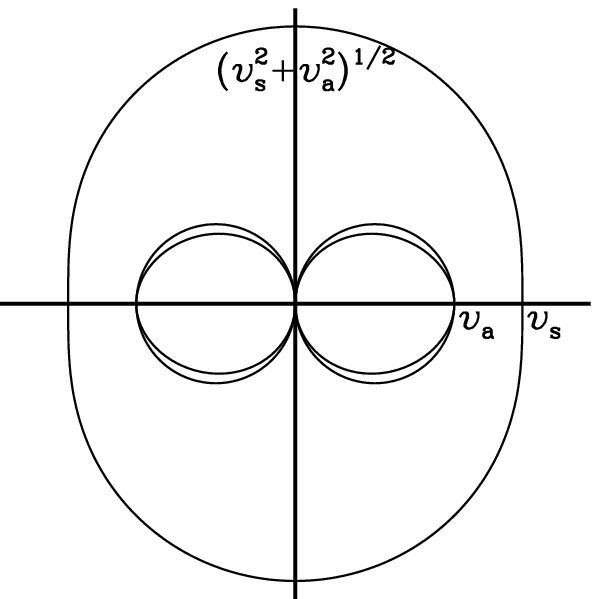}\qquad\epsfysize6cm\epsfbox{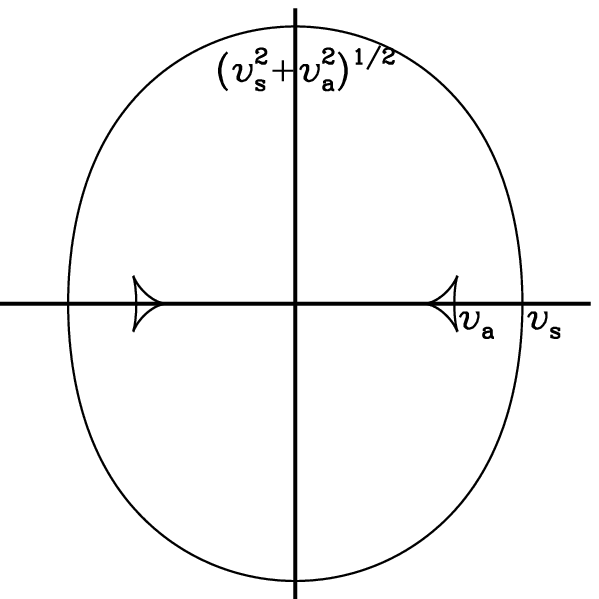}}
\vskip0.5cm
\centerline{\epsfysize6cm\epsfbox{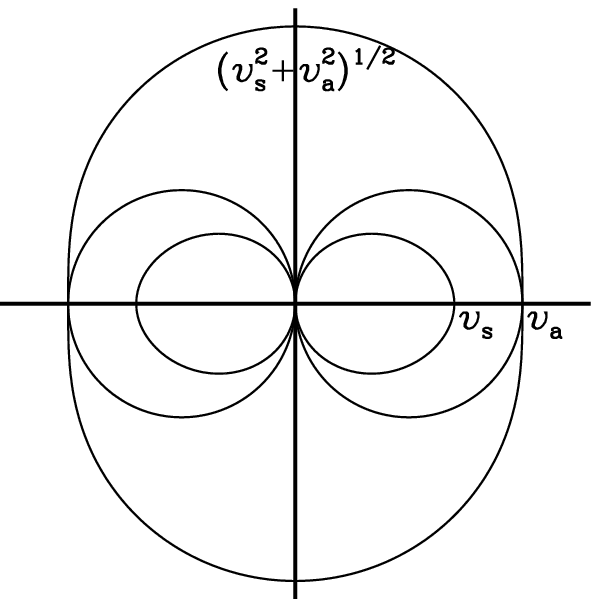}\qquad\epsfysize6cm\epsfbox{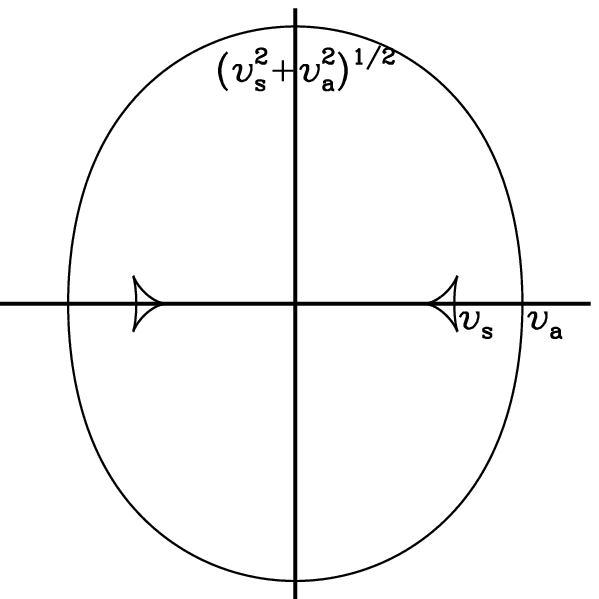}}
\vskip0.5cm
\caption{Polar plots of the phase velocity (left) and group velocity
  (right) of MHD waves for the cases $v_{\rm a}=0.7\,v_{\rm s}$ (top)
  and $v_{\rm s}=0.7\,v_{\rm a}$ (bottom) with a magnetic field in the
  horizontal direction.  [The group velocity plot for the Alfv\'en
  wave consists of the two points $(\pm v_\rma,0)$.]}
\label{f:mhd_waves}
\end{figure}

The phase and group velocity for the full range of $\theta$ are
usually exhibited in \textit{Friedrichs diagrams}\footnote{Kurt Otto
  Friedrichs (1901--1982), German--American.}
(Figure~\ref{f:mhd_waves}), which are polar plots of $v_\rmp(\theta)$
and $v_\rmg(\theta)$.

We can interpret:
\begin{itemize}
\item the fast wave as a quasi-isotropic acoustic-type wave in which
  both gas and magnetic pressure contribute;
\item the slow wave as an acoustic-type wave that is strongly guided
  by the magnetic field;
\item the Alfv\'en wave as analogous to a wave on an elastic string,
propagating by means of magnetic tension and perfectly guided by the
magnetic field.
\end{itemize}

\medskip
\noindent Related example: \ref{e:friedrichs}.

\section{Nonlinear waves, shocks and other discontinuities}

\subsection{One-dimensional gas dynamics}

\subsubsection{Riemann's analysis}

\begin{figure}
  \centerline{\epsfysize9cm\epsfbox{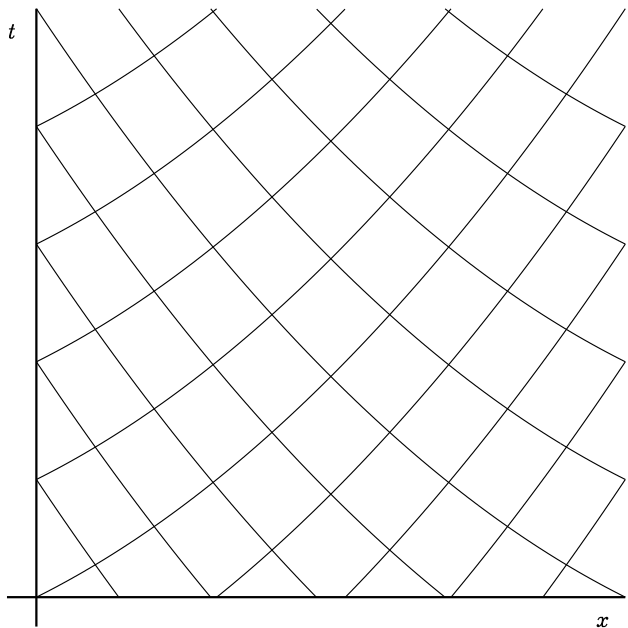}}
  \caption{Characteristic curves in the space-time diagram.}
\label{f:characteristics}
\end{figure}

The equations of mass conservation and motion in one dimension are
\begin{equation}
\begin{split}
  \f{\p\rho}{\p t}+u\f{\p\rho}{\p x}&=-\rho\f{\p u}{\p x},\\
  \f{\p u}{\p t}+u\f{\p u}{\p x}&=-\f{1}{\rho}\f{\p p}{\p x}.
\end{split}
\end{equation}
We assume the gas is homentropic ($s=\cst$) and perfect.  (This
eliminates the entropy wave and leaves only the two sound waves.)
Then $p\propto\rho^\gamma$ and $v_\text{s}^2=\gamma
p/\rho\propto\rho^{\gamma-1}$.  It is convenient to use $v_\text{s}$
as a variable in place of $\rho$ or $p$:
\begin{equation}
  \rmd p=v_\text{s}^2\,\rmd\rho,\qquad
  \rmd\rho=\f{\rho}{v_\text{s}}\left(\f{2\,\rmd v_\text{s}}{\gamma-1}\right).
\end{equation}
Then
\begin{equation}
\begin{split}
  \f{\p u}{\p t}+u\f{\p u}{\p x}+v_\text{s}\f{\p}{\p x}\left(\f{2v_\text{s}}{\gamma-1}\right)&=0,\\
  \f{\p}{\p t}\left(\f{2v_\text{s}}{\gamma-1}\right)+u\f{\p}{\p x}\left(\f{2v_\text{s}}{\gamma-1}\right)+v_\text{s}\f{\p u}{\p x}&=0.
\end{split}
\end{equation}
We add and subtract to obtain
\begin{equation}
  \left[\f{\p}{\p t}+(u+v_\text{s})\f{\p}{\p x}\right]\left(u+\f{2v_\text{s}}{\gamma-1}\right)=0,
\end{equation}
\begin{equation}
  \left[\f{\p}{\p t}+(u-v_\text{s})\f{\p}{\p x}\right]\left(u-\f{2v_\text{s}}{\gamma-1}\right)=0.
\end{equation}
Define the two \textit{Riemann invariants}
\begin{equation}
  R_\pm=u\pm\f{2v_\text{s}}{\gamma-1}.
\end{equation}
Then we deduce that $R_\pm=\cst$ along a \textit{characteristic
  (curve)} of gradient $\rmd x/\rmd t=u\pm v_\text{s}$ in the $(x,t)$
plane.  The $+$ and $-$ characteristics form an interlocking web
covering the space-time diagram (Figure~\ref{f:characteristics}).

Note that both Riemann invariants are needed to reconstruct the
solution ($u$ and $v_\text{s}$).  Half of the information is
propagated along one set of characteristics and half along the other.

In general the characteristics are not known in advance but must be
determined along with the solution.  The $+$ and $-$ characteristics
propagate at the speed of sound to the right and left, respectively,
\textit{with respect to the motion of the fluid}.

This concept generalizes to nonlinear waves the solution of the
classical wave equation for acoustic waves on a uniform and static
background, which is of the form $f(x-v_\text{s}t)+g(x+v_\text{s}t)$.

\subsubsection{Method of characteristics}

A numerical method of solution can be based on the following idea:
\begin{itemize}
\item Start with the initial data ($u$ and $v_\text{s}$) for all
  relevant $x$ at $t=0$.
\item Determine the characteristic slopes at $t=0$.
\item Propagate the $R_\pm$ information for a small increment of time,
  neglecting the variation of the characteristic slopes.
\item Combine the $R_\pm$ information to find $u$ and $v_\text{s}$ at
  each $x$ at the new value of $t$.
\item Re-evaluate the slopes and repeat.
\end{itemize}

\begin{figure}
  \centerline{\epsfysize9cm\epsfbox{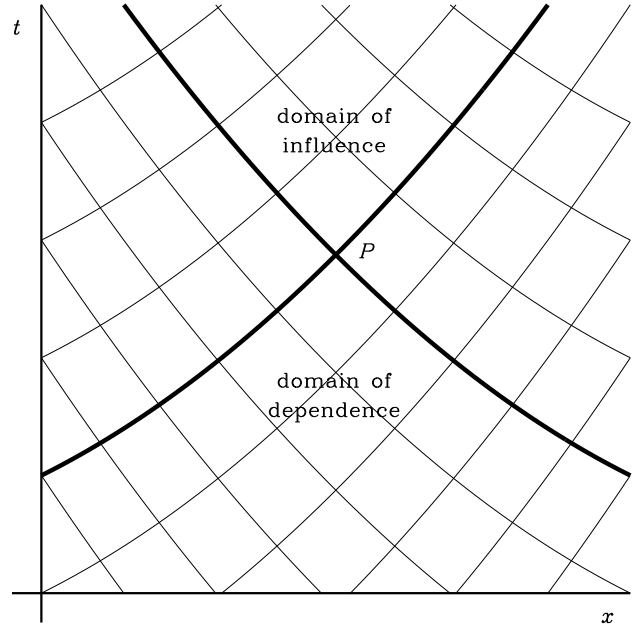}}
  \caption{Domains of dependence and of influence.}
\label{f:domains}
\end{figure}

The \emph{domain of dependence} of a point $P$ in the space-time
diagram is that region of the diagram bounded by the $\pm$
characteristics through $P$ and located in the past of $P$.  The
solution at $P$ cannot depend on anything that occurs outside the
domain of dependence.  Similarly, the \emph{domain of influence} of
$P$ is the region in the future of $P$ bounded by the characteristics
through $P$ (Figure~\ref{f:domains}).

\subsubsection{A simple wave}

Suppose that $R_-$ is uniform, having the same constant value on every
characteristic emanating from an undisturbed region to the right.  Its
value everywhere is that of the undisturbed region:
\begin{equation}
  u-\f{2v_\text{s}}{\gamma-1}=u_0-\f{2v_{\text{s}0}}{\gamma-1}.
\end{equation}
Then, along the $+$ characteristics, both $R_+$ and $R_-$, and
therefore $u$ and $v_\text{s}$, must be constant.  The $+$
characteristics therefore have constant slope $v=u+v_\text{s}$, so
they are straight lines.

The statement that the wave speed $v$ is constant on the family of
straight lines $\rmd x/\rmd t=v$ is expressed by the equation
\begin{equation}
  \f{\p v}{\p t}+v\f{\p v}{\p x}=0.
\end{equation}
This is known as the \emph{inviscid Burgers
  equation}\footnote{Johannes (Jan) Martinus Burgers (1895--1981),
  Dutch.} or the \emph{nonlinear advection equation}.

\begin{figure}
  \centerline{\epsfysize9cm\epsfbox{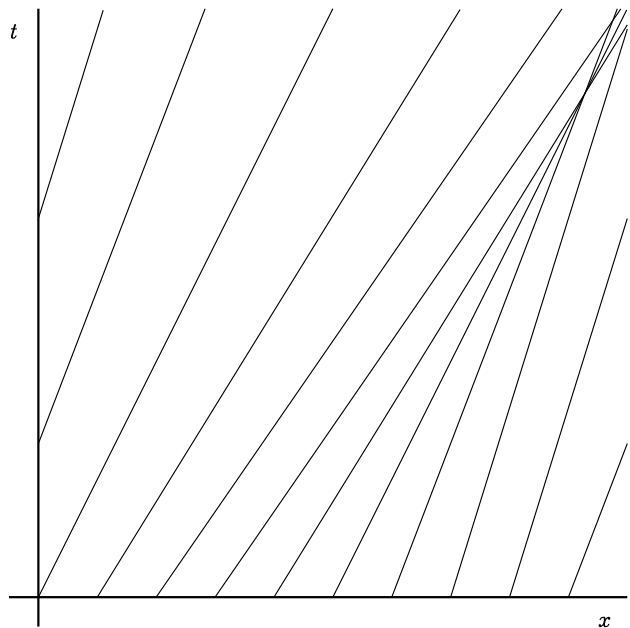}}
  \caption{Formation of a shock from intersecting characteristics.}
\label{f:intersect}
\end{figure}

The inviscid Burgers equation has only one set of characteristics,
with slope $\rmd x/\rmd t=v$.  It is easily solved by the method of
characteristics.  The initial data define $v_0(x)=v(x,0)$ and the
characteristics are straight lines.  In regions where $\rmd v_0/\rmd
x>0$ the characteristics diverge in the future.  In regions where
$\rmd v_0/\rmd x<0$ the characteristics converge and will form a
\textit{shock} at some point.  Contradictory information arrives at
the same point in the space-time diagram, leading to a breakdown of
the solution (Figure~\ref{f:intersect}).

\begin{figure}
  \centerline{\epsfysize9cm\epsfbox{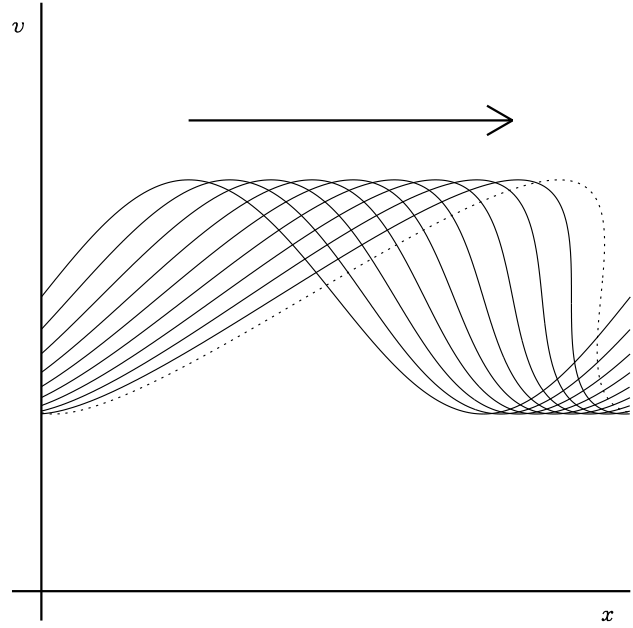}}
  \caption{Wave steepening and shock formation.  The dotted profile is
    multiple-valued and is replaced in practice with a discontinuous
    profile including a shock.}
\label{f:steepen}
\end{figure}

Another viewpoint is that of \emph{wave steepening}.  The graph of $v$
versus $x$ evolves in time by moving each point at its wave speed $v$.
The crest of the wave moves fastest and eventually overtakes the
trough to the right of it.  The profile would become multiple-valued,
but this is physically meaningless and the wave breaks, forming a
discontinuity (Figure~\ref{f:steepen}).

Indeed, the formal solution of the inviscid Burgers equation is
\begin{equation}
  v(x,t)=v_0(x_0)\qquad\textrm{with}\quad x=x_0+v_0(x_0)t.
\end{equation}
By the chain rule, $\p v/\p x=v_0'/(1+v_0't)$, which diverges first at
the breaking time $t=1/\max(-v_0')$.

The crest of a sound wave moves faster than the trough for two reasons.
It is partly because the crest is denser and hotter, so the sound
speed is higher (unless the gas is isothermal), but it is also because
of the self-advection of the wave (recall that the wave speed is
$u+v_\text{s}$).  The breaking time depends on the amplitude and
wavelength of the wave.

\subsection{General analysis of simple nonlinear waves}
\label{s:simple}

Recall the hyperbolic structure of the equations of AFD
(Section~\ref{s:hyperbolic}):
\begin{equation}
  \f{\p\bfU}{\p t}+\bfA_i\f{\p\bfU}{\p x_i}=\bfzero,\qquad
  \bfU=[\rho,p,\bmu,\bmB]^{\rm T}.
\end{equation}
The system is hyperbolic because the eigenvalues of $\bfA_in_i$ are
real for any unit vector $n_i$.  The eigenvalues are identified as
wave speeds, and the corresponding eigenvectors as wave modes.

In a simple wave propagating in the $x$-direction, all physical
quantities are functions of a single variable, the phase
$\varphi(x,t)$.  Then $\bfU=\bfU(\varphi)$ and so
\begin{equation}
  \f{\rmd\bfU}{\rmd\varphi}\f{\p\varphi}{\p t}+\bfA_x\f{\rmd\bfU}{\rmd\varphi}\f{\p\varphi}{\p x}=0.
\end{equation}
This equation is satisfied if $\rmd\bfU/\rmd\varphi$ is an eigenvector
of the hyperbolic system and if
\begin{equation}
  \f{\p\varphi}{\p t}+v\f{\p\varphi}{\p x}=0,
\end{equation}
where $v$ is the corresponding wavespeed.  But since $v=v(\varphi)$ we
again find
\begin{equation}
  \f{\p v}{\p t}+v\f{\p v}{\p x}=0,
\end{equation}
the inviscid Burgers equation.

Wave steepening is therefore generic for simple waves.  However, waves
do not always steepen in practice.  For example, linear dispersion
arising from Coriolis or buoyancy forces (see Section~\ref{s:waves})
can counteract nonlinear wave steepening.  Waves propagating on a
non-uniform background are not simple waves.  In addition, waves may
be damped by diffusive processes (viscosity, thermal conduction or
resistivity) before they can steepen.

Furthermore, even some simple waves do not undergo steepening, in
spite of the above argument.  This happens if the wave speed $v$ does
not depend on the variables that actually vary in the wave mode.  One
example is the entropy wave in hydrodynamics, in which the density
varies but not the pressure or the velocity.  The wave speed is the
fluid velocity, which does not vary in this wave; therefore the
relevant solution of the inviscid Burgers equation is just $v=\cst$.
Another example is the Alfv\'en wave, which involves variations in
transverse velocity and magnetic field components, but whose speed
depends on the longitudinal components and the density.  The slow and
fast magnetoacoustic waves, though, are `genuinely nonlinear' and
undergo steepening.

\subsection{Shocks and other discontinuities}
\label{s:shock}

\subsubsection{Jump conditions}

Discontinuities are resolved in reality by diffusive processes
(viscosity, thermal conduction or resistivity) that become more
important on smaller length-scales.  Properly, we should solve an
enhanced set of equations to resolve the internal structure of a
shock.  This internal solution would then be matched on to the
external solution in which diffusion can be neglected.  However, the
matching conditions can in fact be determined from general principles
without resolving the internal structure.

\begin{figure}
  \vskip0.3cm
  \centerline{\epsfysize4cm\epsfbox{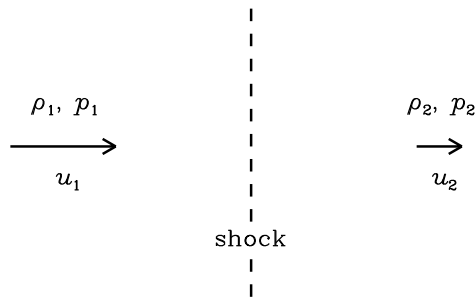}}
  \vskip0.3cm
  \caption{A shock front in its rest frame.}
\label{f:shock}
\end{figure}

Without loss of generality, we consider a shock front at rest at $x=0$
(making a Galilean transformation if necessary).  We look for a
stationary, one-dimensional solution in which gas flows from left to
right.  On the left is upstream, pre-shock material ($\rho_1$, $p_1$,
etc.).  On the right is downstream, post-shock material ($\rho_2$,
$p_2$, etc.) (Figure~\ref{f:shock}).

Consider any equation in conservative form
\begin{equation}
  \f{\p q}{\p t}+\div\bmF=0.
\end{equation}
For a stationary solution in one dimension,
\begin{equation}
  \f{\rmd F_x}{\rmd x}=0,
\end{equation}
which implies that the flux density $F_x$ has the same value on each
side of the shock.  We write the matching condition as
\begin{equation}
  [F_x]_1^2=F_{x2}-F_{x1}=0.
\end{equation}
Including additional physics means that additional diffusive fluxes
(not of mass but of momentum, energy, magnetic flux, etc.) are
present.  These fluxes are negligible outside the shock, so they do
not affect the jump conditions.  This approach is permissible provided
that the new physics does not introduce any source terms in the
equations.  So the total energy is a properly conserved quantity but
\textit{not the entropy} (see later).

From mass conservation:
\begin{equation}
  [\rho u_x]_1^2=0.
\end{equation}
From momentum conservation:
\begin{equation}
  \left[\rho u_x^2+\Pi-\f{B_x^2}{\mu_0}\right]_1^2=0,
\end{equation}
\begin{equation}
  \left[\rho u_xu_y-\f{B_xB_y}{\mu_0}\right]_1^2=0,
\end{equation}
\begin{equation}
  \left[\rho u_xu_z-\f{B_xB_z}{\mu_0}\right]_1^2=0.
\end{equation}
From $\div\bmB=0$:
\begin{equation}
  [B_x]_1^2=0.
\end{equation}
From $\p\bmB/\p t+\curl\bmE=\bfzero$:
\begin{equation}
  [u_xB_y-u_yB_x]_1^2=-[E_z]_1^2=0,
\end{equation}
\begin{equation}
  [u_xB_z-u_zB_x]_1^2=[E_y]_1^2=0.
\end{equation}
(These are the standard electromagnetic conditions at an interface:
the normal component of $\bmB$ and the parallel components of $\bmE$
are continuous.)  From total energy conservation:
\begin{equation}
  \left[\rho u_x(\half u^2+h)+\f{1}{\mu_0}(E_yB_z-E_zB_y)\right]_1^2=0.
\end{equation}

Note that the conservative form of the momentum equation used above is
(cf.\ equation~\ref{stress})
\begin{equation}
  \f{\p}{\p t}(\rho u_i)+\div\left(\rho u_i\bmu+\Pi\,\bme_i-\f{B_i\bmB}{\mu_0}\right)=0.
\end{equation}
Including gravity makes no difference to the shock relations because
$\Phi$ is always continuous (it satisfies $\delsq\Phi=4\pi G\rho$).

Although the entropy in ideal MHD satisfies an equation of conservative form,
\begin{equation}
  \f{\p}{\p t}(\rho s)+\div(\rho s\bmu)=0,
\end{equation}
the dissipation of energy within the shock provides a source term for
entropy.  Therefore the entropy flux is not continuous across the
shock.

\subsubsection{Non-magnetic shocks}

First consider a \emph{normal shock} ($u_y=u_z=0$) with no magnetic
field.  We obtain the \textit{Rankine--Hugionot
  relations}\footnote{William John Macquorn Rankine (1820--1872),
  British.  Pierre-Henri Hugoniot (1851--1887), French.}
\begin{equation}
  [\rho u_x]_1^2=0,\qquad
  [\rho u_x^2+p]_1^2=0,\qquad
  [\rho u_x(\half u_x^2+h)]_1^2=0.
\end{equation}
The specific enthalpy of a perfect gas is
\begin{equation}
  h=\left(\f{\gamma}{\gamma-1}\right)\f{p}{\rho}
\end{equation}
and these equations can be solved algebraically (see Example~\ref{e:shock}).
Introduce the upstream Mach number (the \emph{shock Mach number})
\begin{equation}
  \mathcal{M}_1=\f{u_{x1}}{v_{\text{s}1}}>0.
\end{equation}
Then we find
\begin{equation}
  \f{\rho_2}{\rho_1}=\f{u_{x1}}{u_{x2}}=\f{(\gamma+1)\mathcal{M}_1^2}{(\gamma-1)\mathcal{M}_1^2+2},\qquad
  \f{p_2}{p_1}=\f{2\gamma\mathcal{M}_1^2-(\gamma-1)}{(\gamma+1)},
\end{equation}
and
\begin{equation}
  \mathcal{M}_2^2=\f{2+(\gamma-1)\mathcal{M}_1^2}{2\gamma\mathcal{M}_1^2-(\gamma-1)}.
\end{equation}

Note that $\rho_2/\rho_1$ and $p_2/p_1$ are increasing functions of
$\mathcal{M}_1$.  The case $\mathcal{M}_1=1$ is trivial as it
corresponds to $\rho_2/\rho_1=p_2/p_1=1$.  The other two cases are the
\textit{compression shock} ($\mathcal{M}_1>1$, $\mathcal{M}_2<1$,
$\rho_2>\rho_1$, $p_2>p_1$) and the \textit{rarefaction shock}
($\mathcal{M}_1<1$, $\mathcal{M}_2>1$, $\rho_2<\rho_1$, $p_2<p_1$).

It is shown in Example~\ref{e:shock} that the entropy change in
passing through the shock is positive for compression shocks and
negative for rarefaction shocks.  Therefore \textit{only compression
  shocks are physically realizable.}  Rarefaction shocks are excluded
by the second law of thermodynamics.  All shocks involve dissipation
and irreversibility.

The fact that $\mathcal{M}_1>1$ while $\mathcal{M}_2<1$ means that the
shock travels supersonically relative to the upstream gas and
subsonically relative to the downstream gas.

In the \textit{weak shock} limit $\mathcal{M}_1-1\ll1$ the relative
velocity of the fluid and the shock is close to the sound speed on
both sides.

In the \textit{strong shock} limit $\mathcal{M}_1\gg1$, common in
astrophysical applications, we have
\begin{equation}
  \f{\rho_2}{\rho_1}=\f{u_{x1}}{u_{x2}}\to\f{\gamma+1}{\gamma-1},\qquad
  \f{p_2}{p_1}\gg1,\qquad
  \mathcal{M}_2^2\to\f{\gamma-1}{2\gamma}.
\end{equation}
Note that the compression ratio $\rho_2/\rho_1$ is finite (and equal
to $4$ when $\gamma=5/3$).  In the rest frame of the undisturbed
(upstream) gas the \textit{shock speed} is $u_\text{sh}=-u_{x1}$.
The downstream density, velocity (in that frame) and pressure in the
limit of a strong shock are (as will be used in Section~\ref{s:blast})
\begin{equation}
  \rho_2=\left(\f{\gamma+1}{\gamma-1}\right)\rho_1,\qquad
  u_{x2}-u_{x1}=\f{2u_\text{sh}}{\gamma+1},\qquad
  p_2=\f{2\rho_1u_\text{sh}^2}{\gamma+1}.
\label{strong}
\end{equation}
A significant amount of thermal energy is generated out of kinetic
energy by the passage of a strong shock:
\begin{equation}
  e_2=\f{2u_\text{sh}^2}{(\gamma+1)^2}.
\end{equation}

\subsubsection{Oblique shocks}

When $u_y$ or $u_z$ is non-zero, we have the additional relations
\begin{equation}
  [\rho u_xu_y]_1^2=[\rho u_xu_z]_1^2=0.
\end{equation}
Since $\rho u_x$ is continuous across the shock (and non-zero), we
deduce that $[u_y]_1^2=[u_z]_1^2=0$.  Momentum and energy conservation
apply as before, and we recover the Rankine--Hugoniot relations.  (See
Example~\ref{e:oblique}.)

\subsubsection{Other discontinuities}

The discontinuity is not called a shock if there is no normal flow
($u_x=0$).  In this case we can deduce only that $[p]_1^2=0$.
Arbitrary discontinuities are allowed in $\rho$, $u_y$ and $u_z$.
These are related to the entropy and vortical waves.  If there is a
jump in $\rho$ we have a \textit{contact discontinuity}.  If there is
a jump in $u_y$ or $u_z$ we have a \textit{tangential discontinuity}
or \textit{vortex sheet} (the vorticity being proportional to
$\delta(x)$).  Note that these discontinuities are not produced
naturally by wave steepening, because the entropy and vortical waves
do not steepen.  However they do appear in the Riemann problem
(Section~\ref{s:riemann}) and other situations with discontinuous
initial conditions.

\subsubsection{MHD shocks and discontinuities}

When a magnetic field is included, the jump conditions allow a wider
variety of solutions.  There are different types of discontinuity
associated with the three MHD waves (Alfv\'en, slow and fast), which
we will not discuss here.  Since the parallel components of $\bmB$
need not be continuous, it is possible for them to `switch on' or
`switch off' on passage through a shock.

A \emph{current sheet} is a tangential discontinuity in the magnetic
field.  A classic case would be where $B_y$, say, changes sign across
the interface, with $B_x=0$.  The current density is then proportional
to $\delta(x)$.

\subsubsection{The Riemann problem}
\label{s:riemann}

The Riemann problem is a fundamental initial-value problem for a
hyperbolic system and plays a central role in some numerical methods
for solving the equations of AFD.

The initial condition at $t=0$ consists of two uniform states
separated by a discontinuity at $x=0$.  In the case of one-dimensional
gas dynamics, we have
\begin{equation}
  \rho=\begin{cases}\rho_\mathrm{L},&x<0\\\rho_\mathrm{R},&x>0\end{cases},\qquad
  p=\begin{cases}p_\mathrm{L},&x<0\\p_\mathrm{R},&x>0\end{cases},\qquad
  u_x=\begin{cases}u_\mathrm{L},&x<0\\u_\mathrm{R},&x>0\end{cases},
\end{equation}
where `L' and `R' denote the left and right states.  A simple example
is a `shock tube' problem in which gas at different pressures is at
rest on either side of a partition, which is released at $t=0$.

It can be shown that the initial discontinuity resolves generically
into three simple waves.  The inner one is a contact discontinuity
while the outer ones are shocks or rarefaction waves (see below).

The initial data define no natural length-scale for the Riemann
problem, but they do allow a characteristic velocity scale $c$ to be
defined (although not uniquely).  The result is a \emph{similarity
  solution} in which variables depend on $x$ and $t$ only through the
dimensionless combination $\xi=x/ct$.

\begin{figure}
  \centerline{\epsfysize9cm\epsfbox{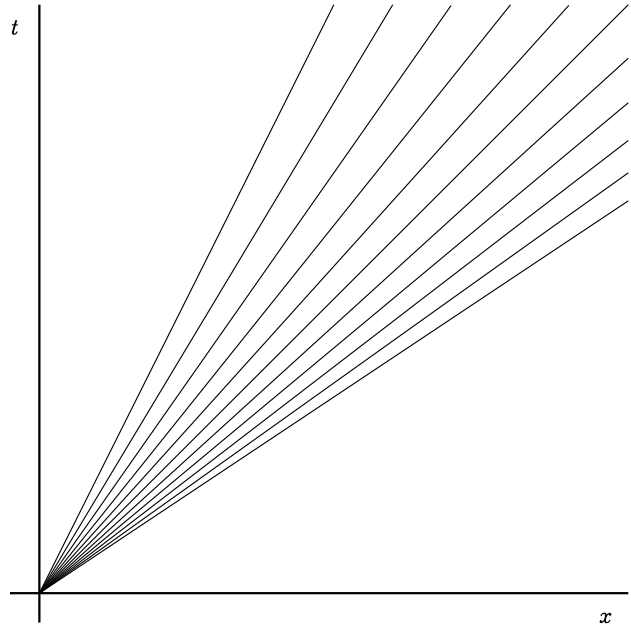}}
  \caption{Expansion fan of characteristics in a rarefaction wave.}
\label{f:expansion}
\end{figure}

Unlike the unphysical rarefaction shock, the \emph{rarefaction wave}
(or \emph{expansion wave}) is a non-dissipative, homentropic,
continuous simple wave in which $\div\bmu>0$.  If we seek a similarity
solution $v=v(\xi)$ of the inviscid Burgers equation $v_t+vv_x=0$ we
find $v=x/t$ (or the trivial solution $v=\cst$).  The characteristics
form an \textit{expansion fan} (Figure~\ref{f:expansion}).

The `$+$' rarefaction wave has $u+v_\rms=x/t$ and
$R_-=u-\f{2}{\gamma-1}v_\rms=\cst$, determined by the undisturbed
right-hand state.  The `$-$' rarefaction wave has $u-v_\rms=x/t$ and
$R_+=u+\f{2}{\gamma-1}v_\rms=\cst$, determined by the undisturbed
left-hand state.  In each case $u$ and $v_\rms$ are linear functions
of $x/t$ and $\div\bmu=\f{2}{\gamma+1}t^{-1}>0$.

\begin{figure}
  \vskip0.5cm
  \centerline{\epsfysize5cm\epsfbox{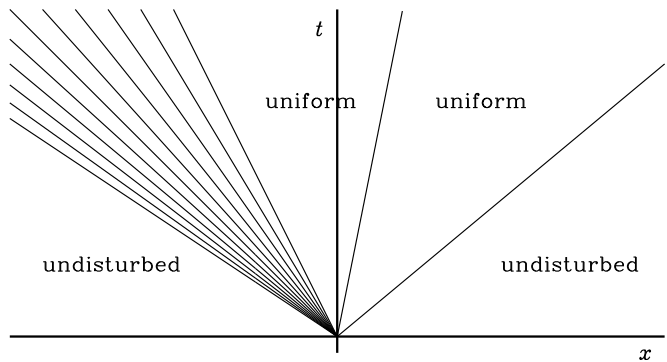}}
  \caption{Typical outcome of a shock-tube problem.  The two uniform
    regions are separated by a contact discontinuity.  The other
    discontinuity is a shock.}
\label{f:shock_tube}
\end{figure}

A typical outcome of a shock-tube problem consists of (from left to
right): undisturbed region, rarefaction wave, uniform region, contact
discontinuity, uniform region, shock, undisturbed
region (Figure~\ref{f:shock_tube}).

In Godunov's method and related computational algorithms, the
equations of AFD are advanced in time by solving (either exactly or
approximately) a Riemann problem at each cell boundary.

\medskip
\noindent Related examples: \ref{e:shock}, \ref{e:oblique},
\ref{e:riemann}, \ref{e:elsasser}.

\section{Spherical blast waves: supernovae}
\label{s:blast}

\noindent\textit{Note: in this section $(r,\theta,\phi)$ are spherical polar coordinates.}

\subsection{Introduction}

In a supernova, an energy of order $10^{51}\,\text{erg}$
($10^{44}\,\text{J}$) is released into the interstellar medium.  An
expanding spherical blast wave is formed as the explosion sweeps up
the surrounding gas.  Several good examples of these supernova
remnants are observed in the Galaxy, e.g.\ Tycho's supernova of 1572
and Kepler's supernova of 1604\footnote{See
  \texttt{http://en.wikipedia.org/wiki/Supernova\_remnant}}.

The effect is similar to a bomb.  When photographs\footnote{See
  \texttt{http://www.atomicarchive.com/Photos/Trinity}} (complete with
length and time scales) were released of the first atomic bomb test in
New Mexico in 1945, both Sedov\footnote{Leonid Ivanovitch Sedov
  (1907--1999), Russian.} in the Soviet Union and Taylor\footnote{Sir
  Geoffrey Ingram Taylor (1886--1975), British.} in the UK were able
to work out the energy of the bomb (equivalent to about 20 kilotons of
TNT), which was supposed to be a secret.

We suppose that an energy $E$ is released at $t=0$, $r=0$ and that the
explosion is spherically symmetric.  The external medium has density
$\rho_0$ and pressure $p_0$.  In the Sedov--Taylor phase of the
explosion, the pressure $p\gg p_0$.  Then a strong shock is formed and
the external pressure $p_0$ can be neglected (formally set to zero).
Gravity is also negligible in the dynamics.

\subsection{Governing equations}

The equations governing the spherically symmetric flow of a perfect
gas, with radial velocity $u_r=u(r,t)$, may be written as
\begin{equation}
\begin{split}
  \left(\f{\p}{\p t}+u\f{\p}{\p r}\right)\rho&=-\f{\rho}{r^2}\f{\p}{\p r}(r^2u),\\
  \left(\f{\p}{\p t}+u\f{\p}{\p r}\right)u&=-\f{1}{\rho}\f{\p p}{\p r},\\
  \left(\f{\p}{\p t}+u\f{\p}{\p r}\right)\ln(p\rho^{-\gamma})&=0.
\end{split}
\label{spherical}
\end{equation}
These imply the total energy equation
\begin{equation}
  \f{\p}{\p t}\left(\f{1}{2}\rho u^2+\f{p}{\gamma-1}\right)+\f{1}{r^2}\f{\p}{\p r}\left[r^2\left(\f{1}{2}\rho u^2+\f{\gamma p}{\gamma-1}\right)u\right]=0.
\label{total}
\end{equation}
The shock is at $r=R(t)$, and the shock speed is $\dot R$.  The
equations are to be solved in $0<r<R$ with the strong shock conditions
(\ref{strong}) at $r=R$:
\begin{equation}
  \rho=\left(\f{\gamma+1}{\gamma-1}\right)\rho_0,\qquad
  u=\f{2\dot R}{\gamma+1},\qquad
  p=\f{2\rho_0\dot R^2}{\gamma+1}.
\end{equation}
The total energy of the explosion is
\begin{equation}
  E=\int_0^R\left(\f{1}{2}\rho u^2+\f{p}{\gamma-1}\right)\,4\pi r^2\,\rmd r=\cst,
\end{equation}
the thermal energy of the external medium being negligible.

\subsection{Dimensional analysis}

The dimensional parameters of the problem on which the solution might
depend are $E$ and $\rho_0$.  Their dimensions are
\begin{equation}
  [E]=ML^2T^{-2},\qquad
  [\rho_0]=ML^{-3}.
\end{equation}
Together, they do not define a characteristic length-scale, so the
explosion is `scale-free' or `self-similar'.  If the dimensional
analysis includes the time $t$ since the explosion, however, we can
find a time-dependent characteristic length-scale.  The radius of the
shock must be
\begin{equation}
  R=\alpha\left(\f{Et^2}{\rho_0}\right)^{1/5},
\end{equation}
where $\alpha$ is a dimensionless constant to be determined.

\subsection{Similarity solution}

The self-similarity of the explosion is expressed using the
dimensionless \textit{similarity variable} $\xi=r/R(t)$.  The
solution has the form
\begin{equation}
  \rho=\rho_0\,\tilde\rho(\xi),\qquad
  u=\dot R\,\tilde u(\xi),\qquad
  p=\rho_0\dot R^2\,\tilde p(\xi),
\end{equation}
where $\tilde\rho(\xi)$, $\tilde u(\xi)$ and $\tilde p(\xi)$ are
dimensionless functions to be determined.  The meaning of this type of
solution is that the graph of $u$ versus $r$, for example, has a
constant shape but both axes of the graph are rescaled as time
proceeds and the shock expands.

\subsection{Dimensionless equations}

We substitute these forms into equations (\ref{spherical}) and cancel
the dimensional factors to obtain
\begin{equation}
\begin{split}
  (\tilde u-\xi)\tilde\rho'&=-\f{\tilde\rho}{\xi^2}\f{\rmd}{\rmd\xi}(\xi^2\tilde u),\\
  (\tilde u-\xi)\tilde u'-\f{3}{2}\tilde u&=-\f{\tilde p'}{\tilde\rho},\\
  (\tilde u-\xi)\left(\f{\tilde p'}{\tilde p}-\f{\gamma\tilde\rho'}{\tilde\rho}\right)-3&=0.
\end{split}
\end{equation}
Similarly, the strong shock conditions are that
\begin{equation}
  \tilde\rho=\f{\gamma+1}{\gamma-1},\qquad
  \tilde u=\f{2}{\gamma+1},\qquad
  \tilde p=\f{2}{\gamma+1}
\end{equation}
at $\xi=1$, while the total energy integral provides a normalization
condition,
\begin{equation}
  1=\f{16\pi}{25}\alpha^5\int_0^1\left(\f{1}{2}\tilde\rho\tilde u^2+\f{\tilde p}{\gamma-1}\right)\,\xi^2\,\rmd\xi,
\end{equation}
that will ultimately determine the value of $\alpha$.

\subsection{First integral}

The one-dimensional conservative form of the total energy
equation~(\ref{total}) is
\begin{equation}
  \f{\p q}{\p t}+\f{\p F}{\p r}=0,
\label{conservative}
\end{equation}
where
\begin{equation}
  q=r^2\left(\f{1}{2}\rho u^2+\f{p}{\gamma-1}\right),\qquad
  F=r^2\left(\f{1}{2}\rho u^2+\f{\gamma p}{\gamma-1}\right)u.
\end{equation}
In dimensionless form,
\begin{equation}
  q=\rho_0R^2\dot R^2\,\tilde q(\xi),\qquad
  F=\rho_0R^2\dot R^3\,\tilde F(\xi),
\label{qf}
\end{equation}
with
\begin{equation}
  \tilde q=\xi^2\left(\f{1}{2}\tilde\rho\tilde u^2+\f{\tilde p}{\gamma-1}\right),\qquad
  \tilde F=\xi^2\left(\f{1}{2}\tilde\rho\tilde u^2+\f{\gamma\tilde p}{\gamma-1}\right)\tilde u.
\end{equation}
We substitute the forms~(\ref{qf}) into the energy equation~(\ref{conservative}) to find
\begin{equation}
  -\xi\tilde q'-\tilde q+\tilde F'=0,
\end{equation}
which implies
\begin{equation}
  \f{\rmd}{\rmd\xi}(\tilde F-\xi\tilde q)=0.
\end{equation}
Thus
\begin{equation}
  \tilde F-\xi\tilde q=\cst=0
\end{equation}
for a solution that is finite at $\xi=0$.  This equation can be solved
for $\tilde p$:
\begin{equation}
  \tilde p=\f{(\gamma-1)\tilde\rho\tilde u^2(\xi-\tilde u)}{2(\gamma\tilde u-\xi)}.
\end{equation}
Note that this solution is compatible with the shock boundary
conditions.  Having found a first integral, we can now dispense with
(e.g.) the thermal energy equation.

Let $\tilde u=v\xi$.  We now have
\begin{equation}
  (v-1)\f{\rmd\ln\tilde\rho}{\rmd\ln\xi}=-\f{\rmd v}{\rmd\ln\xi}-3v,
\end{equation}
\begin{equation}
  (v-1)\f{\rmd v}{\rmd\ln\xi}+\f{1}{\tilde\rho\xi^2}\f{\rmd}{\rmd\ln\xi}\left[\f{(\gamma-1)\tilde\rho\xi^2v^2(1-v)}{2(\gamma v-1)}\right]=\f{3}{2}v.
\end{equation}
Eliminate $\tilde\rho$:
\begin{equation}
  \f{\rmd v}{\rmd\ln\xi}=\f{v(\gamma v-1)[5-(3\gamma-1)v]}{\gamma(\gamma+1)v^2-2(\gamma+1)v+2}.
\end{equation}
Invert and split into partial fractions:
\begin{equation}
  \f{\rmd\ln\xi}{\rmd v}=-\f{2}{5v}+\f{\gamma(\gamma-1)}{(2\gamma+1)(\gamma v-1)}+\f{13\gamma^2-7\gamma+12}{5(2\gamma+1)[5-(3\gamma-1)v]}.
\end{equation}
The solution is
\begin{equation}
  \xi\propto v^{-2/5}(\gamma v-1)^{(\gamma-1)/(2\gamma+1)}[5-(3\gamma-1)v]^{-(13\gamma^2-7\gamma+12)/5(2\gamma+1)(3\gamma-1)}.
\end{equation}
Now
\begin{equation}
  \begin{split}
  \f{\rmd\ln\tilde\rho}{\rmd v}&=-\f{1}{v-1}-\f{3v}{v-1}\f{\rmd\ln\xi}{\rmd v}\\
  &=\f{2}{(2-\gamma)(1-v)}+\f{3\gamma}{(2\gamma+1)(\gamma v-1)}-\f{13\gamma^2-7\gamma+12}{(2-\gamma)(2\gamma+1)[5-(3\gamma-1)v]}.
  \end{split}
\end{equation}
The solution is
\begin{equation}
  \tilde\rho\propto(1-v)^{-2/(2-\gamma)}(\gamma v-1)^{3/(2\gamma+1)}[5-(3\gamma-1)v]^{(13\gamma^2-7\gamma+12)/(2-\gamma)(2\gamma+1)(3\gamma-1)}.
\end{equation}

\begin{figure}
  \centerline{\epsfysize9cm\epsfbox{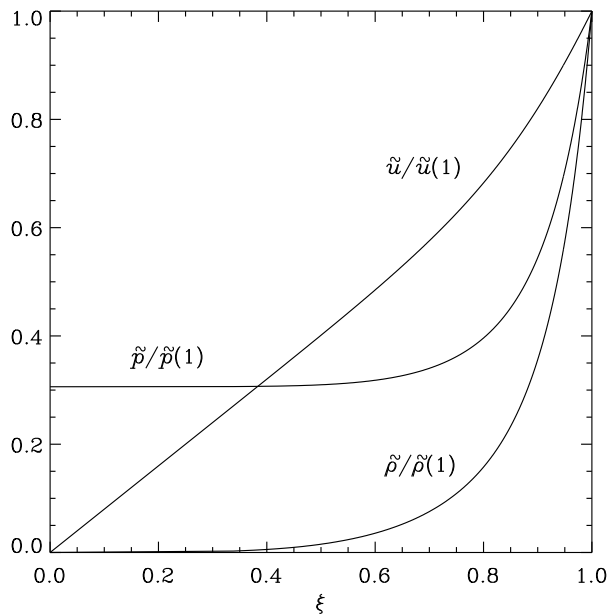}}
  \caption{Sedov's solution for a spherical blast wave in the case $\gamma=5/3$.}
\label{f:sedov}
\end{figure}

For example, in the case $\gamma=5/3$:
\begin{equation}
  \xi\propto v^{-2/5}\left(\f{5v}{3}-1\right)^{2/13}(5-4v)^{-82/195},
\end{equation}
\begin{equation}
  \tilde\rho\propto(1-v)^{-6}\left(\f{5v}{3}-1\right)^{9/13}(5-4v)^{82/13}.
\end{equation}
To satisfy $v=2/(\gamma+1)=3/4$ and $\tilde\rho=(\gamma+1)/(\gamma-1)=4$ at $\xi=1$:
\begin{equation}
  \xi=\left(\f{4v}{3}\right)^{-2/5}\left(\f{20v}{3}-4\right)^{2/13}\left(\f{5}{2}-2v\right)^{-82/195},
\end{equation}
\begin{equation}
  \tilde\rho=4\left(4-4v\right)^{-6}\left(\f{20v}{3}-4\right)^{9/13}\left(\f{5}{2}-2v\right)^{82/13}.
\end{equation}
Then, from the first integral,
\begin{equation}
  \tilde p=\f{3}{4}\left(\f{4v}{3}\right)^{6/5}\left(4-4v\right)^{-5}\left(\f{5}{2}-2v\right)^{82/15}.
\end{equation}
In this solution (Figure~\ref{f:sedov}), $\xi$ ranges from $0$ to $1$,
and $v$ from $3/5$ to $3/4$.  The normalization integral (numerically)
yields $\alpha\approx1.152$.

\subsection{Applications}

Some rough estimates are as follows:
\begin{itemize}
\item For a supernova: $E\sim10^{51}\,\text{erg}$, $\rho_0\sim10^{-24}\,\text{g}\,\text{cm}^{-3}$.  Then
$R\approx6\,\text{pc}$ and $\dot
R\approx2000\,\text{km}\,\text{s}^{-1}$ at $t=1000\,\text{yr}$.
\item For the 1945 New Mexico explosion:
  $E\approx8\times10^{20}\,\text{erg}$,
  $\rho_0\approx1.2\times10^{-3}\,\text{g}\,\text{cm}^{-3}$.  Then
  $R\approx100\,\text{m}$ and $\dot
  R\approx4\,\text{km}\,\text{s}^{-1}$ at $t=0.01\,\text{s}$.
\end{itemize}

The similarity method is useful in a very wide range of nonlinear
problems.  In this case it reduced partial differential equations to
integrable ordinary differential equations.

\medskip
\noindent Related example: \ref{e:blast}.

\section{Spherically symmetric steady flows: stellar winds and accretion}

\noindent\textit{Note: in this section $(r,\theta,\phi)$ are spherical polar coordinates.}

\subsection{Introduction}

Many stars, including the Sun, lose mass through a stellar wind.  The
gas must be sufficiently hot to escape from the star's gravitational
field.  Gravitating bodies can also accrete gas from the interstellar
medium.  The simplest models of these processes neglect the effects of
rotation or magnetic fields and involve a steady, spherically
symmetric flow.

\subsection{Basic equations}

We consider a purely radial flow, either away from or towards a body
of mass $M$.  The gas is perfect and non-self-gravitating, so
$\Phi=-GM/r$.  The fluid variables are functions of $r$ only, and the
only velocity component is $u_r=u(r)$.

Mass conservation for such a flow implies that the mass flux
\begin{equation}
  4\pi r^2\rho u=-\dot M=\cst.
\end{equation}
If $u>0$ (a stellar wind), $-\dot M$ is the mass-loss rate.  If $u<0$
(an accretion flow), $\dot M$ is the mass accretion rate.  We ignore
the secular change in the mass $M$, which would otherwise violate the
steady nature of the flow.

The thermal energy equation (assuming $u\ne0$) implies homentropic
flow:
\begin{equation}
  p=K\rho^\gamma,\qquad
  K=\cst.
\end{equation}

The equation of motion has only one component:
\begin{equation}
  \rho u\f{\rmd u}{\rmd r}=-\rho\f{\rmd\Phi}{\rmd r}-\f{\rmd p}{\rmd r}.
\label{motion}
\end{equation}
Alternatively, we can use the integral form (Bernoulli's equation):
\begin{equation}
  \half u^2+\Phi+h=B=\cst,\qquad
  h=\left(\f{\gamma}{\gamma-1}\right)\f{p}{\rho}=\f{v_\rms^2}{\gamma-1}.
\label{bernoulli}
\end{equation}
In highly subsonic flow the $\half u^2$ term on the left-hand side of
Bernoulli's equation is negligible and the gas is quasi-hydrostatic.
In highly supersonic flow the $h$ term is negligible and the flow is
quasi-ballistic (freely falling).  As discussed below, we are usually
interested in \emph{transonic} solutions that pass smoothly from
subsonic to supersonic flow.

Our aim is to solve for $u(r)$, and to determine $\dot M$ if possible.
At what rate does a star lose mass through a wind, or a black hole
accrete mass from the surrounding medium?

\subsection{First treatment}

We first use the differential form of the equation of motion.  Rewrite
the pressure gradient using the other two equations:
\begin{equation}
  -\f{\rmd p}{\rmd r}=-p\f{\rmd\ln p}{\rmd r}=-\gamma p\f{\rmd\ln\rho}{\rmd r}=\rho v_\text{s}^2\left(\f{2}{r}+\f{1}{u}\f{\rmd u}{\rmd r}\right).
\end{equation}
Equation~(\ref{motion}), multiplied by $u/\rho$, becomes
\begin{equation}
  (u^2-v_\text{s}^2)\f{\rmd u}{\rmd r}=u\left(\f{2v_\text{s}^2}{r}-\f{\rmd\Phi}{\rmd r}\right).
\end{equation}
A critical point (\emph{sonic point}) occurs at any radius $r=r_\rms$
where $|u|=v_\rms$.  For the flow to pass smoothly from subsonic to
supersonic, the right-hand side must vanish at the sonic point:
\begin{equation}
  \f{2v_\text{ss}^2}{r_\text{s}}-\f{GM}{r_\text{s}^2}=0.
\end{equation}
Evaluate Bernoulli's equation~(\ref{bernoulli}) at the sonic point:
\begin{equation}
  \left(\f{1}{2}+\f{1}{\gamma-1}\right)v_\text{ss}^2-\f{GM}{r_\text{s}}=B.
\end{equation}
We deduce that
\begin{equation}
  v_\text{ss}^2=\f{2(\gamma-1)}{(5-3\gamma)}B,\qquad
  r_\rms=\f{(5-3\gamma)}{4(\gamma-1)}\f{GM}{B}.
\end{equation}
There is a unique transonic solution, which exists only for $1\le\gamma<5/3$.  (The case $\gamma=1$ can be treated separately or by taking a limit.)

Now evaluate $\dot M$ at the sonic point:
\begin{equation}
  |\dot M|=4\pi r_\text{s}^2\rho_\text{s}v_\text{ss}.
\end{equation}

\subsection{Second treatment}

We now use Bernoulli's equation instead of the equation of motion.

\begin{figure}
  \centerline{\epsfbox{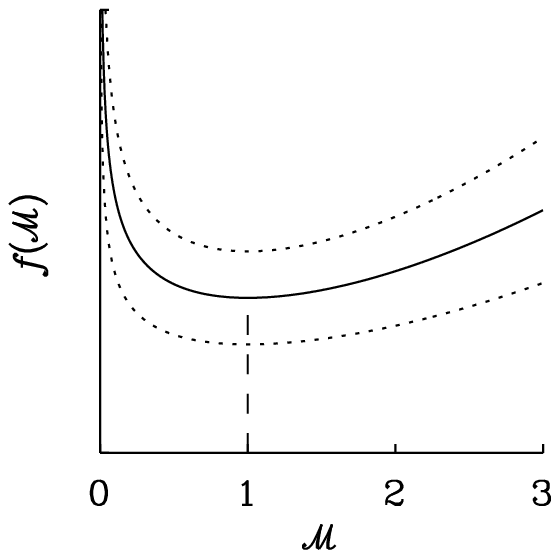}\qquad\epsfbox{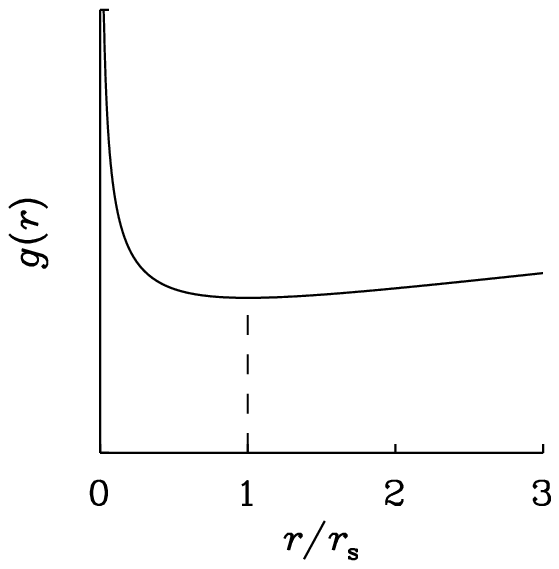}}
  \caption{Shapes of the functions $f(\mathcal{M})$ and $g(r)$ for the
    case $\gamma=4/3$.  Only if $\dot M$ is equal to the critical
    value at which the minima of $f$ and $g$ coincide (solid line,
    left panel) does a smooth transonic solution exist.}
\label{f:bernoulli}
\end{figure}

Introduce the local Mach number $\mathcal{M}=|u|/v_\text{s}$.  Then
\begin{equation}
  4\pi r^2\rho v_\text{s}\mathcal{M}=|\dot M|,\qquad
  v_\text{s}^2=\gamma K\rho^{\gamma-1}.
\end{equation}
Eliminate $\rho$ to obtain
\begin{equation}
  v_\text{s}=(\gamma K)^{1/(\gamma+1)}\left(\f{|\dot M|}{4\pi r^2\mathcal{M}}\right)^{(\gamma-1)/(\gamma+1)}.
\end{equation}
Bernoulli's equation~(\ref{bernoulli}) is
\begin{equation}
  \half v_\text{s}^2\mathcal{M}^2-\f{GM}{r}+\f{v_\text{s}^2}{\gamma-1}=B.
\end{equation}
Substitute for $v_\text{s}$ and separate the variables:
\begin{equation}
\begin{split}
  &(\gamma K)^{2/(\gamma+1)}\left(\f{|\dot M|}{4\pi}\right)^{2(\gamma-1)/(\gamma+1)}\left[\f{\mathcal{M}^{4/(\gamma+1)}}{2}+\f{\mathcal{M}^{-2(\gamma-1)/(\gamma+1)}}{\gamma-1}\right]\\
  &\qquad=Br^{4(\gamma-1)/(\gamma+1)}+GMr^{-(5-3\gamma)/(\gamma+1)}.
\end{split}
\end{equation}
This equation is of the form $f(\mathcal{M})=g(r)$.  Assume that
$1<\gamma<5/3$ and $B>0$.  (If $B<0$ then the flow cannot reach
infinity.)  Then each of $f$ and $g$ is the sum of a positive power
and a negative power, with positive coefficients.  $f(\mathcal{M})$
has a minimum at $\mathcal{M}=1$, while $g(r)$ has a minimum at
\begin{equation}
  r=\f{(5-3\gamma)}{4(\gamma-1)}\f{GM}{B},
\end{equation}
which is the sonic radius $r_\text{s}$ identified previously.  A
smooth passage through the sonic point is possible only if $|\dot M|$
has a special value, so that the minima of $f$ and $g$ are equal.  If
$|\dot M|$ is too large then the solution does not work for all $r$.
If it is too small then the solution remains subsonic (or supersonic)
for all $r$, which may not agree with the boundary
conditions (Figure~\ref{f:bernoulli}).

\begin{figure}
  \centerline{\epsfysize10cm\epsfbox{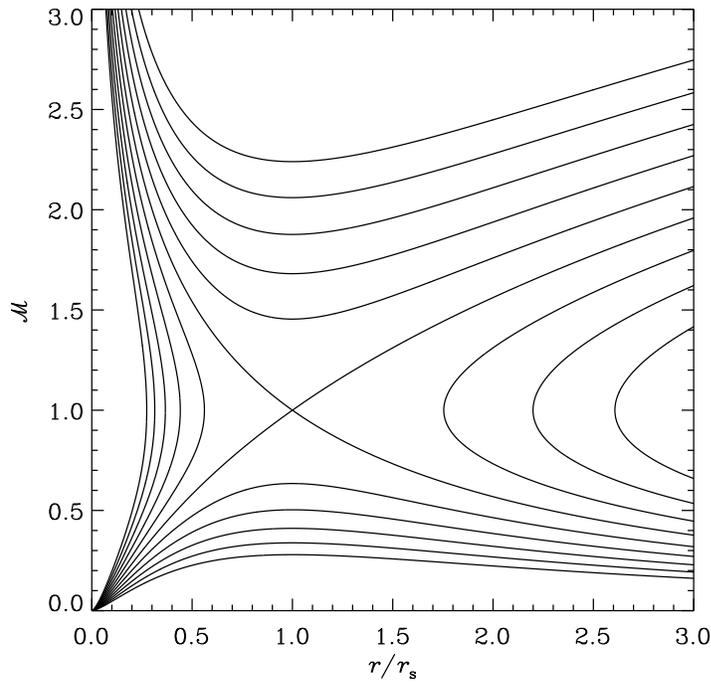}}
  \caption{Solution curves for a stellar wind or accretion flow in the
    case $\gamma=4/3$, showing an $\sfX$-type critical point at the
    sonic radius and at Mach number $\mathcal{M}=1$.}
\label{f:bondi}
\end{figure}

The $(r,\mathcal{M})$ plane shows an $\sfX$-type critical point at
$(r_\text{s},1)$ (Figure~\ref{f:bondi}).

For $r\ll r_\text{s}$ the subsonic solution is close to a hydrostatic
atmosphere.  The supersonic solution is close to free fall.

For $r\gg r_\text{s}$ the subsonic solution approaches a uniform state
($p=\cst$, $\rho=\cst$).  The supersonic solution is close to $u=\cst$
(so $\rho\propto r^{-2}$).

\subsection{Stellar wind}

For a stellar wind the appropriate solution is subsonic
(quasi-hydrostatic) at small $r$ and supersonic (coasting) at large
$r$.  \citet{Parker58}\footnote{Eugene Newman Parker (1927--),
  American.} first presented this simplified model for the solar wind.
The mass-loss rate can be determined from the properties of the
quasi-hydrostatic part of the solution, e.g.\ the density and
temperature at the base of the solar corona.  A completely hydrostatic
solution is unacceptable unless the external medium can provide a
significant non-zero pressure.  Subsonic solutions with $|\dot M|$
less than the critical value are usually unacceptable for similar
reasons.  (In fact the interstellar medium does arrest the supersonic
solar wind in a \textit{termination shock} well beyond Pluto's orbit.)

\subsection{Accretion}

In spherical or \citet{Bondi52}\footnote{Sir Hermann Bondi
  (1919--2005), Austrian--British.} accretion we consider a gas that
is uniform and at rest at infinity (with pressure $p_0$ and density
$\rho_0$).  Then $B=v_{\text{s}0}^2/(\gamma-1)$ and
$v_\text{ss}^2=2v_{\text{s}0}^2/(5-3\gamma)$.  The appropriate
solution is subsonic (uniform) at large $r$ and supersonic (freely
falling) at small $r$.  If the accreting object has a dense surface (a
star rather than a black hole) then the accretion flow will be
arrested by a shock above the surface.

The accretion rate of the critical solution is
\begin{equation}
  \dot M=4\pi r_\text{s}^2\rho_\text{s}v_\text{ss}=4\pi r_\text{s}^2\rho_0v_{\text{s}0}\left(\f{v_\text{ss}}{v_{\text{s}0}}\right)^{(\gamma+1)/(\gamma-1)}=f(\gamma)\dot M_\text{B},
\end{equation}
where
\begin{equation}
  \dot M_\text{B}=\f{\pi G^2M^2\rho_0}{v_{\text{s}0}^3}=4\pi r_\text{a}^2\rho_0v_{\text{s}0}
\end{equation}
is the characteristic Bondi accretion rate and
\begin{equation}
  f(\gamma)=\left(\f{2}{5-3\gamma}\right)^{(5-3\gamma)/2(\gamma-1)}
\end{equation}
is a dimensionless factor.  Here
\begin{equation}
  r_\text{a}=\f{GM}{2v_{\text{s}0}^2}
\end{equation}
is the nominal \emph{accretion radius}, roughly the radius within
which the mass $M$ captures the surrounding medium into a supersonic
inflow.

\textbf{Exercise}: show that
\begin{equation}
  \lim_{\gamma\to1}f(\gamma)=\rme^{3/2},\qquad
  \lim_{\gamma\to5/3}f(\gamma)=1.
\end{equation}
(However, although the case $\gamma=1$ admits a sonic point, the
important case $\gamma=5/3$ does not.)

At different times in its life a star may gain mass from, or lose mass
to, its environment.  Currently the Sun is losing mass at an average
rate of about $2\times10^{-14}\,M_\odot\,\mathrm{yr}^{-1}$.  If it
were not doing so, it could theoretically accrete at the Bondi rate of
about $3\times10^{-15}\,M_\odot\,\mathrm{yr}^{-1}$ from the
interstellar medium.

\medskip
\noindent Related examples: \ref{e:blast}, \ref{e:pw}, \ref{e:bondi}.

\newpage

\section{Axisymmetric rotating magnetized flows: astrophysical jets}

\noindent\textit{Note: in this section $(r,\phi,z)$ are cylindrical polar coordinates.}

\subsection{Introduction}

Stellar winds and jets from accretion discs are examples of outflows
in which rotation and magnetic fields have important or essential
roles.  Using cylindrical polar coordinates $(r,\phi,z)$, we examine
\emph{steady} ($\p/\p t=0$), \emph{axisymmetric}
($\p/\p\phi=0$) models based on the equations of
ideal MHD.

\subsection{Representation of an axisymmetric magnetic field}

\begin{figure}
  \centerline{\epsfysize9cm\epsfbox{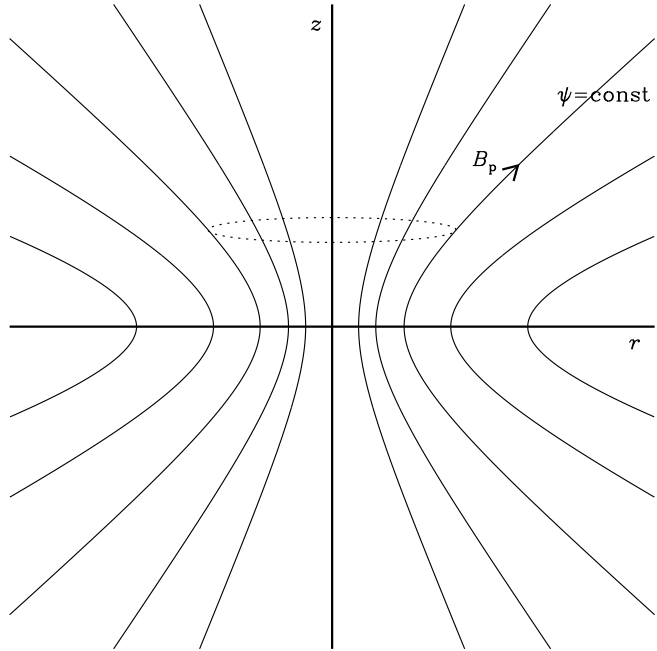}}
  \caption{Magnetic flux function and poloidal magnetic field.}
\label{f:flux_function}
\end{figure}

The solenoidal condition for an axisymmetric magnetic field is
\begin{equation}
  \f{1}{r}\f{\p}{\p r}(rB_r)+
  \f{\p B_z}{\p z}=0.
\end{equation}
We may write
\begin{equation}
  B_r=-\f{1}{r}\f{\p\psi}{\p z},\qquad
  B_z=\f{1}{r}\f{\p\psi}{\p r},
\end{equation}
where $\psi(r,z)$ is the \emph{magnetic flux function}
(Figure~\ref{f:flux_function}).  This is related to the magnetic
vector potential by $\psi=rA_\phi$.  The magnetic flux contained
inside the circle ($r=\cst$, $z=\cst$) is
\begin{equation}
  \int_0^rB_z(r',z)\,2\pi r'\,\rmd r'=2\pi\psi(r,z),
\end{equation}
plus an arbitrary constant that can be set to zero.

Since $\bmB\bcdot\grad\psi=0$, $\psi$ labels the magnetic field lines
or their surfaces of revolution, known as \emph{magnetic
  surfaces}. The magnetic field may be written in the form
\begin{equation}
  \bmB=\grad\psi\btimes\grad\phi+B_\phi\,\bme_\phi=
  \Bigg[-\f{1}{r}\bme_\phi\btimes\grad\psi\Bigg]+
  \Bigg[B_\phi\,\bme_\phi\Bigg].
\end{equation}
The two square brackets represent the \emph{poloidal} (meridional)
and \emph{toroidal} (azimuthal) parts of the magnetic field:
\begin{equation}
  \bmB=\bmB_\rmp+B_\phi\,\bme_\phi.
\end{equation}
Note that
\begin{equation}
  \div\bmB=\div\bmB_\rmp=0.
\end{equation}
Similarly, one can write the velocity in the form
\begin{equation}
  \bmu=\bmu_\rmp+u_\phi\,\bme_\phi,
\end{equation}
although $\div\bmu_\rmp\ne0$ in general.

\subsection{Mass loading and angular velocity}

The steady induction equation in ideal MHD,
\begin{equation}
  \curl(\bmu\btimes\bmB)=\bfzero,
\end{equation}
implies
\begin{equation}
  \bmu\btimes\bmB=-\bmE=\grad\Phi_\rme,
\end{equation}
where $\Phi_\rme$ is the electrostatic potential.  Now
\begin{equation}
\begin{split}
  \bmu\btimes\bmB&=(\bmu_\rmp+u_\phi\,\bme_\phi)\btimes
  (\bmB_\rmp+B_\phi\,\bme_\phi)\\
  &=
  \Bigg[\bme_\phi\btimes(u_\phi\bmB_\rmp-B_\phi\bmu_\rmp)\Bigg]+
  \Bigg[\bmu_\rmp\btimes\bmB_\rmp\Bigg].
\end{split}
\end{equation}
For an axisymmetric solution with $\p\Phi_{\rm
  e}/\p\phi=0$, we have
\begin{equation}
  \bmu_\rmp\btimes\bmB_\rmp=\bfzero,
\end{equation}
i.e.\ the poloidal velocity is parallel to the poloidal magnetic
field.\footnote{It is possible to consider a more general situation in
  which $rE_\phi$ is equal to a non-zero constant.  In this case there
  is a steady drift across the field lines and a steady transport of
  poloidal magnetic flux.  However, such a possibility is best
  considered in the context of non-ideal MHD, which allows both
  advective and diffusive transport of magnetic flux and angular
  momentum.}  Let
\begin{equation}
  \rho\bmu_\rmp=k\bmB_\rmp,
\end{equation}
where $k$ is the \emph{mass loading}, i.e.\ the ratio of mass flux
to magnetic flux.

The steady equation of mass conservation is
\begin{equation}
  0=\div(\rho\bmu)=\div(\rho\bmu_\rmp)=
  \div(k\bmB_\rmp)=\bmB_\rmp\bcdot\grad k.
\end{equation}
Therefore
\begin{equation}
  k=k(\psi),
\end{equation}
i.e.\ $k$ is a \emph{surface function}, constant on each magnetic surface.

We now have
\begin{equation}
  \bmu\btimes\bmB=\bme_\phi\btimes(u_\phi\bmB_\rmp-B_\phi\bmu_\rmp)=
  \left(\f{u_\phi}{r}-\f{kB_\phi}{r\rho}\right)\grad\psi.
\end{equation}
Taking the curl of this equation, we find
\begin{equation}
  \bfzero=\grad\left(\f{u_\phi}{r}-\f{kB_\phi}{r\rho}\right)
  \btimes\grad\psi.
\end{equation}
Therefore
\begin{equation}
  \f{u_\phi}{r}-\f{kB_\phi}{r\rho}=\omega,
\end{equation}
where $\omega(\psi)$ is another surface function, known as the \emph{angular velocity of the magnetic surface}.

The complete velocity field may be written in the form
\begin{equation}
  \bmu=\f{k\bmB}{\rho}+r\omega\,\bme_\phi,
\end{equation}
i.e.\ the total velocity is parallel to the total magnetic field in a
frame of reference rotating with angular velocity $\omega$.  It is
useful to think of the fluid being constrained to move along the field
line like a \emph{bead on a rotating wire}.

\subsection{Entropy}

The steady thermal energy equation,
\begin{equation}
  \bmu\bcdot\grad s=0,
\end{equation}
implies that $\bmB_\rmp\bcdot\grad s=0$ and so
\begin{equation}
  s=s(\psi)
\end{equation}
is another surface function.

\subsection{Angular momentum}

The azimuthal component of the equation of motion is
\begin{equation}
\begin{split}
  \rho\left(\bmu_\rmp\bcdot\grad u_\phi+\f{u_ru_\phi}{r}\right)&=
  \f{1}{\mu_0}\left(\bmB_\rmp\bcdot\grad B_\phi+
  \f{B_rB_\phi}{r}\right)\\
  \f{1}{r}\rho\bmu_\rmp\bcdot\grad(ru_\phi)-
  \f{1}{\mu_0r}\bmB_\rmp\bcdot\grad(rB_\phi)&=0\\
  \f{1}{r}\bmB_\rmp\bcdot\grad
  \left(kru_\phi-\f{rB_\phi}{\mu_0}\right)&=0,
\end{split}
\end{equation}
and so
\begin{equation}
  ru_\phi=\f{rB_{\phi}}{\mu_0k}+\ell,
\end{equation}
where
\begin{equation}
  \ell=\ell(\psi)
\end{equation}
is another surface function, the \emph{angular momentum invariant}.
This is the angular momentum removed in the outflow per unit mass,
although part of the torque is carried by the magnetic field.

\subsection{The Alfv\'en surface}

Define the \emph{poloidal Alfv\'en number} (cf. the Mach number)
\begin{equation}
  A=\f{u_\rmp}{v_\text{ap}}.
\end{equation}
Then
\begin{equation}
  A^2=\f{\mu_0\rho u_\rmp^2}{B_\rmp^2}=\f{\mu_0k^2}{\rho},
\end{equation}
and so $A\propto\rho^{-1/2}$ on each magnetic surface.

Consider the two equations
\begin{equation}
  \f{u_\phi}{r}=\f{kB_\phi}{r\rho}+\omega,\qquad
  ru_\phi=\f{rB_\phi}{\mu_0k}+\ell.
\end{equation}
Eliminate $B_\phi$ to obtain
\begin{equation}
  u_\phi=\f{r^2\omega-A^2\ell}{r(1-A^2)}=\left(\f{1}{1-A^2}\right)r\omega+
  \left(\f{A^2}{A^2-1}\right)\f{\ell}{r}.
\end{equation}
For $A\ll1$ we have
\begin{equation}
  u_\phi\approx r\omega,
\end{equation}
i.e.\ the fluid is in uniform rotation, corotating with the magnetic
surface.  For $A\gg1$ we have
\begin{equation}
  u_\phi\approx\f{\ell}{r},
\end{equation}
i.e.\ the fluid conserves its specific angular momentum.  The point
$r=r_\text{a}(\psi)$ where $A=1$ is the \emph{Alfv\'en point}.  The
locus of Alfv\'en points for different magnetic surfaces forms the
\emph{Alfv\'en surface}.  To avoid a singularity there we require
\begin{equation}
  \ell=r_\text{a}^2\omega.
\end{equation}

\begin{figure}
  \centerline{\epsfysize9cm\epsfbox{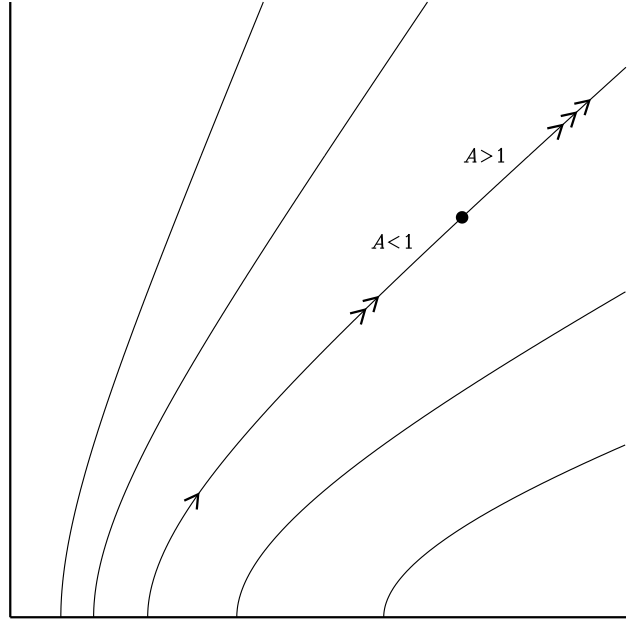}}
  \caption{Acceleration through an Alfv\'en point along a poloidal
    magnetic field line, leading to angular-momentum loss and magnetic
    braking.}
\label{f:magnetic_braking}
\end{figure}

Typically the outflow will start at low velocity in high-density
material, where $A\ll1$.  We can therefore identify $\omega$ as the
angular velocity $u_\phi/r=\Omega_0$ of the footpoint $r=r_0$ of the
magnetic field line at the source of the outflow.  It will then
accelerate smoothly through an Alfv\'en surface and become
super-Alfv\'enic ($A>1$).  If mass is lost at a rate $\dot M$ in the
outflow, angular momentum is lost at a rate $\dot M\ell=\dot
Mr_\text{a}^2\Omega_0$.  In contrast, in a hydrodynamic outflow,
angular momentum is conserved by fluid elements and is therefore lost
at a rate $\dot Mr_0^2\Omega_0$.  A highly efficient removal of
angular momentum occurs if the Alfv\'en radius is large compared to
the footpoint radius.  This effect is the \emph{magnetic lever arm}.
The loss of angular momentum through a stellar wind is called
\emph{magnetic braking} (Figure~\ref{f:magnetic_braking}).  In the
case of the Sun, the Alfv\'en radius is approximately between $20$ and
$30\,R_\odot$.

\subsection{The Bernoulli function}

The total energy equation for a steady flow is
\begin{equation}
  \div\left[\rho\bmu(\half u^2+\Phi+h)+
  \f{\bmE\btimes\bmB}{\mu_0}\right]=0.
\end{equation}
Now since
\begin{equation}
  \bmu=\f{k\bmB}{\rho}+r\omega\,\bme_\phi,
\end{equation}
we have
\begin{equation}
  \bmE=-\bmu\btimes\bmB=-r\omega\,\bme_\phi\btimes\bmB=-r\omega\,\bme_\phi\btimes\bmB_\rmp,
\end{equation}
which is purely poloidal.  Thus
\begin{equation}
  (\bmE\btimes\bmB)_\rmp=\bmE\btimes(B_\phi\,\bme_\phi)=
  -r\omega B_\phi\bmB_\rmp.
\end{equation}
The total energy equation is therefore
\begin{equation}
\begin{split}
  \div\left[k\bmB_\rmp(\half u^2+\Phi+h)-
  \f{r\omega B_\phi}{\mu_0}\bmB_\rmp\right]&=0\\
  \bmB_\rmp\bcdot\grad\left[k\left(\half u^2+\Phi+h-
  \f{r\omega B_\phi}{\mu_0k}\right)\right]&=0\\
  \half u^2+\Phi+h-\f{r\omega B_\phi}{\mu_0k}=
  \varepsilon,
\end{split}
\end{equation}
where
\begin{equation}
  \varepsilon=\varepsilon(\psi)
\end{equation}
is another surface function, the \emph{energy invariant}.

An alternative invariant is
\begin{equation}
  \begin{split}
  \tilde\varepsilon&=\varepsilon-\ell\omega\\
  &=\half u^2+\Phi+h-\f{r\omega B_\phi}{\mu_0k}-
  \left(ru_\phi-\f{rB_{\phi}}{\mu_0k}\right)\omega\\
  &=\half u^2+\Phi+h-ru_\phi\omega\\
  &=\half u_\rmp^2+
  \half(u_\phi-r\omega)^2+\Phi_\mathrm{cg}+h,
  \end{split}
\end{equation}
where
\begin{equation}
  \Phi_\mathrm{cg}=\Phi-\half\omega^2r^2
\end{equation}
is the centrifugal--gravitational potential associated with the
magnetic surface.  One can then see that $\tilde\varepsilon$ is the
Bernoulli function of the flow in the frame rotating with angular
velocity $\omega$.  In this frame the flow is strictly parallel to the
magnetic field and the field therefore does no work because
$\bmJ\btimes\bmB\perp\bmB$ and so
$\bmJ\btimes\bmB\perp(\bmu-r\omega\,\bme_\phi)$.

\subsection{Summary}

We have been able to integrate almost all of the MHD equations,
reducing them to a set of algebraic relations on each magnetic
surface.  If the poloidal magnetic field $\bmB_\rmp$ (or,
equivalently, the flux function $\psi$) is specified in advance, these
algebraic equations are sufficient to determine the complete solution
on each magnetic surface separately, although we must also (i) specify
the initial conditions at the source of the outflow and (ii) ensure
that the solution passes smoothly through critical points where the
flow speed matches the speeds of slow and fast magnetoacoustic waves
(see Example~\ref{e:critical}).

The component of the equation of motion perpendicular to the magnetic
surfaces is the only piece of information not yet used.  This
`transfield' or `Grad--Shafranov' equation ultimately determines the
equilibrium shape of the magnetic surfaces.  It is a very complicated
nonlinear partial differential equation for $\psi(r,z)$ and cannot be
reduced to simple terms.  We do not consider it here.

\subsection{Acceleration from the surface of an accretion disc}

We now consider the launching of an outflow from a thin accretion
disc.  The angular velocity $\Omega(r)$ of the disc corresponds
approximately to circular Keplerian orbital motion around a central
mass~$M$:
\begin{equation}
  \Omega\approx\left(\f{GM}{r^3}\right)^{1/2}
\end{equation}
If the flow starts essentially from rest in high-density material
($A\ll1$), we have
\begin{equation}
  \omega\approx\Omega,
\end{equation}
i.e.\ the angular velocity of the magnetic surface is the angular
velocity of the disc at the footpoint of the field line.  In the
sub-Alfv\'enic region we have
\begin{equation}
  \tilde\varepsilon\approx\half u_\rmp^2+\Phi_\mathrm{cg}+h.
\end{equation}

As in the case of stellar winds, if the gas is hot (comparable to the
escape temperature) an outflow can be driven by thermal pressure.  Of
more interest here is the possibility of a dynamically driven outflow.
For a `cold' wind the enthalpy makes a negligible contribution in this
equation.  Whether the flow accelerates or not above the disc then
depends on the variation of the centrifugal--gravitational potential
along the field line.

Consider a Keplerian disc in a point-mass potential.  Let the
footpoint of the field line be at $r=r_0$, and let the angular
velocity of the field line be
\begin{equation}
  \omega=\Omega_0=\left(\f{GM}{r_0^3}\right)^{1/2},
\end{equation}
as argued above.  Then
\begin{equation}
  \Phi_\mathrm{cg}=-GM(r^2+z^2)^{-1/2}-\f{1}{2}\f{GM}{r_0^3}r^2.
\end{equation}

\begin{figure}
  \centerline{\epsfysize10cm\epsfbox{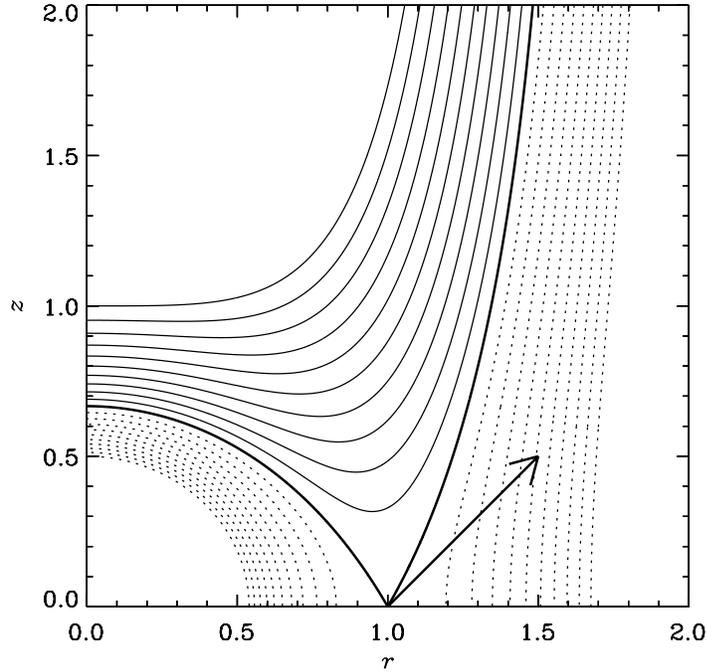}}
  \caption{Contours of $\Phi_\mathrm{cg}$, in units such that $r_0=1$.
    The downhill directions are indicated by dotted contours.  If the
    inclination of the poloidal magnetic field to the vertical
    direction at the surface of the disc exceeds $30^\circ$, gas is
    accelerated along the field lines away from the disc.}
\label{f:potential}
\end{figure}

In units such that $r_0=1$, the equation of the equipotential passing
through the footpoint $(r_0,z)$ is
\begin{equation}
  (r^2+z^2)^{-1/2}+\f{r^2}{2}=\f{3}{2}.
\end{equation}
This can be rearranged into the form
\begin{equation}
  z^2=\f{(2-r)(r-1)^2(r+1)^2(r+2)}{(3-r^2)^2}.
\end{equation}
Close to the footpoint $(1,0)$ we have
\begin{equation}
  z^2\approx3(r-1)^2\qquad\Rightarrow\qquad
  z\approx\pm\sqrt{3}(r-1).
\end{equation}
The footpoint lies at a saddle point of $\Phi_\mathrm{cg}$
(Figure~\ref{f:potential}).  If the inclination of the field line to
the vertical, $i$, at the surface of the disc exceeds $30^\circ$, the
flow is accelerated without thermal assistance.\footnote{A more
  detailed investigation \citep{Ogilvie98} shows that the Keplerian
  rotation of the disc is modified by the Lorentz force.  There is
  then a potential barrier $\propto B^4$ to be overcome by the
  outflow, even when $i>30^\circ$, which means that some thermal
  assistance is required, especially when the disc is strongly
  magnetized.}  This is \emph{magnetocentrifugal acceleration}.

The critical equipotential has an asymptote at $r=r_0\sqrt{3}$.  The
field line must continue to expand sufficiently in the radial
direction in order to sustain the magnetocentrifugal acceleration.

\subsection{Magnetically driven accretion}

To allow a quantity of mass $\Delta M_\text{acc}$ to be accreted from
radius $r_0$, its orbital angular momentum $r_0^2\Omega_0\,\Delta
M_\text{acc}$ must be removed.  The angular momentum removed by a
quantity of mass $\Delta M_\text{jet}$ flowing out in a magnetized jet
from radius $r_0$ is $\ell\,\Delta
M_\text{jet}=r_\text{a}^2\Omega_0\,\Delta M_\text{jet}$.  Therefore
accretion can in principle be driven by an outflow, with
\begin{equation}
  \f{\dot M_\text{acc}}{\dot M_\text{jet}}\approx\f{r_\text{a}^2}{r_0^2}.
\end{equation}
The magnetic lever arm allows an efficient removal of angular momentum
if the Alfv\'en radius is large compared to the footpoint radius.

\medskip
\noindent Related examples: \ref{e:rotating}, \ref{e:critical}.

\section{Lagrangian formulation of ideal MHD}

\subsection{The Lagrangian viewpoint}

In Section~\ref{s:waves} we will discuss waves and instabilities in
differentially rotating astrophysical bodies.  Here we develop a
general theory of disturbances to fluid flows that makes use of the
conserved quantities in ideal fluids and take a Lagrangian approach.

The flow of a fluid can be considered as a time-dependent map,
\begin{equation}
  \bma\mapsto\bmx(\bma,t),
\end{equation}
where $\bma$ is the position vector of a fluid element at some initial
time $t_0$, and $\bmx$ is its position vector at time $t$.  The
Cartesian components of $\bma$ are examples of \emph{Lagrangian
  variables}, labelling the fluid element.  The components of $\bmx$
are \emph{Eulerian variables}, labelling a fixed point in space.  Any
fluid property $X$ (scalar, vector or tensor) can be regarded as a
function of either Lagrangian or Eulerian variables:
\begin{equation}
  X=X^{\rm L}(\bma,t)=X^{\rm E}(\bmx,t).
\end{equation}
The Lagrangian time-derivative is simply
\begin{equation}
  \f{\rmD}{\rmD t}=\left(\f{\p}{\p t}\right)_{\bma},
\end{equation}
and the velocity of the fluid is
\begin{equation}
  \bmu=\f{\rmD\bmx}{\rmD t}=\left(\f{\p\bmx}{\p t}\right)_{\bma}.
\end{equation}

The aim of a Lagrangian formulation of ideal MHD is to derive a
nonlinear evolutionary equation for the function $\bmx(\bma,t)$.  The
dynamics is Hamiltonian in character and can be derived from a
Lagrangian function or action principle.  There are many similarities
with classical field theories.

\subsection{The deformation tensor}

Introduce the \emph{deformation tensor} of the flow,
\begin{equation}
  F_{ij}=\f{\p x_i}{\p a_j},
\end{equation}
and its determinant
\begin{equation}
  F=\det(F_{ij})
\end{equation}
and inverse
\begin{equation}
  G_{ij}=\f{\p a_i}{\p x_j}.
\end{equation}
We note the following properties.  First, the derivative
\begin{equation}
  \f{\p F}{\p F_{ij}}=C_{ij}=FG_{ji}=
  \f{1}{2}\epsilon_{ik\ell}\epsilon_{jmn}F_{km}F_{\ell n}
\label{cij}
\end{equation}
is equal to the cofactor $C_{ij}$ of the matrix element $F_{ij}$.
(This follows from the fact that the determinant can be expanded as
the sum of the products of any row's elements with their cofactors,
which do not depend on that row's elements.)  Second, the matrix of
cofactors has zero divergence on its second index:
\begin{equation}
  \f{\p C_{ij}}{\p a_j}=
  \f{\p}{\p a_j}\left(\f{1}{2}\epsilon_{ik\ell}
  \epsilon_{jmn}\f{\p x_k}{\p a_m}
  \f{\p x_\ell}{\p a_n}\right)=0.
\label{divcij}
\end{equation}
(This follows because the resulting derivative involves the
contraction of the antisymmetric tensor $\epsilon_{jmn}$ with
expressions that are symmetric in either $jm$ or $jn$.)

Now
\begin{equation}
  \f{\rmD F_{ij}}{\rmD t}=
  \f{\p u_i}{\p a_j}
\end{equation}
and, according to equation~(\ref{cij}),
\begin{equation}
  \f{\rmD\ln F}{\rmD t}=
  G_{ji}\f{\rmD F_{ij}}{\rmD t}=
  \f{\p a_j}{\p x_i}\f{\p u_i}{\p a_j}=
  \f{\p u_i}{\p x_i}=\div\bmu.
\end{equation}

\subsection{Geometrical conservation laws}

The equations of ideal MHD comprise the equation of motion and three
`geometrical' conservation laws.  These are the conservation of
specific entropy (thermal energy equation),
\begin{equation}
  \f{\rmD s}{\rmD t}=0,
\label{entropy}
\end{equation}
the conservation of mass,
\begin{equation}
  \f{\rmD\rho}{\rmD t}=-\rho\div\bmu,
\label{mass}
\end{equation}
and the conservation of magnetic flux (induction equation),
\begin{equation}
  \f{\rmD}{\rmD t}\left(\f{\bmB}{\rho}\right)=
  \left(\f{\bmB}{\rho}\right)\bcdot\grad\bmu.
\label{flux}
\end{equation}
These equations describe the pure advection of fluid properties in a
manner equivalent to the advection of various geometrical objects.
The specific entropy is advected as a simple scalar, so that its
numerical value is conserved by material points.  The specific volume
$v=1/\rho$ is advected in the same way as a material volume element
$\rmd V$.  The quantity $\bmB/\rho$ is advected in the same way as a
material line element $\delta\bmx$.  Equivalently, the mass element
$\delta m=\rho\,\delta V$ and the magnetic flux element
$\delta\Phi=\bmB\bcdot\delta\bmS$ satisfy $\rmD\,\delta m/\rmD t=0$
and $\rmD\,\delta\Phi/\rmD t=0$, where $\delta\bmS$ is a material
surface element.  All three conservation laws can be integrated
exactly in Lagrangian variables.

The exact solutions of equations (\ref{entropy}), (\ref{mass}) and
(\ref{flux}) are then
\begin{equation}
  s^{\rm L}(\bma,t)=s_0(\bma),\qquad
  \rho^{\rm L}(\bma,t)=F^{-1}\rho_0(\bma),\qquad
  B_i^{\rm L}(\bma,t)=F^{-1}F_{ij}B_{j0}(\bma),
\end{equation}
where $s_0$, $\rho_0$, and $\bmB_0$ are the initial values at time
$t_0$.  The verification of equation~(\ref{flux}) is
\begin{equation}
  \f{\rmD}{\rmD t}\left(\f{B_i}{\rho}\right)=
  \f{\rmD}{\rmD t}\left(\f{F_{ij}B_{j0}}{\rho_0}\right)=
  \f{\p u_i}{\p a_j}\f{B_{j0}}{\rho_0}=
  \f{\p u_i}{\p x_k}F_{kj}\f{B_{j0}}{\rho_0}=
  \left(\f{B_k}{\rho}\right)\f{\p u_i}{\p x_k}.
\end{equation}
Note that the advected quantities at time $t$ depend only on the
initial values and on the instantaneous mapping $\bma\mapsto\bmx$, not
on the intermediate history of the flow.  The `memory' of an ideal
fluid is perfect.

\subsection{The Lagrangian of ideal MHD}

Newtonian dynamics can be formulated using Hamilton's principle of
stationary action,
\begin{equation}
  \delta\int L\,\rmd t=0,
\end{equation}
where the Lagrangian $L$ is the difference between the kinetic energy
and the potential energy of the system.  By analogy, we may expect the
Lagrangian of ideal MHD to take the form
\begin{equation}
  L=\int{\cal L}\,\rmd V
\end{equation}
where (for a non-self-gravitating fluid)
\begin{equation}
  {\cal L}=\rho\left(\half u^2-\Phi-e-\f{B^2}{2\mu_0\rho}\right)
\end{equation}
is the \emph{Lagrangian density}.

To verify this, we assume that the equation of state can be written in
the form $e=e(v,s)$, where $v=\rho^{-1}$ is the specific volume.
Since $\rmd e=T\,\rmd s-p\,\rmd v$, we have
\begin{equation}
  \left(\f{\p e}{\p v}\right)_s=-p,\qquad
  \left(\f{\p^2e}{\p v^2}\right)_s=\f{\gamma p}{v}
\end{equation}
(strictly, $\gamma$ should be $\Gamma_1$ here).

We then write the action using Lagrangian variables,
\begin{equation}
  S[\bmx]=\iint\tilde{\cal L}(\bmx,\bmu,{\bf F})\,\rmd ^3\bma\,\rmd t,
\end{equation}
with
\begin{equation}
  \tilde{\cal L}=\rho_0\left[\f{1}{2}u^2-\Phi(\bmx)-e(F\rho_0^{-1},s_0)-
  \f{F^{-1}F_{ij}B_{j0}F_{ik}B_{k0}}{2\mu_0\rho_0}\right].
\end{equation}
This uses the fact that $F$ is the Jacobian determinant of the
transformation $\bma\mapsto\bmx$, or, equivalently, that
$\rho\,\rmd^3\bmx=\rho_0\,\rmd^3\bma=\rmd m$ is an invariant mass
measure.  $\tilde{\cal L}$ is now expressed in terms of the function
$\bmx(\bma,t)$ and its derivatives with respect to time ($\bmu$) and
space (${\bf F}$).  The Euler--Lagrange equation for the variational
principle $\delta S=0$ is
\begin{equation}
  \f{\rmD}{\rmD t}\f{\p\tilde{\cal L}}{\p u_i}+
  \f{\p}{\p a_j}\f{\p\tilde{\cal L}}{\p F_{ij}}-
  \f{\p\tilde{\cal L}}{\p x_i}=0.
\end{equation}
The straightforward terms are
\begin{equation}
  \f{\p\tilde{\cal L}}{\p u_i}=\rho_0u_i,\qquad
  \f{\p\tilde{\cal L}}{\p x_i}=
  -\rho_0\f{\p\Phi}{\p x_i}.
\end{equation}
Now
\begin{equation}
  \begin{split}
  \f{\p\tilde{\cal L}}{\p F_{ij}}&=
  \left(p+\f{B^2}{2\mu_0}\right)\f{\p F}{\p F_{ij}}-
  \f{F^{-1}B_{j0}F_{ik}B_{k0}}{\mu_0}\\
  &=C_{ij}\left(p+\f{B^2}{2\mu_0}\right)-
  \f{1}{\mu_0}C_{kj}B_iB_k\\
  &=-C_{kj}V_{ik},
  \end{split}
\end{equation}
where
\begin{equation}
  V_{ik}=-\left(p+\f{B^2}{2\mu_0}\right)\delta_{ik}+\f{B_iB_k}{\mu_0}
\end{equation}
is the stress tensor due to pressure and the magnetic field.

The Euler--Lagrange equation is therefore
\begin{equation}
  \rho_0\f{\rmD u_i}{\rmD t}=
  -\rho_0\f{\p\Phi}{\p x_i}+
  \f{\p}{\p a_j}(C_{kj}V_{ik}).
\end{equation}
Using equation~(\ref{divcij}) we note that
\begin{equation}
  \f{\p}{\p a_j}(C_{kj}V_{ik})=
  C_{kj}\f{\p V_{ik}}{\p a_j}=
  FG_{jk}\f{\p V_{ik}}{\p a_j}=
  F\f{\p V_{ik}}{\p x_k}.
\end{equation}
On dividing through by $F$, the Euler--Lagrange equation becomes the
desired equation of motion,
\begin{equation}
  \rho\f{\rmD\bmu}{\rmD t}=-\rho\grad\Phi+\div{\bf V}.
\end{equation}

In this construction, the fluid flow is viewed as a field
$\bmx(\bma,t)$ on the initial state space.  Ideal MHD is seen as a
nonlinear field theory derived from an action principle.  When
considering stability problems, it is useful to generalize this
concept and to view a perturbed flow as a field on an unperturbed
flow.

\subsection{The Lagrangian displacement}

Now consider two different flows, $\bmx(\bma,t)$ and $\hat\bmx(\bma,t)$, for
which the initial values of the advected quantities,
$s_0$, $\rho_0$ and $\bmB_0$, are the same.
The two deformation tensors are related by the chain rule,
\begin{equation}
  \hat F_{ij}=J_{ik}F_{kj},
\end{equation}
where
\begin{equation}
  J_{ik}=\f{\p\hat x_i}{\p x_k}
\end{equation}
is the Jacobian matrix of the map $\bmx\mapsto\hat\bmx$.  Similarly,
\begin{equation}
  \hat F=JF,
\end{equation}
where
\begin{equation}
  J=\det(J_{ij})
\end{equation}
is the Jacobian determinant.  The advected quantities in the two flows
are therefore related by the composition of maps,
\begin{equation}
\begin{split}
  \hat s^{\rm L}(\bma,t)&=s^{\rm L}(\bma,t),\\
  \hat\rho^{\rm L}(\bma,t)&=J^{-1}\rho^{\rm L}(\bma,t),\\
  \hat B_i^{\rm L}(\bma,t)&=J^{-1}J_{ij}B_j^{\rm L}(\bma,t).
\end{split}
\end{equation}

\begin{figure}
  \centerline{\epsfbox{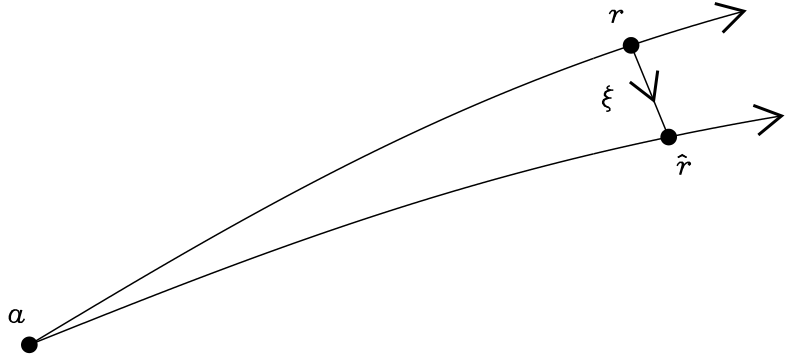}}
\caption{The Lagrangian displacement of a fluid element.}
\label{f:displacement}
\end{figure}

The \emph{Lagrangian displacement} (Figure~\ref{f:displacement}) is the relative displacement of
the fluid element in the two flows,
\begin{equation}
  \bxi=\hat\bmx-\bmx.
\end{equation}

Thus (with $\xi_{i,j}=\p\xi_i/\p x_j$)
\begin{equation}
\begin{split}
  J_{ij}&=\delta_{ij}+\xi_{i,j},\\
  J&=\f{1}{6}\epsilon_{ijk}\epsilon_{lmn}J_{il}J_{jm}J_{kn}\\
  &=\f{1}{6}\epsilon_{ijk}\epsilon_{lmn}(\delta_{il}+\xi_{i,l})(\delta_{jm}+\xi_{j,m})(\delta_{kn}+\xi_{k,n})\\
  &=\f{1}{6}\epsilon_{ijk}\epsilon_{ijk}+\f{1}{2}\epsilon_{ijk}\epsilon_{ijn}\xi_{k,n}+\f{1}{2}\epsilon_{ijk}\epsilon_{imn}\xi_{j,m}\xi_{k,n}+O(\xi^3)\\
  &=1+\xi_{k,k}+\f{1}{2}(\xi_{j,j}\xi_{k,k}-\xi_{j,k}\xi_{k,j})+O(\xi^3).
\end{split}
\end{equation}
From the binomial theorem,
\begin{equation}
  J^{-1}=1-\xi_{k,k}+\f{1}{2}(\xi_{j,j}\xi_{k,k}+\xi_{j,k}\xi_{k,j})+O(\xi^3).
\end{equation}

\subsection{The Lagrangian for a perturbed flow}

We now use the action principle to construct a theory for the
displacement as a field on the unperturbed flow: $\bxi=\bxi(\bmx,t)$.
The action for the perturbed flow is
\begin{equation}
  \hat S[\bxi]=\int\!\!\int\hat{\cal L}\left(\bxi,
  \f{\p\bxi}{\p t},\grad\bxi\right)
  \,\rmd ^3\bmx\,\rmd t,
\end{equation}
where
\begin{equation}
\begin{split}
  \hat{\cal L}&=\rho\left(\f{1}{2}\hat u^2-\hat\Phi-\hat e-
  \f{\hat B^2}{2\mu_0\hat\rho}\right)\\
  &=\rho\left[\f{1}{2}\hat u^2-\Phi(\bmx+\bxi)-
  e(J\rho^{-1},s)-\f{J^{-1}J_{ij}B_jJ_{ik}B_k}{2\mu_0\rho}\right],
\end{split}
\end{equation}
with
\begin{equation}
  \hat\bmu=\f{\rmD\hat\bmx}{\rmD t}=
  \bmu+\f{\p\bxi}{\p t}+\bmu\bcdot\grad\bxi.
\end{equation}
The Euler--Lagrange equation for the variational principle $\delta\hat S=0$
is
\begin{equation}
  \f{\p}{\p t}\f{\p\hat{\cal L}}{\p\dot\xi_i}+\f{\p}{\p x_j}\f{\p\hat{\cal L}}{\p\xi_{i,j}}-\f{\p\hat{\cal L}}{\p\xi_i}=0,
\end{equation}
where $\dot\xi_i=\p\xi_i/\p t$.  We expand the various terms of
$\hat{\cal L}$ in powers of $\bxi$:
\begin{equation}
\begin{split}
  &\f{1}{2}\rho\hat u^2=\f{1}{2}\rho u^2+
  \rho u_i\left(\f{\p\xi_i}{\p t}+
  u_j\f{\p\xi_i}{\p x_j}\right)+
  \f{1}{2}\rho\left(\f{\p\xi_i}{\p t}+
  u_j\f{\p\xi_i}{\p x_j}\right)
  \left(\f{\p\xi_i}{\p t}+
  u_k\f{\p\xi_i}{\p x_k}\right),\\
  &-\rho\Phi(\bmx+\bxi)=-\rho\left[\Phi(\bmx)+
  \xi_i\f{\p\Phi}{\p x_i}+
  \f{1}{2}\xi_i\xi_j\f{\p^2\Phi}{\p x_i\p x_j}+
  O(\xi^3)\right],\\
  &-\rho\left(\hat e+\f{\hat B^2}{2\mu_0\hat\rho}\right)=
  -\left(\rho e+\f{B^2}{2\mu_0}\right)-
  V_{ij}\f{\p\xi_i}{\p x_j}-
  \f{1}{2}V_{ijk\ell}\f{\p\xi_i}{\p x_j}
  \f{\p\xi_k}{\p x_\ell}+O(\xi^3).
\end{split}
\end{equation}
This last expression uses the fact that $\hat e$, $\hat\bmB$ and
$\hat\rho$ depend only on $\grad\bxi$ (through $J$ and $J_{ij}$) and
can therefore be expanded in a Taylor series in this quantity.  A
short calculation of this expansion gives
\begin{equation}
  V_{ij}=-\left(p+\f{B^2}{2\mu_0}\right)\delta_{ij}+
  \f{B_iB_j}{\mu_0},
\end{equation}
which is the stress tensor used above, and
\begin{equation}
  V_{ijk\ell}=\left[(\gamma-1)p+\f{B^2}{2\mu_0}\right]
  \delta_{ij}\delta_{k\ell}+
  \left(p+\f{B^2}{2\mu_0}\right)\delta_{i\ell}\delta_{jk}+
  \f{1}{\mu_0}B_jB_\ell\delta_{ik}-
  \f{1}{\mu_0}(B_iB_j\delta_{k\ell}+B_kB_\ell\delta_{ij}),
\end{equation}
which has the symmetry
\begin{equation}
  V_{ijk\ell}=V_{k\ell ij}
\end{equation}
necessitated by its function in the Taylor series.  We now have
\begin{equation}
\begin{split}
  \f{\p\hat{\cal L}}{\p\dot\xi_i}&=\rho u_i+
  \rho\f{\rmD\xi_i}{\rmD t},\\
  \f{\p\hat{\cal L}}{\p\xi_{i,j}}&=\rho u_iu_j+
  \rho u_j\f{\rmD\xi_i}{\rmD t}-
  V_{ij}-V_{ijk\ell}\f{\p\xi_k}{\p x_\ell}+O(\xi^2),\\
  \f{\p\hat{\cal L}}{\p\xi_i}&=
  -\rho\f{\p\Phi}{\p x_i}-
  \rho\xi_j\f{\p^2\Phi}{\p x_i\p x_j}+O(\xi^2).
\end{split}
\end{equation}

Since $\bxi=\bfzero$ must be a solution of the Euler--Lagrange
equation, it is no surprise that the terms independent of $\bxi$
cancel by virtue of the equation of motion of the unperturbed flow,
\begin{equation}
  \f{\p}{\p t}(\rho u_i)+\f{\p}{\p x_j}
  (\rho u_iu_j-V_{ij})+\rho\f{\p\Phi}{\p x_i}=0.
\end{equation}
The remaining terms are
\begin{equation}
  \f{\p}{\p t}
  \left(\rho\f{\rmD\xi_i}{\rmD t}\right)+
  \f{\p}{\p x_j}
  \left(\rho u_j\f{\rmD\xi_i}{\rmD t}-
  V_{ijk\ell}\f{\p\xi_k}{\p x_\ell}\right)+
  \rho\xi_j\f{\p^2\Phi}{\p x_i\p x_j}+O(\xi^2)=0,
\end{equation}
or (making use of the equation of mass conservation)
\begin{equation}
  \rho\f{\rmD^2\xi_i}{\rmD t^2}=
  \f{\p}{\p x_j}
  \left(V_{ijk\ell}\f{\p\xi_k}{\p x_\ell}\right)-
  \rho\xi_j\f{\p^2\Phi}{\p x_i\p x_j}+O(\xi^2).
\end{equation}

This equation, which can be extended to any order in $\bxi$, provides
the basis for a nonlinear perturbation theory for any flow in ideal
MHD.  In a linear theory we would neglect terms of $O(\xi^2)$.

\subsection{Notes on linear perturbations}

The \emph{Lagrangian perturbation} $\Delta X$ of a quantity $X$ is
the difference in the values of the quantity in the two flows \emph{for
the same fluid element},
\begin{equation}
  \Delta X=\hat X^{\rm L}(\bma,t)-X^{\rm L}(\bma,t).
\end{equation}
It follows that
\begin{equation}
  \Delta s=0,\qquad
  \Delta\rho=-\rho\f{\p\xi_i}{\p x_i}+O(\xi^2),\qquad
  \Delta B_i=B_j\f{\p\xi_i}{\p x_j}-
  B_i\f{\p\xi_j}{\p x_j}+O(\xi^2),
\end{equation}
and
\begin{equation}
  \Delta u_i=\f{\rmD\xi_i}{\rmD t}.
\end{equation}

In linear theory, $\grad\bxi$ is small and terms higher than the first
order are neglected.  Thus
\begin{equation}
  \Delta s=0,\qquad
  \Delta\rho=-\rho\div\bxi,\qquad
  \Delta\bmB=\bmB\bcdot\grad\bxi-(\div\bxi)\bmB.
\end{equation}
In linear theory, $\Delta s=0$ implies
\begin{equation}
  \Delta p=\f{\gamma p}{\rho}\Delta\rho=-\gamma p\div\bxi.
\end{equation}

The \emph{Eulerian perturbation} $\delta X$ of a quantity $X$ is the
difference in the values of the quantity in the two flows \emph{at the
  same point in space},
\begin{equation}
  \delta X=\hat X^{\rm E}(\bmx,t)-X^{\rm E}(\bmx,t).
\end{equation}
By Taylor's theorem,
\begin{equation}
  \Delta X=\delta X+\bxi\bcdot\grad X+O(\xi^2),
\end{equation}
and so, in linear theory,
\begin{equation}
  \delta X=\Delta X-\bxi\bcdot\grad X.
\end{equation}
Thus
\begin{equation}
\begin{split}
  \delta\rho&=-\rho\div\bxi-\bxi\bcdot\grad\rho,\\
  \delta p&=-\gamma p\div\bxi-\bxi\bcdot\grad p,\\
  \delta\bmB&=\bmB\bcdot\grad\bxi-\bxi\bcdot\grad\bmB-(\div\bxi)\bmB,
\end{split}
\end{equation}
exactly as was obtained in Section~\ref{s:linear} for perturbations of
magnetostatic equilibria.

The relation
\begin{equation}
  \delta\bmu=\f{\rmD\bxi}{\rmD t}-\bxi\bcdot\grad\bmu
\end{equation}
can be used to introduce the Lagrangian displacement into a linear
theory derived using Eulerian perturbations.  Only in the case of a
static basic state, $\bmu=\bfzero$, does this reduce to the simple
relation $\delta\bmu=\p\bxi/\p t$.

\newpage

\section{Waves and instabilities in stratified rotating astrophysical bodies}

\label{s:waves}

\subsection{The energy principle}

\label{s:energy}

For linear perturbations to a static equilibrium ($\bmu=\bfzero$), the
displacement satisfies
\begin{equation}
  \rho\f{\p^2\xi_i}{\p t^2}=-\rho\f{\p\,\delta\Phi}{\p x_i}-\rho\xi_j\f{\p^2\Phi}{\p x_i\p x_j}+\f{\p}{\p x_j}\left(V_{ijkl}\f{\p\xi_k}{\p x_l}\right),
\label{xiddot}
\end{equation}
where we now allow for self-gravitation through
\begin{equation}
  \delsq\,\delta\Phi=4\pi G\,\delta\rho=-4\pi G\div(\rho\bxi).
\end{equation}

We may write equation~(\ref{xiddot}) in the form
\begin{equation}
  \f{\p^2\bxi}{\p t^2}=\mathcal{F}\bxi,
\end{equation}
where $\mathcal{F}$ is a linear differential operator (or
integro-differential if self-gravitation is taken into account).  The
force operator $\mathcal{F}$ can be shown to be self-adjoint with
respect to the inner product,
\begin{equation}
  \langle\bmeta,\bxi\rangle=\int\rho\bmeta^*\bcdot\bxi\,\rmd V
\end{equation}
if appropriate boundary conditions apply to the vector fields $\bxi$
and $\bmeta$.  Let $\delta\Psi$ be the gravitational potential
perturbation associated with the displacement $\bmeta$, so
$\delsq\delta\Psi=-4\pi G\div(\rho\bmeta)$.  Then
\begin{equation}
\begin{split}
  \langle\bmeta,\mathcal{F}\bxi\rangle&=\int\left[-\rho\eta_i^*\f{\p\,\delta\Phi}{\p x_i}-\rho\eta_i^*\xi_j\f{\p^2\Phi}{\p x_i\p x_j}+\eta_i^*\f{\p}{\p x_j}\left(V_{ijkl}\f{\p\xi_k}{\p x_l}\right)\right]\,\rmd V\\
  &=\int\left[-\delta\Phi\f{\delsq\delta\Psi^*}{4\pi G}-\rho\xi_i\eta_j^*\f{\p^2\Phi}{\p x_i\p x_j}-V_{ijkl}\f{\p\xi_k}{\p x_l}\f{\p\eta_i^*}{\p x_j}\right]\,\rmd V\\
  &=\int\left[\f{\grad(\delta\Phi)\bcdot\grad(\delta\Psi^*)}{4\pi G}-\rho\xi_i\eta_j^*\f{\p^2\Phi}{\p x_i\p x_j}+\xi_k\f{\p}{\p x_l}\left(V_{ijkl}\f{\p\eta_i^*}{\p x_j}\right)\right]\,\rmd V\\
  &=\int\left[-\delta\Psi^*\f{\delsq\delta\Phi}{4\pi G}-\rho\xi_i\eta_j^*\f{\p^2\Phi}{\p x_i\p x_j}+\xi_i\f{\p}{\p x_j}\left(V_{klij}\f{\p\eta_k^*}{\p x_l}\right)\right]\,\rmd V\\
  &=\int\left[-\rho\xi_i\f{\p\,\delta\Psi^*}{\p x_i}-\rho\xi_i\eta_j^*\f{\p^2\Phi}{\p x_i\p x_j}+\xi_i\f{\p}{\p x_j}\left(V_{ijkl}\f{\p\eta_k^*}{\p x_l}\right)\right]\,\rmd V\\
  &=\langle\mathcal{F}\bmeta,\bxi\rangle.
\end{split}
\end{equation}
Here the integrals are over all space.  We assume that the exterior of
the body is a medium of zero density in which the force-free limit of
MHD holds and $\bmB$ decays sufficiently fast as $|\bmx|\to\infty$
that we may integrate freely by parts (using the divergence theorem)
and ignore surface terms.  We also assume that the body is isolated
and self-gravitating, so that $\delta\Phi=O(r^{-1})$, or in fact
$O(r^{-2})$ if $\delta M=0$.  We have used the symmetry properties of
$\p^2\Phi/\p x_i\p x_j$ and $V_{ijkl}$.

The functional
\begin{equation}
  W[\bxi]=-\f{1}{2}\langle\bxi,\mathcal{F}\bxi\rangle
  =\f{1}{2}\int\left(-\f{|\grad\delta\Phi|^2}{4\pi G}+\rho\xi_i^*\xi_j\f{\p^2\Phi}{\p x_i\p x_j}+V_{ijkl}\f{\p\xi_i^*}{\p x_j}\f{\p\xi_k}{\p x_l}\right)\,\rmd V
\end{equation}
is therefore real and represents the change in potential energy
associated with the displacement $\bxi$.

If the basic state is static, we may consider normal-mode solutions of
the form
\begin{equation}
  \bxi=\real\left[\tilde\bxi(\bmx)\exp(-\rmi\omega t)\right],
\end{equation}
for which we obtain
\begin{equation}
  -\omega^2\tilde\bxi=\mathcal{F}\tilde\bxi
\end{equation}
and
\begin{equation}
  \omega^2=-\f{\langle\tilde\bxi,\mathcal{F}\tilde\bxi\rangle}{\langle\tilde\bxi,\tilde\bxi\rangle}=\f{2W[\tilde\bxi]}{\langle\tilde\bxi,\tilde\bxi\rangle}.
\end{equation}
Therefore $\omega^2$ is real and we have either
oscillations ($\omega^2>0$) or instability ($\omega^2<0$).

The above expression for $\omega^2$ satisfies the usual Rayleigh--Ritz
variational principle for self-adjoint eigenvalue problems.  The
eigenvalues $\omega^2$ are the stationary values of
$2W[\bxi]/\langle\bxi,\bxi\rangle$ among trial displacements $\bxi$
satisfying the boundary conditions.  In particular, the lowest
eigenvalue is the global minimum value of
$2W[\bxi]/\langle\bxi,\bxi\rangle$.  Therefore the equilibrium is
unstable if and only if $W[\bxi]$ can be made negative by a trial
displacement~$\bxi$ satisfying the boundary conditions.  This is
called the \emph{energy principle}.

This discussion is incomplete because it assumes that the
eigenfunctions form a complete set.  In general a continuous spectrum,
not associated with square-integrable modes, is also present.  However,
it can be shown that a necessary and sufficient condition for
instability is that $W[\bxi]$ can be made negative as described above.
Consider the equation for twice the energy of the perturbation,
\begin{equation}
\begin{split}
  \f{\rmd}{\rmd t}\left(\langle\dot\bxi,\dot\bxi\rangle+2W[\bxi]\right)&=\langle\ddot\bxi,\dot\bxi\rangle+\langle\dot\bxi,\ddot\bxi\rangle-\langle\dot\bxi,\mathcal{F}\bxi\rangle-\langle\bxi,\mathcal{F}\dot\bxi\rangle\\
  &=\langle\mathcal{F}\bxi,\dot\bxi\rangle+\langle\dot\bxi,\mathcal{F}\bxi\rangle-\langle\dot\bxi,\mathcal{F}\bxi\rangle-\langle\mathcal{F}\bxi,\dot\bxi\rangle\\
  &=0.
\end{split}
\end{equation}
Therefore
\begin{equation}
  \langle\dot\bxi,\dot\bxi\rangle+2W[\bxi]=2E=\cst,
\end{equation}
where $E$ is determined by the initial data $\bxi_0$ and
$\dot\bxi_0$.  If $W$ is positive definite then the equilibrium is
stable because $\bxi$ is limited by the constraint $W[\bxi]\le E$.

Suppose that a (real) trial displacement $\bmeta$ can be found for
which
\begin{equation}
  \f{2W[\bmeta]}{\langle\bmeta,\bmeta\rangle}=-\gamma^2,
\end{equation}
where $\gamma>0$.  Then let the initial conditions be $\bxi_0=\bmeta$ and
$\dot\bxi_0=\gamma\bmeta$ so that
\begin{equation}
  \langle\dot\bxi,\dot\bxi\rangle+2W[\bxi]=2E=0.
\end{equation}
Now let
\begin{equation}
  a(t)=\ln\left(\f{\langle\bxi,\bxi\rangle}{\langle\bmeta,\bmeta\rangle}\right)
\end{equation}
so that
\begin{equation}
  \f{\rmd a}{\rmd t}=\f{2\langle\bxi,\dot\bxi\rangle}{\langle\bxi,\bxi\rangle}
\end{equation}
and
\begin{equation}
\begin{split}
  \f{\rmd^2a}{\rmd t^2}&=\f{2(\langle\bxi,\mathcal{F}\bxi\rangle+\langle\dot\bxi,\dot\bxi\rangle)\langle\bxi,\bxi\rangle-4\langle\bxi,\dot\bxi\rangle^2}{\langle\bxi,\bxi\rangle^2}\\
  &=\f{2(-2W[\bxi]+\langle\dot\bxi,\dot\bxi\rangle)\langle\bxi,\bxi\rangle-4\langle\bxi,\dot\bxi\rangle^2}{\langle\bxi,\bxi\rangle^2}\\
  &=\f{4(\langle\dot\bxi,\dot\bxi\rangle\langle\bxi,\bxi\rangle-\langle\bxi,\dot\bxi\rangle^2)}{\langle\bxi,\bxi\rangle^2}\\
  &\ge0
\end{split}
\end{equation}
by the Cauchy--Schwarz inequality.  Thus
\begin{equation}
\begin{split}
  \f{\rmd a}{\rmd t}&\ge\dot a_0=2\gamma\\
  a&\ge2\gamma t+a_0=2\gamma t.
\end{split}
\end{equation}
Therefore the disturbance with these initial conditions grows at least
as fast as $\exp(\gamma t)$ and the equilibrium is unstable.

\subsection{Spherically symmetric star}

The simplest model of a star neglects rotation and magnetic fields and
assumes a spherically symmetric hydrostatic equilibrium in which
$\rho(r)$ and $p(r)$ satisfy
\begin{equation}
  \f{\rmd p}{\rmd r}=-\rho g,
\end{equation}
with inward radial gravitational acceleration
\begin{equation}
  g(r)=\f{\rmd\Phi}{\rmd r}=\f{G}{r^2}\int_0^r\rho(r')\,4\pi r^{\prime2}\,\rmd r'.
\end{equation}
The stratification induced by gravity provides a non-uniform background for wave propagation.

In this case the linearized equation of motion is (cf.\ Section~\ref{s:linear})
\begin{equation}
  \rho\f{\p^2\bxi}{\p t^2}=-\rho\grad\delta\Phi-\delta\rho\grad\Phi-\grad\delta p,
\end{equation}
with $\delta\rho=-\div(\rho\bxi)$, $\delsq\delta\Phi=4\pi
G\,\delta\rho$ and $\delta p=-\gamma p\div\bxi-\bxi\bcdot\grad p$.
For normal modes $\propto\exp(-\rmi\omega t)$,
\begin{equation}
\begin{split}
  \rho\omega^2\bxi&=\rho\grad\delta\Phi+\delta\rho\grad\Phi+\grad\delta p\\
  \omega^2\int_V\rho|\bxi|^2\,\rmd V&=\int_V\bxi^*\bcdot(\rho\grad\delta\Phi+\delta\rho\grad\Phi+\grad\delta p)\,\rmd V,
\end{split}
\end{equation}
where $V$ is the volume of the star.  At the surface $S$ of the star,
we assume that $\rho$ and $p$ vanish.  Then $\delta p$ also vanishes
on $S$ (assuming that $\bxi$ and its derivatives are bounded).

The $\delta p$ term can be integrated by parts as follows:
\begin{equation}
\begin{split}
  \int_V\bxi^*\bcdot\grad\delta p\,\rmd V&=-\int_V(\div\bxi)^*\delta p\,\,\rmd V\\
  &=\int_V\f{1}{\gamma p}(\delta p+\bxi\bcdot\grad p)^*\delta p\,\,\rmd V\\
  &=\int_V\left[\f{|\delta p|^2}{\gamma p}+\f{1}{\gamma p}(\bxi^*\bcdot\grad p)(-\bxi\bcdot\grad p-\gamma p\div\bxi)\right]\,\rmd V.
\end{split}
\end{equation}
The $\delta\rho$ term partially cancels with the above:
\begin{equation}
\begin{split}
  \int_V\bxi^*\bcdot(\delta\rho\grad\Phi)&=\int_V(-\bxi^*\bcdot\grad p)\f{\delta\rho}{\rho}\,\rmd V\\
  &=\int_V(\bxi^*\bcdot\grad p)(\div\bxi+\bxi\bcdot\grad\ln\rho)\,\rmd V.
\end{split}
\end{equation}
Finally, the $\delta\Phi$ term can be transformed as in Section~\ref{s:energy} to give
\begin{equation}
  \int_V\rho\bxi^*\bcdot\grad\delta\Phi\,\rmd V=-\int_\infty\f{|\grad\delta\Phi|^2}{4\pi G}\,\rmd V,
\end{equation}
where the integral on the right-hand side is over all space.  Thus
\begin{equation}
\begin{split}
  \omega^2\int_V\rho|\bxi|^2\,\rmd V&=-\int_\infty\f{|\grad\delta\Phi|^2}{4\pi G}\,\rmd V+\int_V\left[\f{|\delta p|^2}{\gamma p}-(\bxi^*\bcdot\grad p)\bcdot\left(\f{1}{\gamma}\bxi\bcdot\grad\ln p-\bxi\bcdot\grad\ln\rho\right)\right]
  \,\rmd V\\
  &=-\int_\infty\f{|\grad\delta\Phi|^2}{4\pi G}\,\rmd V+\int_V\left(\f{|\delta p|^2}{\gamma p}+\rho N^2|\xi_r|^2\right)\,\rmd V,
\end{split}
\end{equation}
where $N(r)$ is the \emph{Brunt--V\"ais\"al\"a frequency}\footnote{Sir
  David Brunt (1886--1965), British.  Vilho V\"ais\"al\"a
  (1889--1969), Finnish.} (or \emph{buoyancy frequency}) given by
\begin{equation}
  N^2=g\left(\f{1}{\gamma}\f{\rmd\ln p}{\rmd r}-\f{\rmd\ln\rho}{\rmd r}\right)\propto g\f{\rmd s}{\rmd r}.
\end{equation}
$N$ is the frequency of oscillation of a fluid element that is
displaced vertically in a stably stratified atmosphere if it maintains
pressure equilibrium with its surroundings.  The stratification is
stable if the specific entropy increases outwards.

The integral expression for $\omega^2$ satisfies the energy principle.
There are three contributions to $\omega^2$: the self-gravitational
term (destabilizing), the acoustic term (stabilizing) and the buoyancy
term (stabilizing if $N^2>0$).

If $N^2<0$ for any interval of $r$, a trial displacement can always be
found such that $\omega^2<0$.  This is done by localizing $\xi_r$ in
that interval and arranging the other components of $\bxi$ such that
$\delta p=0$.  Therefore the star is unstable if $\p s/\p r<0$
anywhere.  This is \emph{Schwarzschild's criterion}\footnote{Karl Schwarzschild (1873--1916), German.} for convective
instability.

\subsection{Modes of an incompressible sphere}

\noindent\textit{Note: in this subsection $(r,\theta,\phi)$ are spherical polar coordinates.}

Analytical solutions can be obtained in the case of a homogeneous
incompressible `star' of mass $M$ and radius $R$ which has
\begin{equation}
  \rho=\left(\f{3M}{4\pi R^3}\right)H(R-r),
\end{equation}
where $H$ is the Heaviside step function.  For $r\le R$ we have
\begin{equation}
  g=\f{GMr}{R^3},\qquad
  p=\f{3GM^2(R^2-r^2)}{8\pi R^6}.
\end{equation}
For an incompressible fluid,
\begin{equation}
  \div\bxi=0,
\end{equation}
\begin{equation}
  \delta\rho=-\bxi\bcdot\grad\rho=\xi_r\left(\f{3M}{4\pi R^3}\right)\delta(r-R)
\end{equation}
and
\begin{equation}
  \delsq\delta\Phi=4\pi G\,\delta\rho=\xi_r\left(\f{3GM}{R^3}\right)\delta(r-R),
\label{delsqdeltaphi}
\end{equation}
while $\delta p$ is indeterminate and is a variable independent of
$\bxi$.  The linearized equation of motion is
\begin{equation}
  -\rho\omega^2\bxi=-\rho\grad\delta\Phi-\grad\delta p.
\end{equation}
Therefore we have potential flow: $\bxi=\grad U$, with $\delsq U=0$
and $-\rho\omega^2 U=-\rho\,\delta\Phi-\delta p$ in $r\le R$.
Appropriate solutions of Laplace's equation regular at $r=0$ are the
solid spherical harmonics (with arbitrary normalization)
\begin{equation}
  U=r^\ell Y_\ell^m(\theta,\phi),
\end{equation}
where $\ell$ and $m$ are integers with $\ell\ge|m|$.
Equation~(\ref{delsqdeltaphi}) also implies
\begin{equation}
  \delta\Phi=
  \begin{cases}
    Ar^\ell Y_\ell^m,&r<R,\\
    Br^{-\ell-1}Y_\ell^m,&r>R,
  \end{cases}
\end{equation}
where $A$ and $B$ are constants to be determined.  The matching
conditions from equation~(\ref{delsqdeltaphi}) at $r=R$ are
\begin{equation}
  [\delta\Phi]=0,\qquad
  \left[\f{\p\,\delta\Phi}{\p r}\right]=\xi_r\left(\f{3GM}{R^3}\right).
\end{equation}
Thus
\begin{equation}
  BR^{-\ell-1}-AR^\ell=0,\qquad
  -(\ell+1)BR^{-\ell-2}-\ell AR^{\ell-1}=\ell R^{\ell-1}\left(\f{3GM}{R^3}\right),
\end{equation}
with solution
\begin{equation}
  A=-\f{\ell}{2\ell+1}\left(\f{3GM}{R^3}\right),\qquad B=AR^{2\ell+1}.
\end{equation}
At $r=R$ the Lagrangian pressure perturbation should vanish:
\begin{equation}
\begin{split}
  \Delta p&=\delta p+\bxi\bcdot\grad p=0\\
  \left(\f{3M}{4\pi R^3}\right)&\left[\omega^2R^\ell+\left(\f{\ell}{2\ell+1}\right)\left(\f{3GM}{R^3}\right)R^\ell\right]-\f{3GM^2}{4\pi R^5}\ell R^{\ell-1}=0\\
  \omega^2&=\left(\ell-\f{3\ell}{2\ell+1}\right)\f{GM}{R^2}=\f{2\ell(\ell-1)}{2\ell+1}\f{GM}{R^3}.
\end{split}
\end{equation}
This result was obtained by Lord Kelvin.  Since $\omega^2\ge0$ the
star is stable.  Note that $\ell=0$ corresponds to $\bxi=\bfzero$ and
$\ell=1$ corresponds to $\bxi=\cst$, which is a translational mode of
zero frequency.  The remaining modes are non-trivial and are called
\emph{f~modes} (fundamental modes).  These can be thought of as
surface gravity waves, related to ocean waves for which $\omega^2=gk$.
In the first expression for $\omega^2$ above, the first term in
brackets derives from surface gravity, while the second derives from
self-gravity.

\subsection{The plane-parallel atmosphere}

The local dynamics of a stellar atmosphere can be studied in a
Cartesian (`plane parallel') approximation.  The gravitational
acceleration is taken to be constant (appropriate to an atmosphere)
and in the $-z$ direction.  For hydrostatic equilibrium,
\begin{equation}
  \f{\rmd p}{\rmd z}=-\rho g.
\end{equation}
A simple example is an \emph{isothermal atmosphere} in which $p=c_\text{s}^2\rho$ with $c_\text{s}=\cst$:
\begin{equation}
  \rho=\rho_0\,\rme^{-z/H},\qquad
  p=p_0\,\rme^{-z/H}.
\end{equation}
$H=c_\text{s}^2/g$ is the \emph{isothermal scale-height}.  The Brunt--V\"ais\"al\"a frequency in an isothermal atmosphere is given by
\begin{equation}
  N^2=g\left(\f{1}{\gamma}\f{\rmd\ln p}{\rmd z}-\f{\rmd\ln\rho}{\rmd z}\right)=\left(1-\f{1}{\gamma}\right)\f{g}{H},
\end{equation}
which is constant and is positive for $\gamma>1$.  An isothermal
atmosphere is \emph{stably (subadiabatically) stratified} if $\gamma>1$ and \emph{neutrally (adiabatically) stratified} if $\gamma=1$.

A further example is a \emph{polytropic atmosphere} in which
$p\propto\rho^{1+1/m}$ in the undisturbed state, where $m$ is a
positive constant.  In general $1+1/m$ differs from the adiabatic
exponent $\gamma$ of the gas.  For hydrostatic equilibrium,
\begin{equation}
  \rho^{1/m}\f{\rmd\rho}{\rmd z}\propto-\rho g\quad\Rightarrow\quad
  \rho^{1/m}\propto-z,
\end{equation}
if the top of the atmosphere is located at $z=0$, with vacuum above.
Let
\begin{equation}
  \rho=\rho_0\left(-\f{z}{H}\right)^m,
\end{equation}
for $z<0$, where $\rho_0$ and $H$ are constants.  Then
\begin{equation}
  p=p_0\left(-\f{z}{H}\right)^{m+1},
\end{equation}
where
\begin{equation}
  p_0=\f{\rho_0gH}{m+1}
\end{equation}
to satisfy $\rmd p/\rmd z=-\rho g$.  In this case
\begin{equation}
  N^2=\left(m-\f{m+1}{\gamma}\right)\f{g}{-z}.
\end{equation}

We return to the linearized equations, looking for solutions of the form
\begin{equation}
  \bxi=\real\left[\tilde\bxi(z)\exp(-\rmi\omega t+\rmi\bmk_\text{h}\bcdot\bmx)\right],\qquad\text{etc.}
\end{equation}
where `h' denotes a horizontal vector (having only $x$ and $y$
components).  Then
\begin{equation}
\begin{split}
  -\rho\omega^2\bxi_\text{h}&=-\rmi\bmk_\text{h}\,\delta p,\\
  -\rho\omega^2\xi_z&=-g\,\delta\rho-\f{\rmd\,\delta p}{\rmd z},\\
  \delta\rho&=-\xi_z\f{\rmd\rho}{\rmd z}-\rho\Delta,\\
  \delta p&=-\xi_z\f{\rmd p}{\rmd z}-\gamma p\Delta,
\end{split}
\end{equation}
where
\begin{equation}
  \Delta=\div\bxi=\rmi\bmk_\text{h}\bcdot\bxi_\text{h}+\f{\rmd\xi_z}{\rmd z}.
\end{equation}
The self-gravitation of the perturbation is neglected in the
atmosphere: $\delta\Phi=0$ (the Cowling approximation).  Note that
only two $z$-derivatives of perturbation quantities occur:
$\rmd\,\delta p/\rmd z$ and $\rmd\xi_z/\rmd z$.  This is a
second-order system of ordinary differential equations (ODEs),
combined with algebraic equations.

We can easily eliminate $\bxi_\rmh$ to obtain
\begin{equation}
  \Delta=-\f{k_\rmh^2}{\rho\omega^2}\,\delta p+\f{\rmd\xi_z}{\rmd z},
\end{equation}
where $k_\text{h}=|\bmk_\text{h}|$, and eliminate $\delta\rho$ to
obtain
\begin{equation}
  -\rho\omega^2\xi_z=g\xi_z\f{\rmd\rho}{\rmd z}+\rho g\Delta-\f{\rmd\,\delta p}{\rmd z}.
\end{equation}
We consider these two differential equations in combination with the
remaining algebraic equation
\begin{equation}
  \delta p=\rho g\xi_z-\gamma p\Delta.
\end{equation}

A first approach is to solve the algebraic equation for $\Delta$ and
substitute to obtain the two coupled ODEs
\begin{equation}
\begin{split}
  \f{\rmd\xi_z}{\rmd z}&=\f{g}{v_\rms^2}\xi_z+\f{1}{\rho v_\rms^2}\left(\f{v_\rms^2k_\rmh^2}{\omega^2}-1\right)\delta p,\\
  \f{\rmd\,\delta p}{\rmd z}&=\rho(\omega^2-N^2)\xi_z-\f{g}{v_\rms^2}\delta p.
\end{split}
\end{equation}
Note that $v_\rms^2k_\rmh^2$ is the square of the `Lamb frequency',
i.e.\ the ($z$-dependent) frequency of a horizontal sound wave of
wavenumber $k_\rmh$.  In a short-wavelength (WKB) approximation, where
$\xi_z\propto\exp\left[\rmi\int k_z(z)\,\rmd z\right]$ with $k_z\gg
g/v_\rms^2$, the local dispersion relation derived from these ODEs is
\begin{equation}
  v_\rms^2k_z^2=(\omega^2-N^2)\left(1-\f{v_\rms^2k_\rmh^2}{\omega^2}\right).
\end{equation}
Propagating waves ($k_z^2>0$) are possible when
\begin{equation}
  \text{either}\quad\omega^2>\max(v_\rms^2k_\rmh^2,N^2)\qquad
  \text{or}\quad0<\omega^2<\min(v_\rms^2k_\rmh^2,N^2).
\end{equation}
The high-frequency branch describes \emph{p modes} (acoustic waves:
`p' for pressure) while the low-frequency branch describes \emph{g
  modes} (internal gravity waves: `g' for gravity).

There is a special incompressible solution in which $\Delta=0$, i.e.\
$\delta p=\rho g\xi_z$.  This satisfies
\begin{equation}
  \f{\rmd\xi_z}{\rmd z}=\f{gk_\rmh^2}{\omega^2}\xi_z,\qquad
  \f{\rmd\xi_z}{\rmd z}=\f{\omega^2}{g}\xi_z.
\end{equation}
For compatibility of these equations,
\begin{equation}
  \f{gk_\text{h}^2}{\omega^2}=\f{\omega^2}{g}\quad\Rightarrow\quad
  \omega^2=\pm gk_\text{h}.
\end{equation}
The acceptable solution in which $\xi_z$ decays with depth is
\begin{equation}
  \omega^2=gk_\text{h},\qquad\xi_z\propto\exp(k_\text{h}z).
\end{equation}
This is a \emph{surface gravity wave} known in stellar oscillations as
the \emph{f mode} (fundamental mode).  It is vertically evanescent.

The other wave solutions (p and g modes) can be found analytically in
the case of a polytropic atmosphere.\footnote{\citet{Lamb32},
  Art.~312.}  We now eliminate variables in favour of $\Delta$.  First
use the algebraic relation to eliminate $\delta p$:
\begin{equation}
  \Delta=-\f{gk_\rmh^2}{\omega^2}\xi_z+\f{v_\rms^2k_\rmh^2}{\omega^2}\Delta+\f{\rmd\xi_z}{\rmd z},
\end{equation}
\begin{equation}
  -\rho\omega^2\xi_z=\rho g\Delta-\rho g\f{\rmd\xi_z}{\rmd z}+\f{\rmd(\gamma p\Delta)}{\rmd z}.
\end{equation}
Then eliminate $\rmd\xi_z/\rmd z$:
\begin{equation}
  -\rho\omega^2\xi_z=-\rho g\left(\f{gk_\rmh^2}{\omega^2}\xi_z-\f{v_\rms^2k_\rmh^2}{\omega^2}\Delta\right)+\f{\rmd(\gamma p\Delta)}{\rmd z}.
\end{equation}
Thus we have
\begin{equation}
 \f{\rmd\xi_z}{\rmd z}-\f{gk_\rmh^2}{\omega^2}\xi_z=\left(1-\f{v_\rms^2k_\rmh^2}{\omega^2}\right)\Delta,
\end{equation}
\begin{equation}
  -\omega^2\left(1-\f{g^2k_\rmh^2}{\omega^4}\right)\xi_z=\f{1}{\rho}\f{\rmd(\rho v_\rms^2\Delta)}{\rmd z}+v_\rms^2\f{gk_\rmh^2}{\omega^2}\Delta.
\end{equation}
Combine, eliminating $\xi_z$:
\begin{equation}
  \left(\f{\rmd}{\rmd z}-\f{gk_\rmh^2}{\omega^2}\right)\left[v_\rms^2\f{\rmd\Delta}{\rmd z}+\f{1}{\rho}\f{\rmd(\rho v_\rms^2)}{\rmd z}\Delta+v_\rms^2\f{gk_\rmh^2}{\omega^2}\Delta\right]+\omega^2\left(1-\f{g^2k_\rmh^2}{\omega^4}\right)\left(1-\f{v_\rms^2k_\rmh^2}{\omega^2}\right)\Delta=0.
\end{equation}
Expand out:
\begin{equation}
\begin{split}
  &v_\rms^2\f{\rmd^2\Delta}{\rmd z^2}+\left[\f{\rmd v_\rms^2}{\rmd z}+\f{1}{\rho}\f{\rmd(\rho v_\rms^2)}{\rmd z}\right]\f{\rmd\Delta}{\rmd z}+\bigg[\f{\rmd}{\rmd z}\left(\f{1}{\rho}\f{\rmd(\rho v_\rms^2)}{\rmd z}\right)+\f{\rmd v_\rms^2}{\rmd z}\f{gk_\rmh^2}{\omega^2}-\f{gk_\rmh^2}{\omega^2}\f{1}{\rho}\f{\rmd(\rho v_\rms^2)}{\rmd z}\\
  &\qquad+\omega^2\left(1-\f{g^2k_\rmh^2}{\omega^4}-\f{v_\rms^2k_\rmh^2}{\omega^2}\right)\bigg]\Delta=0.
\end{split}
\end{equation}
In the case of a polytropic atmosphere, $v_\rms^2\propto z$ and $\rho v_\rms^2\propto z^{m+1}$:
\begin{equation}
  v_\rms^2\f{\rmd^2\Delta}{\rmd z^2}+(m+2)\f{v_\rms^2}{z}\f{\rmd\Delta}{\rmd z}+\left[-m\f{gk_\rmh^2}{\omega^2}\f{v_\rms^2}{z}+\omega^2\left(1-\f{g^2k_\rmh^2}{\omega^4}-\f{v_\rms^2k_\rmh^2}{\omega^2}\right)\right]\Delta=0.
\end{equation}
In fact, $v_\rms^2/z=-\gamma g/(m+1)$.  Divide through by this factor:
\begin{equation}
  z\f{\rmd^2\Delta}{\rmd z^2}+(m+2)\f{\rmd\Delta}{\rmd z}-\left[m\f{gk_\rmh^2}{\omega^2}+\f{(m+1)}{\gamma g}\omega^2\left(1-\f{g^2k_\rmh^2}{\omega^4}\right)+k_\rmh^2z\right]\Delta=0.
\end{equation}
Finally,
\begin{equation}
  z\f{\rmd^2\Delta}{\rmd z^2}+(m+2)\f{\rmd\Delta}{\rmd z}-(A+k_\text{h}z)k_\text{h}\Delta=0,
\end{equation}
where
\begin{equation}
  A=\f{(m+1)}{\gamma}\f{\omega^2}{gk_\text{h}}+\left(m-\f{m+1}{\gamma}\right)\f{gk_\text{h}}{\omega^2}
\end{equation}
is a dimensionless constant.  Let $\Delta=w(z)\,\rme^{k_\text{h}z}$:
\begin{equation}
  z\f{\rmd^2w}{\rmd z^2}+(m+2+2k_\text{h}z)\f{\rmd w}{\rmd z}-(A-m-2)k_\text{h}w=0.
\end{equation}
This is related to the \emph{confluent hypergeometric equation} and
has a regular singular point at $z=0$.  Using the method of Frobenius,
we seek power-series solutions
\begin{equation}
  w=\sum_{r=0}^\infty a_rz^{\sigma+r},
\end{equation}
where $\sigma$ is a number to be determined and $a_0\ne0$.
The indicial equation is
\begin{equation}
  \sigma(\sigma+m+1)=0
\end{equation}
and the regular solution has $\sigma=0$.  The recurrence relation is then
\begin{equation}
  \f{a_{r+1}}{a_r}=\f{(A-m-2-2r)k_\text{h}}{(r+1)(r+m+2)}.
\end{equation}
In the case of an infinite series, $a_{r+1}/a_r\sim-2k_\text{h}/r$ as
$r\to\infty$, so $w$ behaves like $\rme^{-2k_\text{h}z}$ and $\Delta$
diverges like $\rme^{-k_\text{h}z}$ as $z\to-\infty$.  Solutions in
which $\Delta$ decays with depth are those for which the series
terminates and $w$ is a polynomial.  For a polynomial of degree $n-1$
($n\ge1$),
\begin{equation}
  A=2n+m.
\end{equation}
This gives a quadratic equation for $\omega^2$:
\begin{equation}
  \f{(m+1)}{\gamma}\left(\f{\omega^2}{gk_\text{h}}\right)^2-(2n+m)\left(\f{\omega^2}{gk_\text{h}}\right)+\left(m-\f{m+1}{\gamma}\right)=0.
\end{equation}
A negative root for $\omega^2$ exists if and only if
$m-(m+1)/\gamma<0$, i.e.\ $N^2<0$, as expected from Schwarzschild's
criterion for stability.

\begin{figure}
  \centerline{\epsfysize10cm\epsfbox{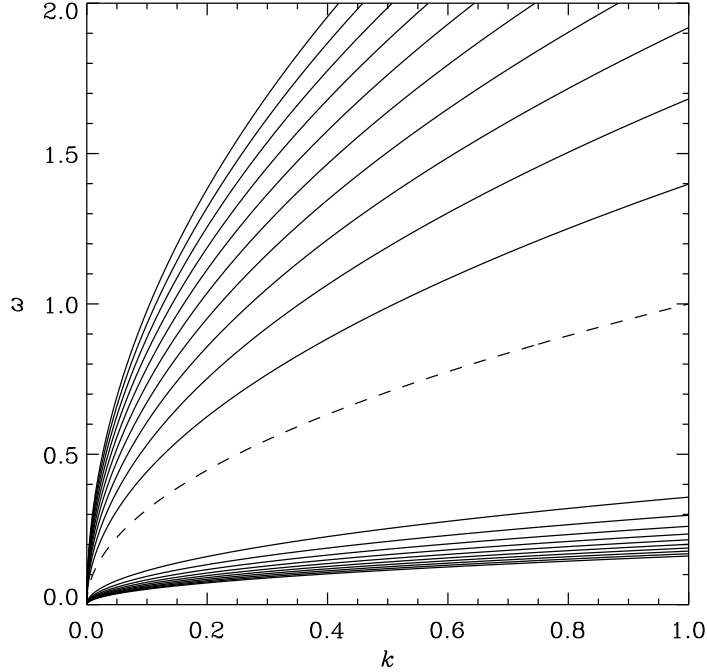}}
  \caption{Dispersion relation, in arbitrary units, for a stably
    stratified plane-parallel polytropic atmosphere with $m=3$ and
    $\gamma=5/3$.  The dashed line is the f~mode.  Above it are the
    first ten p~modes and below it are the first ten g~modes.  Each
    curve is a parabola.}
\label{f:dispersion}
\end{figure}

For $n\gg1$, the large root is
\begin{equation}
  \f{\omega^2}{gk_\text{h}}\sim\f{2n\gamma}{m+1}\qquad\text{(p~modes, $\omega^2\propto v_\text{s}^2$)}
\end{equation}
and the small root is
\begin{equation}
  \f{\omega^2}{gk_\text{h}}\sim\f{1}{2n}\left(m-\f{m+1}{\gamma}\right)\qquad\text{(g~modes, $\omega^2\propto N^2$)}.
\end{equation}
The f~mode is the `trivial' solution $\Delta=0$.  p~modes (`p' for
pressure) are \emph{acoustic waves}, which rely on compressibility.
g~modes are \emph{gravity waves}, which rely on buoyancy.  Typical
branches of the dispersion relation are illustrated in
Figure~\ref{f:dispersion}.

In solar-type stars the inner part (radiative zone) is convectively
stable ($N^2>0$) and the outer part (convective zone) is unstable
($N^2<0$).  However, the convection is so efficient that only a very
small entropy gradient is required to sustain the convective heat
flux, so $N^2$ is very small and negative in the convective zone.
Although g~modes propagate in the radiative zone at frequencies
smaller than $N$, they cannot reach the surface.  Only f and p~modes
(excited by convection) are observed at the solar surface.

In more massive stars the situation is reversed.  Then f, p and
g~modes can be observed, in principle, at the surface.  g~modes are
particularly well observed in certain classes of white dwarf.

\begin{figure}
  \centerline{\epsfbox{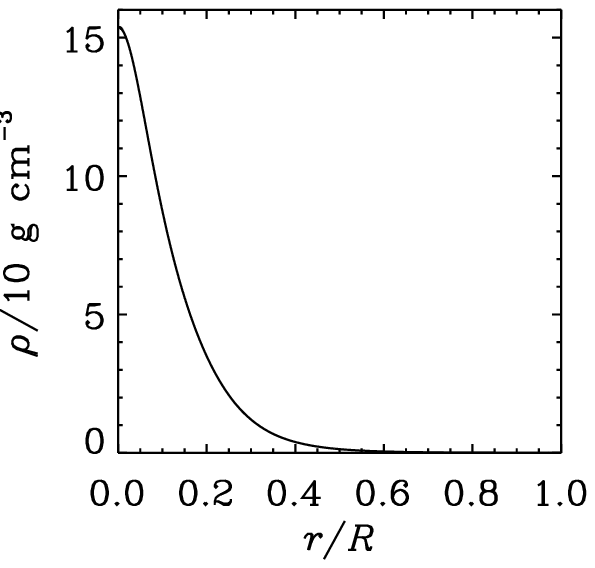}\qquad\epsfbox{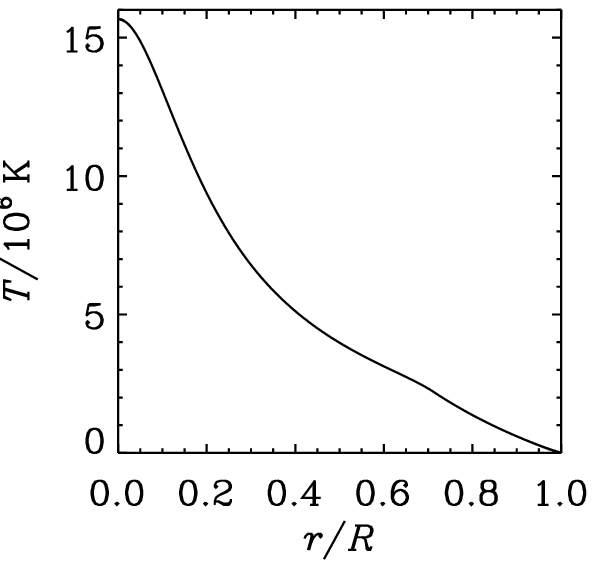}}
  \centerline{\epsfbox{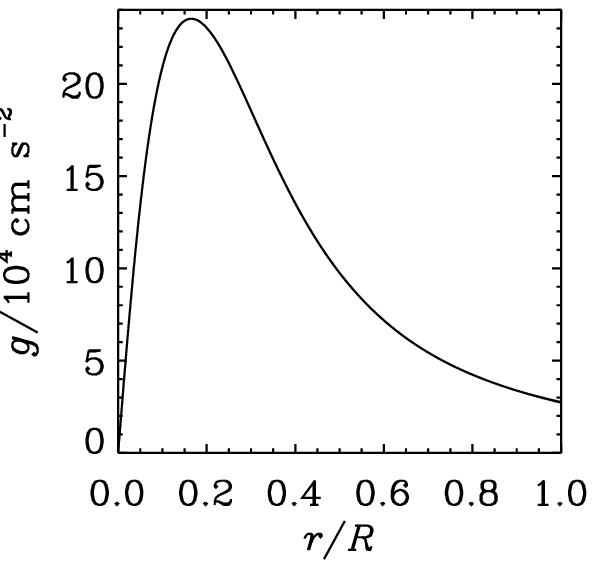}\qquad\epsfbox{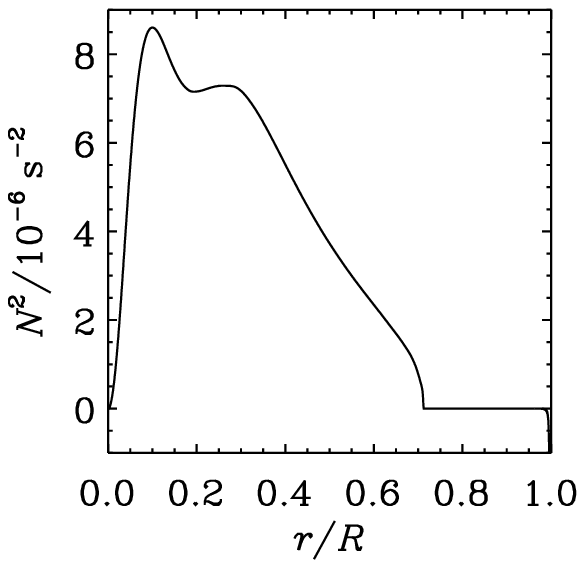}}
  \caption{A standard model of the present Sun, up to the photosphere.
    Density, temperature, gravity and squared buoyancy frequency are
    plotted versus fractional radius.}
\end{figure}

\begin{figure}
  \centerline{\epsfbox{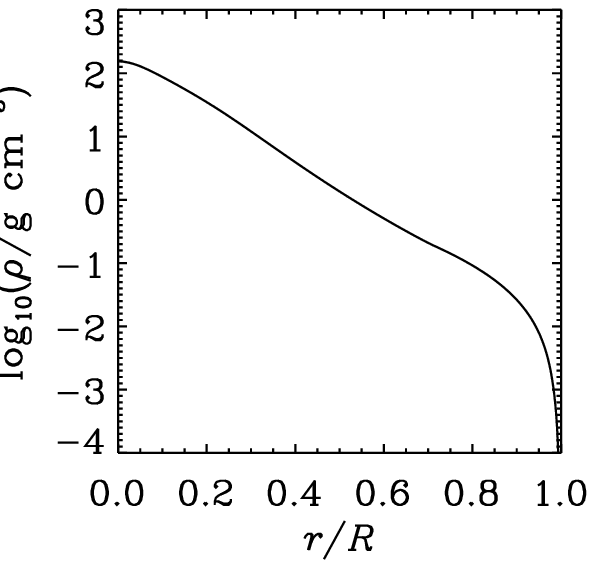}\qquad\epsfbox{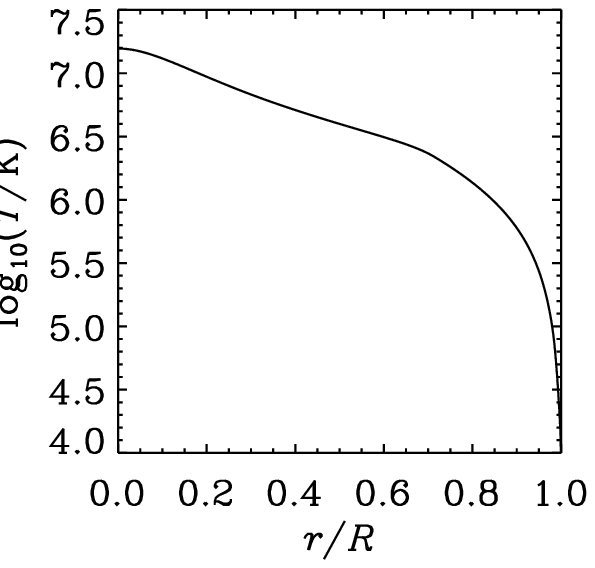}}
  \centerline{\epsfbox{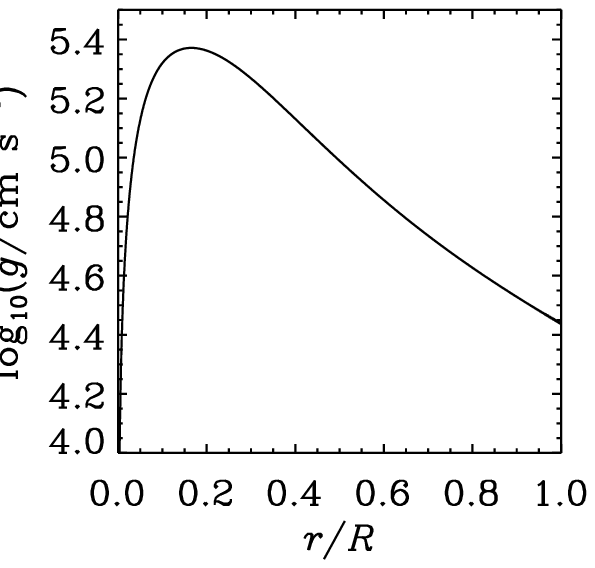}\qquad\epsfbox{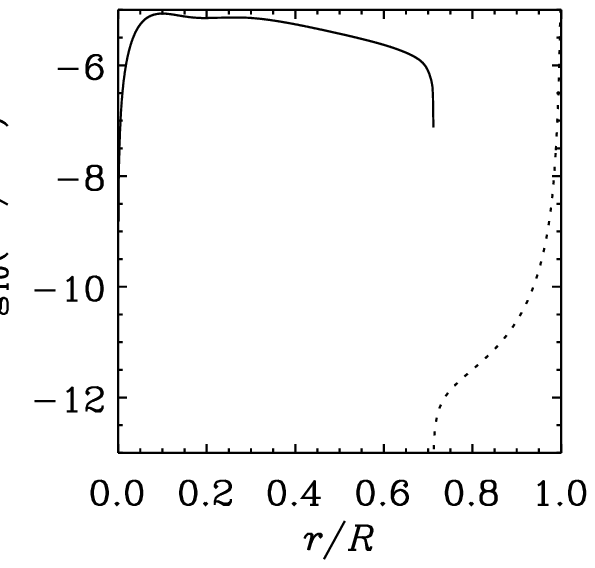}}
\caption{The same model plotted on a logarithmic scale.  In the convective region where $N^2<0$, the dotted line shows $-N^2$ instead.}
\end{figure}

\medskip
\noindent Related examples: \ref{e:radial}, \ref{e:isothermal},
\ref{e:slab}, \ref{e:buoyancy}.

\subsection{Tidally forced oscillations}

When astrophysical fluid bodies such as stars and planets orbit
sufficiently close to one another, they deform each other in ways that
can cause irreversible evolution of their spin and orbital motion over
astronomical time-scales.  We consider here some of the simplest
aspects of this problem.

Consider a binary star (or star--planet or planet--moon system, etc.)
with a circular orbit.  Let the orbital separation be $a$ and the
orbital (angular) frequency
\begin{equation}
  \Omega_\rmo=\left(\f{GM}{a^3}\right)^{1/2},
\end{equation}
where $M=M_1+M_2$ is the combined mass of the two bodies.  Let
$\bmR_1(t)$ and $\bmR_2(t)$ be the position vectors of the centres of
mass of the two bodies, and $\bmd=\bmR_2-\bmR_1$ their separation.

The gravitational potential due to body~2 (treated as a point mass or
spherical mass) at position $\bmR_1+\bmx$ within body~1 is
\begin{equation}
\begin{split}
  -\f{GM_2}{|\bmd-\bmx|}&=-GM_2\left(|\bmd|^2-2\bmd\bcdot\bmx+|\bmx|^2\right)^{-1/2}\\
  &=-\f{GM_2}{|\bmd|}\left(1-\f{2\bmd\bcdot\bmx}{|\bmd|^2}+\f{|\bmx|^2}{|\bmd|^2}\right)^{-1/2}\\
  &=-\f{GM_2}{|\bmd|}\left[1-\f{1}{2}\left(-\f{2\bmd\bcdot\bmx}{|\bmd|^2}+\f{|\bmx|^2}{|\bmd|^2}\right)+\f{3}{8}\left(-\f{2\bmd\bcdot\bmx}{|\bmd|^2}+\f{|\bmx|^2}{|\bmd|^2}\right)^2+\cdots\right]\\
  &=-\f{GM_2}{|\bmd|}\left[1+\f{\bmd\bcdot\bmx}{|\bmd|^2}+\f{3(\bmd\bcdot\bmx)^2-|\bmd|^2|\bmx|^2}{2|\bmd|^4}+O\left(\f{|\bmx|^3}{|\bmd|^3}\right)\right].
\end{split}
\end{equation}
In this Taylor expansion, the term independent of $\bmx$ is a uniform
potential that has no effect.  The term linear in $\bmx$ gives rise to
a uniform acceleration $GM\bmd/|\bmd|^3$, which causes the orbital
motion of body~1.  The remaining terms constitute the \textit{tidal
  potential} $\Psi$; the quadratic terms written here are the tidal
potential in the \textit{quadrupolar approximation}.

For a circular orbit, the coordinate system can be chosen such that
\begin{equation}
  \bmd=(a\cos\Omega_\rmo t,a\sin\Omega_\rmo t,0).
\end{equation}
Introduce spherical polar coordinates within body~1 such that
\begin{equation}
  \bmx=(r\sin\theta\cos\phi,r\sin\theta\sin\phi,r\cos\phi).
\end{equation}
Then
\begin{equation}
  \bmd\bcdot\bmx=ar\sin\theta\cos(\phi-\Omega_\rmo t),
\end{equation}
\begin{equation}
\begin{split}
  \Psi&=\f{GM_2r^2}{2a^3}\left[1-3\sin^2\theta\cos^2(\phi-\Omega_\rmo t)\right]\\
  &=\f{GM_2r^2}{4a^3}\left[2-3\sin^2\theta-3\sin^2\theta\cos(2\phi-2\Omega_\rmo t)\right].
\end{split}
\end{equation}
The first two terms are static; the remaining oscillatory part can be
written as
\begin{equation}
  \mathrm{Re}\left[-\f{3GM_2r^2\sin^2\theta}{4a^3}\,\rme^{2\rmi(\phi-\Omega_\rmo t)}\right],
\end{equation}
which involves the spherical harmonic function $Y_2^2(\theta,\phi)\propto\sin^2\theta\,\rme^{2\rmi\phi}$.

The tidal frequency in a non-rotating frame is $2\Omega_\rmo$.  In a
frame rotating with the spin angular velocity $\Omega_\rms$ of body~1,
the tidal frequency is $2(\Omega_\rmo-\Omega_\rms)$, owing to an
angular Doppler shift.

If the tidal frequency is sufficiently small, it might be assumed that
body~1 responds hydrostatically to the tidal potential.  Under this
assumption of an \textit{equilibrium tide}, body~1 is deformed into a
spheroid with a tidal bulge that points instantaneously towards
body~2, and no tidal torque is exerted.

We can allow for a more general linear response, including dissipation
and wavelike disturbances (a \textit{dynamical tide}) as follows.  The
most important aspect of the tidally deformed body is its exterior
gravitational potential perturbation $\delta\Phi$, because it is only
through gravity that the bodies communicate and exchange energy and
angular momentum.  We write the linear response as
\begin{equation}
  \delta\Phi=\mathrm{Re}\left[-k\f{3GM_2\sin^2\theta}{4a^3}\f{R_1^5}{r^3}\,\rme^{2\rmi(\phi-\Omega_\rmo t)}\right],
\end{equation}
where $R_1$ is the radius of body~1 (or some appropriate measure of
its radius if it is deformed by its rotation) and $k$ is the
\textit{potential Love number}, a dimensionless complex number that
describes the amplitude and phase of the tidal response.  Note that
$\delta\Phi$ involves the same frequency and the same spherical harmonic
$Y_2^2(\theta,\phi)$, but combined with $r^{-3}$ rather than $r^2$ to
make it a valid solution of Laplace's equation in the exterior of
body~1.  The factor of $R_1^5$ is introduced so that $k$ is
dimensionless and measures the ratio of $\delta\Phi$ and $\Psi$ at the
surface of body~1.

The imaginary part of $k$ determines the part of the tidal response
that is out of phase with the tidal forcing, and which is associated
with dissipation and irreversible evolution.  The torque acting on the
orbit of body~2 is
\begin{equation}
\begin{split}
  -T&=M_2r\sin\theta\left(-\f{1}{r\sin\theta}\f{\p\,\delta\Phi}{\p\phi}\right)\Bigg|_{r=a,\,\theta=\pi/2,\,\phi=\Omega_\rmo t}\\
  &=\mathrm{Re}\left(M_2k\f{3GM_2}{4a^3}\f{R_1^5}{a^3}2\rmi\right)\\
  &=-\mathrm{Im}(k)\f{3GM_2^2R_1^5}{2a^6}.
\end{split}
\end{equation}
By Newton's Third Law, there is an equal and opposite torque, $+T$,
acting on the spin of body~1.

\begin{figure}
  \centerline{\epsfysize9cm\epsfbox{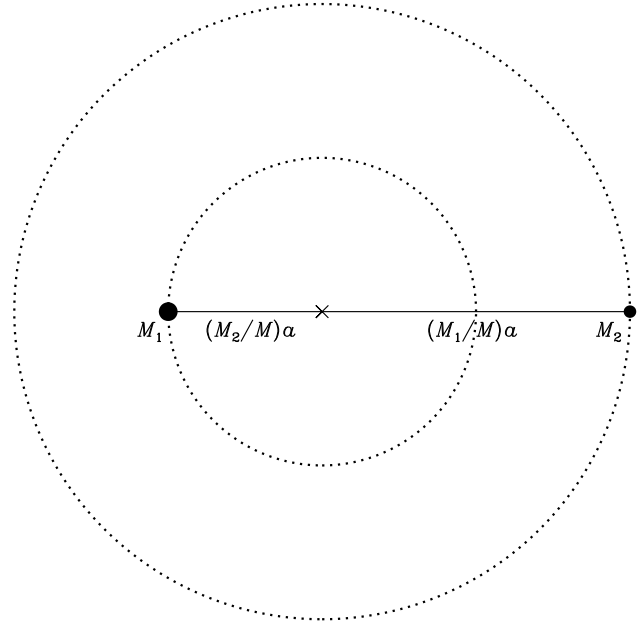}}
  \caption{A binary star with two components in circular orbital
    motion about the centre of mass.}
\label{f:binary}
\end{figure}

The orbital angular momentum about the centre of mass is
\begin{equation}
  L_\rmo=\mu(GMa)^{1/2},
\end{equation}
where $\mu=M_1M_2/M$ is the reduced mass of the system.  [This result
can be obtained by considering (Figure~\ref{f:binary})
\begin{equation}
  M_1\left(\f{M_2a}{M}\right)^2\Omega_\rmo+M_2\left(\f{M_1a}{M}\right)^2\Omega_\rmo=\f{M_1M_2}{M}(GMa)^{1/2}.]
\end{equation}
It evolves according to
\begin{equation}
  \f{\rmd L_\rmo}{\rmd t}=-T,
\end{equation}
which determines the rate of orbital migration:
\begin{equation}
\begin{split}
  \f{1}{2}\f{M_1M_2}{M}(GMa)^{1/2}\f{1}{a}\f{\rmd a}{\rmd t}&=-\mathrm{Im}(k)\f{3GM_2^2R_1^5}{2a^6}\\
  -\f{1}{a}\f{\rmd a}{\rmd t}&=3\,\mathrm{Im}(k)\f{M_2}{M_1}\left(\f{R_1}{a}\right)^5\Omega_\rmo.
\end{split}
\end{equation}
The orbital energy
\begin{equation}
  E_\rmo=-\mu\f{GM}{2a}
\end{equation}
evolves according to
\begin{equation}
  \f{\rmd E_\rmo}{\rmd t}=\mu\f{GM}{2a}\f{1}{a}\f{\rmd a}{\rmd t}=-\Omega_\rmo T.
\end{equation}
The spin angular momentum $L_\rms=I_1\Omega_\rms$ and spin energy
$E_\rms=\half I_1\Omega_\rms^2$, where $I_1$ is the moment of inertia
of body~1, evolve according to
\begin{equation}
  \f{\rmd L_\rms}{\rmd t}=T,\qquad
  \f{\rmd E_\rms}{\rmd t}=\Omega_\rms T.
\end{equation}
The total energy therefore satisfies
\begin{equation}
  \f{\rmd}{\rmd t}(E_\rmo+E_\rms)=(\Omega_\rms-\Omega_\rmo)T=-D,
\end{equation}
where $D>0$ is the rate of dissipation of energy.  To ensure $D>0$,
the sign of $\mathrm{Im}(k)$ should be the same as the sign of the
tidal frequency $2(\Omega_\rmo-\Omega_\rms)$.

In a dissipative spin--orbit coupling, the tidal torque $T$ tries to
bring about an equalization of the spin and orbital angular
velocities.  Its action, mediated by gravity, is comparable to a
frictional interaction between differential rotating components in a
mechanical system.

\begin{figure}
  \centerline{\epsfysize9cm\epsfbox{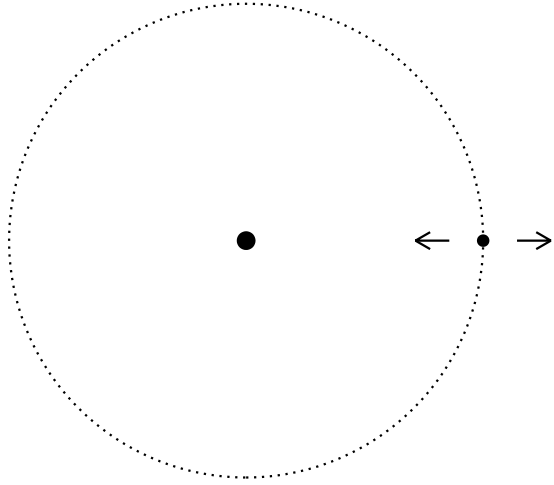}}
  \caption{Orbital migration away from the synchronous orbit driven by
    tidal dissipation in a system of extreme mass ratio.}
\label{f:migration}
\end{figure}

In binary stars, and other cases in which the spin angular momentum is
small compared to the orbital angular momentum, there is indeed a
tendency towards \textit{synchronization} of the spin with the orbital
motion (as the Moon is synchronized with its orbit around the Earth).
However, in systems of extreme mass ratio in which the spin of the
large body contains most of the angular momentum, the tidal torque
instead causes orbital migration away from the synchronous orbit at
which $\Omega=\Omega_\rms$ (Figure~\ref{f:migration}).

This situation applies to the moons of solar-system planets, most of
which migrate outwards, and to extrasolar planets in close orbits
around their host stars, where the migration is usually inward and may
lead to the destruction of the planet.

\newpage

\subsection{Rotating fluid bodies}

\noindent\textit{Note: in this subsection $(r,\phi,z)$ are cylindrical polar coordinates.}

\subsubsection{Equilibrium}

The equations of ideal gas dynamics in cylindrical polar coordinates are
\begin{equation}
\begin{split}
  &\f{\rmD u_r}{\rmD t}-\f{u_\phi^2}{r}=-\f{\p\Phi}{\p r}-\f{1}{\rho}\f{\p p}{\p r},\\
  &\f{\rmD u_\phi}{\rmD t}+\f{u_ru_\phi}{r}=-\f{1}{r}\f{\p\Phi}{\p\phi}-\f{1}{\rho r}\f{\p p}{\p\phi},\\
  &\f{\rmD u_z}{\rmD t}=-\f{\p\Phi}{\p z}-\f{1}{\rho}\f{\p p}{\p z},\\
  &\f{\rmD\rho}{\rmD t}=-\rho\left[\f{1}{r}\f{\p}{\p r}(r u_r)+\f{1}{r}\f{\p u_\phi}{\p\phi}+\f{\p u_z}{\p z}\right],\\
  &\f{\rmD p}{\rmD t}=-\gamma p\left[\f{1}{r}\f{\p}{\p r}(r u_r)+\f{1}{r}\f{\p u_\phi}{\p\phi}+\f{\p u_z}{\p z}\right],
\end{split}
\end{equation}
with
\begin{equation}
  \f{\rmD}{\rmD t}=\f{\p}{\p t}+u_r\f{\p}{\p r}+\f{u_\phi}{r}\f{\p}{\p\phi}+u_z\f{\p}{\p z}.
\end{equation}
Consider a steady, axisymmetric basic state with density $\rho(r,z)$, pressure $p(r,z)$, gravitational potential $\Phi(r,z)$ and with differential rotation
\begin{equation}
  \bmu=r\Omega(r,z)\,\bme_\phi.
\end{equation}
For equilibrium we require
\begin{equation}
  -r\Omega^2\,\bme_r=-\grad\Phi-\f{1}{\rho}\grad p.
\end{equation}
Take the curl to obtain
\begin{equation}
  -r\f{\p\Omega^2}{\p z}\,\bme_\phi=\grad p\btimes\grad\left(\f{1}{\rho}\right)=\grad T\btimes\grad s.
\end{equation}
This is just the vorticity equation in a steady state.  It is
sometimes called the \emph{thermal wind equation}.  The equilibrium is
called \emph{barotropic} if $\grad p$ is parallel to $\grad\rho$,
otherwise it is called \emph{baroclinic}.  In a barotropic state the
angular velocity is independent of $z$: $\Omega=\Omega(r)$.  This is a
version of the \emph{Taylor--Proudman theorem}\footnote{Joseph
  Proudman (1888--1975), British.} which states that under certain
conditions the velocity in a rotating fluid is independent of height.

We can also write
\begin{equation}
  \f{1}{\rho}\grad p=\bmg=-\grad\Phi+r\Omega^2\,\bme_r,
\end{equation}
where $\bmg$ is the \emph{effective gravitational acceleration},
including the centrifugal force associated with the (non-uniform)
rotation.

In a barotropic state with $\Omega(r)$ we can write
\begin{equation}
  \bmg=-\grad\Phi_\mathrm{cg},\qquad
  \Phi_\mathrm{cg}=\Phi(r,z)+\Psi(r),\qquad
  \Psi=-\int r\Omega^2\,\rmd r.
\end{equation}
Also, since $p=p(\rho)$ in the equilibrium state, we can define the
\emph{pseudo-enthalpy} $\tilde h(\rho)$ such that $\rmd\tilde h=\rmd
p/\rho$.  An example is a polytropic model for which
\begin{equation}
  p=K\rho^{1+1/m},\qquad
  \tilde h=(m+1)K\rho^{1/m}.
\end{equation}
($\tilde h$ equals the true enthalpy only if the equilibrium is
homentropic.)  The equilibrium condition then reduces to
\begin{equation}
  \bfzero=-\grad\Phi_\mathrm{cg}-\grad\tilde h
\end{equation}
or
\begin{equation}
  \Phi+\Psi+\tilde h=C=\cst.
\label{phipsi}
\end{equation}

An example of a rapidly and differentially rotating equilibrium is an
accretion disc around a central mass $M$.  For a non-self-gravitating
disc $\Phi=-GM(r^2+z^2)^{-1/2}$.  Assume the disc is barotropic and
let the arbitrary additive constant in $\tilde h$ be defined (as in
the polytropic example above) such that $\tilde h=0$ at the surfaces
$z=\pm H(r)$ of the disc where $\rho=p=0$.  Then
\begin{equation}
  -GM(r^2+H^2)^{-1/2}+\Psi(r)=C,
\end{equation}
from which
\begin{equation}
  r\Omega^2=-\f{\rmd}{\rmd r}\left[GM(r^2+H^2)^{-1/2}\right].
\end{equation}
For example, if $H=\epsilon r$ with $\epsilon=\cst$ being the aspect
ratio of the disc, then
\begin{equation}
  \Omega^2=(1+\epsilon^2)^{-1/2}\f{GM}{r^3}.
\end{equation}
The thinner the disc is, the closer it is to Keplerian rotation.  Once
we have found the relation between $\Omega(r)$ and $H(r)$,
equation~(\ref{phipsi}) then determines the spatial distribution of
$\tilde h$ (and therefore of $\rho$ and $p$) within the disc.

\subsubsection{Linear perturbations}

The basic state is independent of $t$ and $\phi$, allowing us to
consider linear perturbations of the form
\begin{equation}
  \real\left[\delta u_r(r,z)\exp(-\rmi\omega t+\rmi m\phi)\right],\qquad\text{etc.}
\end{equation}
where $m$ is the azimuthal wavenumber (an integer).  The linearized
equations in the Cowling approximation are
\begin{equation}
\begin{split}
  &-\rmi\hat\omega\,\delta u_r-2\Omega\,\delta u_\phi=-\f{1}{\rho}\f{\p\,\delta p}{\p r}+\f{\delta\rho}{\rho^2}\f{\p p}{\p r},\\
  &-\rmi\hat\omega\,\delta u_\phi+\f{1}{r}\delta\bmu\bcdot\grad(r^2\Omega)=-\f{\rmi m\,\delta p}{\rho r},\\
  &-\rmi\hat\omega\,\delta u_z=-\f{1}{\rho}\f{\p\,\delta p}{\p z}+\f{\delta\rho}{\rho^2}\f{\p p}{\p z},\\
  &-\rmi\hat\omega\,\delta\rho+\delta\bmu\bcdot\grad\rho=-\rho\left[\f{1}{r}\f{\p}{\p r}(r\,\delta u_r)+\f{\rmi m\,\delta u_\phi}{r}+\f{\p\,\delta u_z}{\p z}\right],\\
  &-\rmi\hat\omega\,\delta p+\delta\bmu\bcdot\grad p=-\gamma p\left[\f{1}{r}\f{\p}{\p r}(r\,\delta u_r)+\f{\rmi m\,\delta u_\phi}{r}+\f{\p\,\delta u_z}{\p z}\right],
\end{split}
\end{equation}
where
\begin{equation}
  \hat\omega=\omega-m\Omega
\end{equation}
is the intrinsic frequency, i.e.\ the angular frequency of the wave
measured in a frame of reference that rotates with the local angular
velocity of the fluid.

Eliminate $\delta u_\phi$ and $\delta\rho$ to obtain
\begin{equation}
\begin{split}
  (\hat\omega^2-A)\,\delta u_r-B\,\delta u_z&=
  -\f{{\rm i}\hat\omega}{\rho}\left(\f{\p\,\delta p}{\p r}-
  \f{\p p}{\p r}\f{\delta p}{\gamma p}\right)+
  2\Omega\f{{\rm i}m\,\delta p}{\rho r},\\
  -C\,\delta u_r+(\hat\omega^2-D)\,\delta u_z&=
  -\f{{\rm i}\hat\omega}{\rho}\left(\f{\p\,\delta p}{\p z}-
  \f{\p p}{\p z}\f{\delta p}{\gamma p}\right),
\end{split}
\end{equation}
where
\begin{equation}
\begin{split}
  A&=\f{2\Omega}{r}\f{\p}{\p r}(r^2\Omega)-
  \f{1}{\rho}\f{\p p}{\p r}
  \left(\f{1}{\gamma p}\f{\p p}{\p r}-
  \f{1}{\rho}\f{\p\rho}{\p r}\right),\\
  B&=\f{2\Omega}{r}\f{\p}{\p z}(r^2\Omega)-
  \f{1}{\rho}\f{\p p}{\p r}
  \left(\f{1}{\gamma p}\f{\p p}{\p z}-
  \f{1}{\rho}\f{\p\rho}{\p z}\right),\\
  C&=-\f{1}{\rho}\f{\p p}{\p z}
  \left(\f{1}{\gamma p}\f{\p p}{\p r}-
  \f{1}{\rho}\f{\p\rho}{\p r}\right),\\
  D&=-\f{1}{\rho}\f{\p p}{\p z}
  \left(\f{1}{\gamma p}\f{\p p}{\p z}-
  \f{1}{\rho}\f{\p\rho}{\p z}\right).
\end{split}
\end{equation}
Note that $A$, $B$, $C$ and $D$ involve radial and vertical
deriavtives of the specific angular momentum $r^2\Omega$ and the
specific entropy $s$.  The thermal wind equation implies
\begin{equation}
  B=C,
\end{equation}
so the matrix
\begin{equation}
  \bfM=\begin{bmatrix}A&B\\C&D\end{bmatrix}=\begin{bmatrix}A&B\\B&D\end{bmatrix}
\end{equation}
is symmetric.

\subsubsection{The H\o iland criteria}

It can be useful to introduce the Lagrangian displacement $\bxi$ such
that
\begin{equation}
  \Delta\bmu=\delta\bmu+\bxi\bcdot\grad\bmu=\f{\rmD\bxi}{\rmD t},
\end{equation}
i.e.
\begin{equation}
  \delta u_r=-\rmi\hat\omega\xi_r,\qquad
  \delta u_\phi=-\rmi\hat\omega\xi_\phi-r\bxi\bcdot\grad\Omega,\qquad
  \delta u_z=-\rmi\hat\omega\xi_z.
\end{equation}
Note that
\begin{equation}
  \f{1}{r}\f{\p}{\p r}(r\,\delta u_r)+\f{\rmi m\,\delta u_\phi}{r}+\f{\p\,\delta u_z}{\p z}=-\rmi\hat\omega\left[\f{1}{r}\f{\p}{\p r}(r\xi_r)+\f{\rmi m\xi_\phi}{r}+\f{\p\xi_z}{\p z}\right].
\end{equation}

The linearized equations constitute an eigenvalue problem for $\omega$
but it is not self-adjoint except when $m=0$.  We specialize to the
case $m=0$ (axisymmetric perturbations).
Then
\begin{equation}
\begin{split}
  (\omega^2-A)\xi_r-B\xi_z&=
  \f{1}{\rho}\left(\f{\p\,\delta p}{\p r}-
  \f{\p p}{\p r}\f{\delta p}{\gamma p}\right),\\
  -B\xi_r+(\omega^2-D)\xi_z&=
  \f{1}{\rho}\left(\f{\p\,\delta p}{\p z}-
  \f{\p p}{\p z}\f{\delta p}{\gamma p}\right),
\end{split}
\label{abcd}
\end{equation}
with
\begin{equation}
  \delta p=-\gamma p\div\bxi-\bxi\bcdot\grad p.
\end{equation}
Multiply the first of equations~(\ref{abcd}) by $\rho\xi_r^*$ and the
second by $\rho\xi_z^*$ and integrate over the volume $V$ of the fluid
(using the boundary condition $\delta p=0$) to obtain
\begin{equation}
\begin{split}
  \omega^2\int_V\rho(|\xi_r|^2+|\xi_z|^2)\,\rmd V&=\int_V\left[\rho Q(\bxi)+\bxi^*\bcdot\grad\delta p-\f{\delta p}{\gamma p}\bxi^*\bcdot\grad p\right]\,\rmd V\\
  &=\int_V\left[\rho Q(\bxi)-\f{\delta p}{\gamma p}(\gamma p\div\bxi^*+\bxi^*\bcdot\grad p)\right]\,\rmd V\\
  &=\int_V\left(\rho Q(\bxi)+\f{|\delta p|^2}{\gamma p}\right)\,\rmd V,
\end{split}
\end{equation}
where
\begin{equation}
  Q(\bxi)=A|\xi_r|^2+B(\xi_r^*\xi_z+\xi_z^*\xi_r)+D|\xi_z|^2=\begin{bmatrix}\xi_r^*&\xi_z^*\end{bmatrix}\begin{bmatrix}A&B\\B&D\end{bmatrix}\begin{bmatrix}\xi_r\\\xi_z\end{bmatrix}
\end{equation}
is the (real) Hermitian form associated with the matrix $\bfM$.

Note that this integral involves only the meridional components of the
displacement.  If we had not made the Cowling approximation there
would be the usual negative definite contribution to $\omega^2$ from
self-gravitation.

The above integral relation therefore shows that $\omega^2$ is real,
and a variational property ensures that instability \emph{to
axisymmetric perturbations} occurs if and only if the integral on the
right-hand side can be made negative by a suitable trial displacement.
If $Q$ is positive definite then $\omega^2>0$ and we have stability.
Now the characteristic equation of the matrix $\bfM$ is
\begin{equation}
  \lambda^2-(A+D)\lambda+AD-B^2=0.
\end{equation}
The eigenvalues $\lambda_\pm$ are both positive if and only if
\begin{equation}
  A+D>0\qquad\text{and}\qquad AD-B^2>0.
\end{equation}
If these conditions are satisfied throughout the fluid then $Q>0$,
which implies $\omega^2>0$, so the fluid is stable to axisymmetric
perturbations (neglecting self-gravitation).  These conditions are
also necessary for stability.  If one of the eigenvalues is negative
in some region of the meridional plane, then a trial displacement can
be found which is localized in that region, has $\delta p=0$ and
$Q<0$, implying instability.  (By choosing $\bxi$ in the correct
direction and tuning $\div\bxi$ appropriately, it is possible to
arrange for $\delta p$ to vanish.)

Using $\ell=r^2\Omega$ (specific angular momentum) and
$s=c_p(\gamma^{-1}\ln p-\ln\rho)+\cst$ (specific entropy) for a
perfect ideal gas, we have
\begin{equation}
\begin{split}
  A&=\f{1}{r^3}\f{\p\ell^2}{\p r}-\f{g_r}{c_p}\f{\p s}{\p r},\\
  B&=\f{1}{r^3}\f{\p\ell^2}{\p z}-\f{g_r}{c_p}\f{\p s}{\p z}=-\f{g_z}{c_p}\f{\p s}{\p r},\\
  D&=-\f{g_z}{c_p}\f{\p s}{\p z},
\end{split}
\end{equation}
so the two conditions become
\begin{equation}
  \f{1}{r^3}\f{\p\ell^2}{\p r}-\f{1}{c_p}\bmg\bcdot\grad s>0
\end{equation}
and
\begin{equation}
  -g_z\left(\f{\p\ell^2}{\p r}\f{\p s}{\p z}-\f{\p\ell^2}{\p z}\f{\p s}{\p r}\right)>0.
\end{equation}
These are the \emph{H\o iland stability criteria}\footnote{Einar H\o iland (1907--1974), Norwegian}.

(If the criteria are marginally satisfied a further investigation may
be required.)

Consider first the non-rotating case $\ell=0$.  The first criterion
reduces to the \emph{Schwarzschild criterion} for convective
stability,
\begin{equation}
   -\f{1}{c_p}\bmg\bcdot\grad s\equiv N^2>0.
\end{equation}
In the homentropic case $s=\cst$ (which is a barotropic model) they
reduce to the \emph{Rayleigh criterion}\footnote{John William Strutt, Lord Rayleigh (1842--1919), British} for centrifugal (inertial)
stability,
\begin{equation}
  \f{\rmd\ell^2}{\rmd r}>0.
\end{equation}
which states that the specific angular momentum should increase with
$r$ for stability.

The second H\o iland criterion is equivalent to
\begin{equation}
  (\bme_r\btimes(-\bmg))\bcdot(\grad\ell^2\btimes\grad s)>0
\end{equation}
In other words the vectors $\bme_r\btimes(-\bmg)$ and
$\grad\ell^2\btimes\grad s$ should be parallel (rather than
antiparallel).  In a rotating star, for stability we require that the
specific angular momentum should increase with $r$ on each surface of
constant entropy.

\medskip
\noindent Related example: \ref{e:inertial}.

\newpage

\appendix

\section{Examples}
\label{s:examples}

\subsection{Validity of a fluid approach}
\label{e:collisions}

The Coulomb cross-section for `collisions' (i.e.\ large-angle
scatterings) between electrons and protons is
$\sigma\approx1\times10^{-4}(T/\text{K})^{-2}\,\text{cm}^2$.  Why does
it depend on the inverse square of the temperature?

Using the numbers quoted in Section~\ref{s:validity} (or elsewhere),
estimate the order of magnitude of the mean free path and the
collision frequency in (i) the centre of the Sun, (ii) the solar
corona, (iii) a molecular cloud and (iv) the hot phase of the
interstellar medium.  Is a fluid approach likely to be valid in these
systems?

\subsection{Vorticity equation}
\label{e:vorticity}

Show that the vorticity $\bomega=\curl\bmu$ of an ideal fluid without
a magnetic field satisfies the equation
\begin{equation}
  \f{\p\bomega}{\p t}=\curl(\bmu\btimes\bomega)+\grad p\btimes\grad v,
\end{equation}
where $v=1/\rho$ is the specific volume.  Explain why the last term,
which acts as a source of vorticity, can also be written as $\grad
T\btimes\grad s$.  Under what conditions does this `baroclinic' source
term vanish, and in what sense(s) can the vorticity then be said to be
`conserved'?

Show that the (Rossby--Ertel) \textit{potential vorticity}
$\f{1}{\rho}\,\bomega\bcdot\grad s$
is conserved, as a material invariant, even when the baroclinic term
is present.

\subsection{Homogeneous expansion or contraction}
\label{e:expansion}

(This question explores a very simple fluid flow in which
compressibility and self-gravity are important.)

A homogeneous perfect gas of density $\rho=\rho_0(t)$ occupies the
region $|\bmx|<R(t)$, surrounded by a vacuum.  The pressure is
$p=p_0(t)(1-|\bmx|^2/R^2)$ and the velocity field is $\bmu=A(t)\bmx$,
where $A=\dot R/R$.

Using either Cartesian or spherical polar coordinates, show that the
equations of Newtonian gas dynamics and the boundary conditions are
satisfied provided that
\begin{equation}
  \rho_0\propto R^{-3},\qquad
  p_0\propto R^{-3\gamma},\qquad
  \ddot R=-\f{4\pi G\rho_0R}{3}+\f{2p_0}{\rho_0R}.
\end{equation}
Deduce the related energy equation
\begin{equation}
\f{1}{2}\dot R^2-\f{4\pi
  G\rho_0R^2}{3}+\f{2p_0}{3(\gamma-1)\rho_0}=\cst,
\end{equation}
and interpret the three contributions.  Discuss the dynamics
qualitatively in the two cases $\gamma>4/3$ and
$1<\gamma<4/3$.\footnote{This flow is similar in form to the
  cosmological `Hubble flow' and can be seen as a homogeneous
  expansion or contraction centred on any point, if a Galilean
  transformation is made.  In the limit $R\to\infty$ (for
  $\gamma>4/3$), or if the pressure is negligible, the equations
  derived here correspond to the Friedmann equations for a `dust'
  universe (i.e.\ negligible relativistic pressure $p\ll\rho c^2$)
  with a scale factor $a\propto R$, $\f{\ddot a}{a}=-\f{4\pi
    G\rho_0}{3}$ and $\f{\dot a^2+\cst}{a^2}=\f{8\pi G\rho_0}{3}$.
  See \citet{Bondi60} for a discussion of Newtonian cosmology.}

\subsection{Dynamics of ellipsoidal bodies}
\label{e:ellipsoidal}

(This question uses Cartesian tensor notation and the summation
convention.)

A fluid body occupies a time-dependent ellipsoidal volume centred on
the origin.  Let $f(\bmx,t)=1-S_{ij}x_ix_j$, where $S_{ij}(t)$ is a
symmetric tensor with positive eigenvalues, such that the body
occupies the region $0<f\le1$ with a free surface at $f=0$.  The
velocity field is $u_i=A_{ij}x_j$, where $A_{ij}(t)$ is a tensor that
is not symmetric in general.  Assume that the gravitational potential
inside the body has the form $\Phi=B_{ij}x_ix_j+\cst$, where
$B_{ij}(t)$ is a symmetric tensor.

Show that the equations of Newtonian gas dynamics and the boundary
conditions are satisfied if the density and pressure are of the form
\begin{equation}
  \rho=\rho_0(t)\hat\rho(f),\qquad
  p=\rho_0(t)T(t)\hat p(f),
\end{equation}
where the dimensionless functions $\hat p(f)$ and $\hat\rho(f)$ are
related by $\hat p'(f)=\hat\rho(f)$ with the normalization
$\hat\rho(1)=1$ and the boundary condition $\hat p(0)=0$, provided
that the coefficients evolve according to
\begin{equation}
\begin{split}
  &\dot S_{ij}+S_{ik}A_{kj}+S_{jk}A_{ki}=0,\\
  &\dot A_{ij}+A_{ik}A_{kj}=-2B_{ij}+2TS_{ij},\\
  &\dot\rho_0=-\rho_0A_{ii},\\
  &\dot T=-(\gamma-1)TA_{ii}.
\end{split}
\end{equation}
Examples of the spatial structure are the homogeneous body:
$\hat\rho=1$, $\hat p=f$, and the polytrope of index $n$:
$\hat\rho=f^n$, $\hat p=f^{n+1}/(n+1)$.  Show that Poisson's equation
cannot be satisfied if the body is inhomogeneous.\footnote{It can be
  shown that the self-gravity of a homogeneous ellipsoid generates an
  interior gravitational potential of the assumed form.  The behaviour
  of self-gravitating, homogeneous, incompressible ellipsoids was
  investigated by many great mathematicians, including Maclaurin,
  Jacobi, Dirichlet, Dedekind, Riemann and Poincar\'e, illustrating
  the equilibrium and stability of rotating and tidally deformed
  astrophysical bodies \citep{Chandrasekhar69}.}

Show how the results of the previous question are recovered in the
case of a homogeneous, spherically symmetric body.

\subsection{Resistive MHD}
\label{e:resistive}

Ohm's Law for a medium of electrical conductivity $\sigma$ is
$\bmJ=\sigma\bmE$, where $\bmE$ is the electric field measured in the
rest frame of the conductor.  Show that, in the presence of a finite
and uniform conductivity, the ideal induction equation is modified to
\begin{equation}
  \f{\p\bmB}{\p t}=\curl(\bmu\btimes\bmB)+\eta\delsq\bmB,
\end{equation}
where $\eta=1/(\mu_0\sigma)$ is the magnetic diffusivity, proportional
to the resistivity of the fluid.  Hence argue that the effects of
finite conductivity are small if the magnetic Reynolds number
$Rm=LU/\eta$ is large, where $L$ and $U$ are characteristic scales of
length and velocity for the fluid flow.\footnote{The magnetic
  diffusivity in a fully ionized plasma is on the order of
  $10^{13}(T/\mathrm{K})^{-3/2}\,\mathrm{cm}^2\,\mathrm{s}^{-1}$.
  Simple estimates imply that $Rm\gg1$ for observable solar
  phenomena.}

\subsection{Flux freezing}
\label{e:euler}

Consider a magnetic field that is defined in terms of two \emph{Euler
  potentials} $\alpha$ and $\beta$ by
\begin{equation}
  \bmB=\grad\alpha\btimes\grad\beta.
\end{equation}
(This is sometimes called a \emph{Clebsch representation}.)  Show that
  a vector potential of the form $\bmA=\alpha\grad\beta+\grad\gamma$
  generates this magnetic field via $\bmB=\curl\bmA$, and that the
  magnetic field lines are the intersections of the families of
  surfaces $\alpha=\cst$ and $\beta=\cst$.  Show also that
\begin{equation}
  \f{\p\bmB}{\p t}-\curl(\bmu\btimes\bmB)=
  \grad\left(\f{\rmD\alpha}{\rmD t}\right)\btimes\grad\beta+
  \grad\alpha\btimes\grad\left(\f{\rmD\beta}{\rmD t}\right).
\end{equation}
Deduce that the ideal induction equation is satisfied if the families
of surfaces $\alpha=\cst$ and $\beta=\cst$ are material surfaces, in
which case the magnetic field lines can also be identified with
material curves.

\subsection{Equilibrium of a solar prominence}
\label{e:prominence}

A simple model for a prominence or filament in the solar atmosphere
involves a two-dimensional magnetostatic equilibrium in the $(x,z)$
plane with uniform gravity $\bmg=-g\,\bme_z$.  The gas is isothermal
with isothermal sound speed $c_\text{s}$.  The density and magnetic
field depend only on $x$ and the field lines become straight as
$|x|\to\infty$.

Show that the solution is of the form
\begin{equation}
  B_z=B_0\tanh(kx),
\end{equation}
where $k$ is a constant to be determined.  Sketch the field lines and
find the density distribution.

\subsection{Equilibrium of a magnetic star}
\label{e:equilibrium}

A star contains an axisymmetric and purely toroidal magnetic field
$\bmB=B(r,z)\,\bme_\phi$, where $(r,\phi,z)$ are cylindrical
polar coordinates.  Show that the equation of magnetostatic
equilibrium can be written in the form
\begin{equation}
  \bfzero=-\rho\grad\Phi-\grad p-\f{B}{\mu_0r}\grad(rB).
\end{equation}
Assuming that the equilibrium is barotropic such that $\grad p$ is
everywhere parallel to $\grad\rho$, show that the magnetic field must
be of the form
\begin{equation}
  B=\f{1}{r}f(r^2\rho),
\end{equation}
where $f$ is an arbitrary function.  Sketch the topology of the
contour lines of $r^2\rho$ in a star and show that a magnetic field of
this form is confined to the interior.

\subsection{Force-free magnetic fields}
\label{e:fff}

(a) Show that an axisymmetric force-free magnetic field satisfies
\begin{equation}
  B_\phi=\f{f(\psi)}{r},
\end{equation}
where $\psi$ is the poloidal magnetic flux function, $r$ is the
cylindrical radius and $f$ is an arbitrary function.  Show also
that $\psi$ satisfies the equation
\begin{equation}
  r^2\div(r^{-2}\grad\psi)+f\f{\rmd f}{\rmd\psi}=0.
\end{equation}

\medskip

(b) Let $V$ be a fixed volume bounded by a surface $S$. Show that the
rate of change of the magnetic energy in $V$ is
\begin{equation}
  \f{1}{\mu_0}\int_S[(\bmu\bcdot\bmB)\bmB-B^2\bmu]\bcdot\rmd\bmS-\int_V\bmu\bcdot\bmF_\rmm\,\rmd V,
\end{equation}
where $\bmF_\rmm$ is the Lorentz force per unit volume.  If $V$ is an
axisymmetric volume containing a magnetic field that remains
axisymmetric and force-free, and if the velocity on $S$ consists of a
differential rotation $\bmu=r\Omega(r,z)\,\bme_\phi$, deduce that the
instantaneous rate of change of the magnetic energy in $V$ is
\begin{equation}
  \f{2\pi}{\mu_0}\int f(\psi)\Delta\Omega(\psi)\,\rmd\psi,
\end{equation}
where $\Delta\Omega(\psi)$ is the difference in angular velocity of
the two endpoints on $S$ of the field line labelled by $\psi$, and the
range of integration is such as to cover $S$ once.

\subsection{Helicity}
\label{e:helicity}

The magnetic helicity in a volume $V$ is
\begin{equation}
  H_\text{m}=\int_V\bmA\bcdot\bmB\,\rmd V.
\end{equation}
A \textit{thin, untwisted magnetic flux tube} is a thin tubular
structure consisting of the neighbourhood of a smooth curve $C$, such
that the magnetic field is confined within the tube and is parallel to
$C$.

\medskip

(a) Consider a simple example of a single, closed, untwisted
magnetic flux tube such that
\begin{equation}
  \bmB=B(r,z)\,\bme_\phi,
\end{equation}
where $(r,\phi,z)$ are cylindrical polar coordinates and $B(r,z)$ is a
positive function localized near $(r=a,z=0)$.  The tube is contained
entirely within $V$.  Show that the magnetic helicity of this field is
uniquely defined and equal to zero.

\medskip

(b) Use the fact that $H_\text{m}$ is conserved in ideal MHD
to argue that the magnetic helicity of any single, closed, untwisted
and unknotted flux tube contained within $V$ is also zero.

\medskip

(c) Consider a situation in which $V$ contains two such
flux tubes $T_1$ and $T_2$.  Let $F_1$ and $F_2$ be the magnetic
fluxes associated with $T_1$ and $T_2$.  By writing $\bmB=\bmB_1+\bmB_2$,
etc., and assuming that the tubes are thin, show that
\begin{equation}
  H_\text{m}=\pm2F_1F_2
\end{equation}
if the tubes are simply interlinked, while $H_\text{m}=0$ if they are unlinked.

\subsection{Variational principles}
\label{e:variational}

The magnetic energy in a volume $V$ bounded by a surface $S$ is
\begin{equation}
  E_\text{m}=\int_V{\frac{B^2}{2\mu_0}}\,\rmd V.
\end{equation}

\medskip

(a) Making use of the representation $\bmB=\curl\bmA$ of
the magnetic field in terms of a magnetic vector potential, show that
the magnetic field that minimizes $E_\text{m}$, subject to the
tangential components of $\bmA$ being specified on $S$, is a potential
field.  Argue that this constraint corresponds to specifying the normal
component of $\bmB$ on $S$.

\medskip

(b) Making use of the representation
$\bmB=\grad\alpha\btimes\grad\beta$ of the magnetic field in terms of
Euler potentials, show that the magnetic field that minimizes
$E_\text{m}$, subject to $\alpha$ and $\beta$ being specified on $S$,
is a force-free field.  Argue that this constraint corresponds to
specifying the normal component of $\bmB$ on $S$ and also the way in
which points on $S$ are connected by magnetic field lines.

\subsection{Friedrichs diagrams}
\label{e:friedrichs}

The dispersion relations $\omega(\bmk)$ for Alfv\'en and
magnetoacoustic waves in a uniform medium are given by
\begin{equation}
  v_\rmp^2=v_\text{a}^2\cos^2\theta,
\end{equation}
\begin{equation}
  v_\rmp^4-(v_\text{s}^2+v_\text{a}^2)v_\rmp^2+
  v_\text{s}^2v_\text{a}^2\cos^2\theta=0,
\end{equation}
where $v_\rmp=\omega/k$ is the phase velocity and $\theta$ is the
angle between $\bmk$ and $\bmB$.  Use the form of $v_\rmp(\theta)$ for
each mode to calculate the group velocities
$\bmv_\text{g}=\p\omega/\p\bmk$, determining their components parallel
and perpendicular to $\bmB$.

Sketch the phase and group diagrams by tracking
$\bmv_\rmp=v_\rmp\hat\bmk$ and $\bmv_\text{g}$, respectively,
over the full range of $\theta$.  Treat the cases
$v_\text{s}>v_\text{a}$ and $v_\text{s}<v_\text{a}$ separately.  By
analysing the limit $\theta\to\pi/2$, show that the group diagram for
the slow wave has a cusp at speed
$v_\text{s}v_\text{a}(v_\text{s}^2+v_\text{a}^2)^{-1/2}$.

\subsection{Shock relations}
\label{e:shock}

The Rankine--Hugoniot relations in the rest frame of a non-magnetic
shock are
\begin{equation}
  [\rho u_x]_1^2=0,
\end{equation}
\begin{equation}
  [\rho u_x^2+p]_1^2=0,
\end{equation}
\begin{equation}
  [\rho u_x(\half u_x^2+h)]_1^2=0,
\end{equation}
where $u_x>0$ and $[Q]_1^2=Q_2-Q_1$ is the difference between the
downstream and upstream values of any quantity $Q$.  Show that the
velocity, density and pressure ratios
\begin{equation}
  U=\f{u_2}{u_1},\qquad
  D=\f{\rho_2}{\rho_1},\qquad
  P=\f{p_2}{p_1}
\end{equation}
across a shock in a perfect gas are given by
\begin{equation}
  D=\f{1}{U}=\f{(\gamma+1){\cal M}_1^2}{(\gamma-1){\cal M}_1^2+2},\qquad
  P=\frac{2\gamma{\cal M}_1^2-(\gamma-1)}{(\gamma+1)},
\end{equation}
where ${\cal M}=u_x/v_\text{s}$ is the Mach number, and also that
\begin{equation}
  {\cal M}_2^2=\frac{(\gamma-1){\cal M}_1^2+2}{2\gamma{\cal M}_1^2-(\gamma-1)}.
\end{equation}

Show that the entropy change in passing through the shock is given by
\begin{equation}
  \frac{[s]_1^2}{c_v}=\ln P-\gamma\ln\left[\frac{(\gamma+1)P+(\gamma-1)}{(\gamma-1)P+(\gamma+1)}\right]
\end{equation}
and deduce that only compression shocks ($D>1$, $P>1$) are
physically realizable.

\subsection{Oblique shocks}
\label{e:oblique}

For a hydrodynamic shock, let $u_{X2}$ and $u_{Y2}$ be the downstream
velocity components parallel and perpendicular, respectively, to the
upstream velocity vector $\bmu_1$.  In the limit of a strong shock,
$\mathcal{M}_1\gg1$, derive the relation
\begin{equation}
  u_{Y2}^2=(|\bmu_1|-u_{X2})\left[u_{X2}-\left(\f{\gamma-1}{\gamma+1}\right)|\bmu_1|\right].
\end{equation}
Sketch this relation in the $(u_{X2},u_{Y2})$ plane. Hence show that
the maximum angle through which the velocity vector can be deflected
on passing through a stationary strong shock is $\arcsin(1/\gamma)$.

\subsection{The Riemann problem}
\label{e:riemann}

A perfect gas flows in one dimension in the absence of boundaries,
gravity and magnetic fields.

\medskip

(a) Determine all possible smooth local solutions of the equations of
one-dimensional gas dynamics that depend only on the variable
$\xi=x/t$ for $t>0$. Show that one such solution is a rarefaction wave
in which $\rmd u/\rmd\xi=2/(\gamma+1)$. How do the adiabatic sound
speed and specific entropy vary with $\xi$?

\medskip

(b) At $t=0$ the gas is initialized with uniform density $\rho_\rmL$,
pressure $p_\rmL$ and velocity $u_\rmL$ in the region $x<0$ and with
uniform density $\rho_\rmR$, pressure $p_\rmR$ and velocity $u_\rmR$
in the region $x>0$.  Explain why the subsequent flow is of the
similarity form described in part~(a).  What constraints must be
satisfied by the initial values if the subsequent evolution is to
involve only two uniform states connected by a rarefaction wave? Give
a non-trivial example of such a solution.

\medskip

(c) Explain why, for more general choices of the initial values, the
solution cannot have the simple form described in part~(b), even if
$u_\rmR>u_\rmL$. What other features will appear in the solution?
(Detailed calculations are not required.)

\subsection{Nonlinear waves in incompressible MHD}
\label{e:elsasser}

Show that the equations of ideal MHD in the case of an incompressible
fluid of uniform density $\rho$ can be written in the symmetrical form
\begin{equation}
  \frac{\partial\bmz_\pm}{\partial t}+\bmz_\mp\bcdot\grad\bmz_\pm=
  -\grad\psi,
\end{equation}
\begin{equation}
  \div\bmz_\pm=0,
\end{equation}
where
\begin{equation}
  \bmz_\pm=\bmu\pm\bmv_\text{a}
\end{equation}
are the \textit{Els\"asser variables},
$\bmv_\text{a}=(\mu_0\rho)^{-1/2}\bmB$ is the vector Alfv\'en
velocity, and $\psi=\Phi+(\Pi/\rho)$ is a modified pressure.

Consider a static basic state in which the magnetic field is uniform
and $\psi=\cst$.  Write down the exact equations governing
perturbations $(\bmz_\pm',\psi')$ (i.e.\ without performing a
linearization).  Hence show that there are special solutions in which
disturbances of arbitrary amplitude propagate along the magnetic field
lines in one direction or other without change of form.  How do these
relate to the MHD wave modes of a compressible fluid?  Why does the
general argument for wave steepening not apply to these nonlinear
simple waves?

\subsection{Spherical blast waves}
\label{e:blast}

A supernova explosion of energy $E$ occurs at time $t=0$ in an
unmagnetized perfect gas of adiabatic exponent $\gamma$. The
surrounding medium is initially cold and has non-uniform density
$Cr^{-\beta}$, where $C$ and $\beta$ are constants (with $0<\beta<3$)
and $r$ is the distance from the supernova.

\medskip

(a) Explain why a self-similar spherical blast wave may be expected to
occur, and deduce that the radius $R(t)$ of the shock front increases as
a certain power of $t$.

\medskip

(b) Write down the self-similar form of the velocity, density and
pressure for $0<r<R(t)$ in terms of three undetermined dimensionless
functions of $\xi=r/R(t)$. Obtain a system of dimensionless ordinary
differential equations governing these functions, and formulate the
boundary conditions on the dimensionless functions at the strong shock
front $\xi=1.$

\medskip

(c) Show that special solutions exist in which the radial velocity and
the density are proportional to $r$ for $r<R(t)$, if
\begin{equation}
  \beta=\f{7-\gamma}{\gamma+1}.
\end{equation}
For the case $\gamma=5/3$ express the velocity, density and pressure
for this special solution in terms of the original dimensional
variables.

\subsection{Accretion on to a black hole}
\label{e:pw}

Write down the equations of steady, spherical accretion of a perfect
gas in an arbitrary gravitational potential $\Phi(r)$.

Accretion on to a black hole can be approximated within a Newtonian
theory by using the \textit{Paczy\'nski--Wiita potential}
\begin{equation}
  \Phi=-\frac{GM}{r-r_\text{h}},
\end{equation}
where $r_\text{h}=2GM/c^2$ is the radius of the event horizon and $c$
is the speed of light.

Show that the sonic radius $r_\text{s}$ is related to $r_\text{h}$ and
the nominal accretion radius $r_\text{a}=GM/2v_{\text{s}0}^2$ (where
$v_{\text{s}0}$ is the sound speed at infinity) by
\begin{equation}
  2r_\text{s}^2-\left[(5-3\gamma)r_\text{a}+4r_\text{h}\right]r_\text{s}+2r_\text{h}^2-4(\gamma-1)r_\text{a}r_\text{h}=0.
\end{equation}
Argue that the accretion flow passes through a unique sonic point for any
value of $\gamma>1$.  Assuming that $v_{\text{s}0}\ll c$, find
approximations for $r_\text{s}$ in the cases (i) $\gamma<5/3$, (ii)
$\gamma=5/3$ and (iii) $\gamma>5/3$.

\subsection{Spherical flow in a power-law potential}
\label{e:bondi}

For steady, spherically symmetric, adiabatic flow in a gravitational
potential $\Phi=-Ar^{-\beta}$, where $A$ and $\beta$ are positive
constants, show that a necessary condition for either (i) an inflow
that starts from rest at $r=\infty$ or (ii) an outflow that reaches $r=\infty$
to pass through a sonic point is
\begin{equation}
   \gamma<f(\beta),
\end{equation}
where $\gamma>1$ is the adiabatic exponent and $f(\beta)$ is a
function to be determined.

Assuming that this condition is satisfied, calculate the accretion
rate of a transonic accretion flow in terms of $A$, $\beta$, $\gamma$
and the density and sound speed at $r=\infty$.  Evaluate your
expression in each of the limits $\gamma\to1$ and $\gamma\to
f(\beta)$.  (You may find it helpful to define $\delta=\gamma-1$.)

\subsection{Rotating outflows}
\label{e:rotating}

The wind from a rotating star can be modelled as a steady,
axisymmetric, adiabatic flow in which the magnetic field is
neglected. Let $\psi(r,z)$ be the mass flux function, such that
\begin{equation}
  \rho\bmu_\rmp=\grad\psi\btimes\grad\phi,
\end{equation}
where $(r,\phi,z)$ are cylindrical polar coordinates and $\bmu_\rmp$ is
the poloidal part of the velocity. Show that the specific entropy, the
specific angular momentum and the Bernoulli function are constant
along streamlines, giving rise to three functions $s(\psi)$,
$\ell(\psi)$ and $\varepsilon(\psi)$.  Use the remaining dynamical
equation to show that $\psi$ satisfies the partial differential
equation
\begin{equation}
  \f{1}{\rho}\div\left(\f{1}{\rho r^2}\grad\psi\right)=\f{\rmd\varepsilon}{\rmd\psi}-T\f{\rmd s}{\rmd\psi}-\f{\ell}{r^2}\f{\rmd\ell}{\rmd\psi}.
\end{equation}

\subsection{Critical points of magnetized outflows}
\label{e:critical}

The integrals of the equations of ideal MHD for a steady axisymmetric
outflow are
\begin{equation}
  \bmu=\f{k\bmB}{\rho}+r\omega\,\bme_\phi,
\label{integral1}
\end{equation}
\begin{equation}
  u_\phi-\f{B_\phi}{\mu_0k}=\f{\ell}{r},
\label{integral2}
\end{equation}
\begin{equation}
  s=s(\psi),
\label{integral3}
\end{equation}
\begin{equation}\tag{4}
  \half|\bmu-r\omega\,\bme_\phi|^2+\Phi-
  \half r^2\omega^2+h=\tilde\varepsilon,
\label{integral4}
\end{equation}
where $k(\psi)$, $\omega(\psi)$, $\ell(\psi)$, $s(\psi)$ and
$\tilde\varepsilon(\psi)$ are surface functions.  Assume that the magnetic
flux function $\psi(r,z)$ is known from a solution of the
Grad--Shafranov equation, and let the cylindrical radius $r$ be used
as a parameter along each magnetic field line.  Then the poloidal
magnetic field $\bmB_\text{p}=\grad\psi\btimes\grad\phi$ is a known
function of $r$ on each field line.  Assume further that the surface
functions $k(\psi)$, $\omega(\psi)$, $\ell(\psi)$, $s(\psi)$ and
$\tilde\varepsilon(\psi)$ are known.

Show that equations (\ref{integral1})--(\ref{integral3}) can then be
used, in principle, and together with the equation of state, to
determine the velocity $\bmu$ and the specific enthalpy $h$ as
functions of $\rho$ and $r$ on each field line.  Deduce that equation
(\ref{integral4}) has the form
\begin{equation}
  f(\rho,r)=\tilde\varepsilon=\cst
\end{equation}
on each field line.

Show that
\begin{equation}
  -\rho\f{\p f}{\p\rho}=\f{u_\text{p}^4-(v_\text{s}^2+v_\text{a}^2)u_\text{p}^2+v_\text{s}^2v_\text{ap}^2}{u_\text{p}^2-v_\text{ap}^2},
\end{equation}
where $v_\text{s}$ is the adiabatic sound speed, $v_\text{a}$ is the
(total) Alfv\'en speed and the subscript `p' denotes the poloidal
(meridional) component.  Deduce that the flow has critical points
where $u_\text{p}$ equals the phase speed of axisymmetric fast or slow
magnetoacoustic waves.  What condition must be satisfied by $\p f/\p
r$ for the flow to pass through these critical points?

\subsection{Radial oscillations of a star}
\label{e:radial}

Show that purely radial (i.e.\ spherically symmetric) oscillations
of a spherical star satisfy the Sturm--Liouville equation
\begin{equation}
  \f{\rmd}{\rmd r}\left[\f{\gamma p}{r^2}\f{\rmd}{\rmd r}(r^2\xi_r)\right]-
  \f{4}{r}\f{\rmd p}{\rmd r}\xi_r+\rho\omega^2\xi_r=0.
\end{equation}
How should $\xi_r$ behave near the centre of the star and near the
surface $r=R$ at which $p=0$?

Show that the associated variational principle can be written in the
equivalent forms
\begin{equation}
\begin{split}
  \omega^2\int_0^R\rho|\xi_r|^2\,r^2\,\rmd r
  &=\int_0^R\left[\f{\gamma p}{r^2}\left|\f{\rmd}{\rmd r}(r^2\xi_r)\right|^2+
  4r\f{\rmd p}{\rmd r}|\xi_r|^2\right]\,\rmd r\\
  &=\int_0^R\left[\gamma pr^4\left|\f{\rmd}{\rmd r}
  \left(\f{\xi_r}{r}\right)\right|^2+(4-3\gamma)r\f{\rmd p}{\rmd r}|\xi_r|^2\right]\,\rmd r,
\end{split}
\end{equation}
where $\gamma$ is assumed to be independent of $r$.  Deduce that the
star is unstable to purely radial perturbations if and only if
$\gamma<4/3$.  Why does it not follow from the first form of the
variational principle that the star is unstable for all values of
$\gamma$?

Can you reach the same conclusion using only the virial theorem?

\subsection{Waves in an isothermal atmosphere}
\label{e:isothermal}

Show that linear waves of frequency $\omega$ and horizontal wavenumber
$k_\text{h}$ in a plane-parallel isothermal atmosphere satisfy the
equation
\begin{equation}
  \f{\rmd^2\xi_z}{\rmd z^2}-\f{1}{H}\f{\rmd\xi_z}{\rmd z}+\f{(\gamma-1)}{\gamma^2H^2}\xi_z+(\omega^2-N^2)\left(\f{1}{v_\text{s}^2}-\f{k_\text{h}^2}{\omega^2}\right)\xi_z=0,
\end{equation}
where $H$ is the isothermal scale-height, $N$ is the
Brunt--V\"ais\"al\"a frequency and $v_\text{s}$ is the adiabatic
sound speed.

Consider solutions of the vertically wavelike form
\begin{equation}
  \xi_z\propto\rme^{z/2H}\exp(\rmi k_zz),
\end{equation}
where $k_z$ is real, so that the wave energy density (proportional to
$\rho|\bxi|^2$) is independent of~$z$.  Obtain the dispersion relation
connecting $\omega$ and $\bmk$.  Assuming that $N^2>0$, show that
propagating waves exist in the limits of high and low frequencies, for
which
\begin{equation}
  \omega^2\approx v_\text{s}^2k^2\quad\text{(acoustic waves)}\qquad\text{and}\qquad
  \omega^2\approx\f{N^2k_\text{h}^2}{k^2}\quad\text{(gravity waves)}
\end{equation}
respectively.  Show that the minimum frequency at which acoustic
waves propagate is $v_\text{s}/2H$.

Explain why the linear approximation must break down above some
height in the atmosphere.

\subsection{Gravitational instability of a slab}
\label{e:slab}

An isothermal ideal gas of sound speed $c_\rms$ forms a
self-gravitating slab in hydrostatic equilibrium with density $\rho(z)$,
where $(x,y,z)$ are Cartesian coordinates.

\medskip

(a) Verify that
\begin{equation}
  \rho\propto\sech^2\left(\f{z}{H}\right),
\end{equation}
and relate the scalehheight $H$ to the surface density
\begin{equation}
  \Sigma=\int_{-\infty}^\infty\rho\,\rmd z.
\end{equation}

\medskip

(b) Assuming that the perturbations are also isothermal, derive the
linearized equations governing displacements of the form
\begin{equation}
  \mathrm{Re}\left[\bxi(z)\,\rme^{\rmi(kx-\omega t)}\right],
\end{equation}
where $k$ is a real wavenumber.  Show that $\omega^2$ is real for
disturbances satisfying appropriate conditions as $|z|\to\infty$.

\medskip

(c) For a marginally stable mode with $\omega^2=0$, derive the
associated Legendre equation
\begin{equation}
  \f{\rmd}{\rmd\tau}\left[(1-\tau^2)\f{\rmd\,\delta\Phi}{\rmd\tau}\right]+\left(2-\f{\nu^2}{1-\tau^2}\right)\delta\Phi=0,
\end{equation}
where $\tau=\tanh(z/H)$, $\nu=kH$ and $\delta\Phi$ is the Eulerian
perturbation of the gravitational potential.  Verify that two solutions
of this equation are
\begin{equation}
  \left(\f{1+\tau}{1-\tau}\right)^{\nu/2}(\nu-\tau)\quad\hbox{and}\quad\left(\f{1-\tau}{1+\tau}\right)^{\nu/2}(\nu+\tau).
\end{equation}
Deduce that the marginally stable mode has $|k|=1/H$ and
$\delta\Phi\propto\sech(z/H)$. Would you expect the unstable modes to
have wavelengths greater or less than $2\pi H$?

\subsection{Magnetic buoyancy instabilities}
\label{e:buoyancy}

A perfect gas forms a static atmosphere in a uniform gravitational
field $-g\,\bme_z$, where $(x,y,z)$ are Cartesian coordinates.  A
horizontal magnetic field $B(z)\,\bme_y$ is also present.

Derive the linearized equations governing small displacements of the
form
\begin{equation}
  \real\left[\bxi(z)\,\exp(-\rmi\omega t+\rmi k_xx+\rmi k_yy)\right],
\end{equation}
where $k_x$ and $k_y$ are real horizontal wavenumbers, and show that
\begin{equation}
\begin{split}
  &\omega^2\int_a^b\rho|\bxi|^2\,\rmd z=[\xi_z^*\,\delta\Pi]_a^b\\
  &\qquad+\int_a^b\left(\f{|\delta\Pi|^2}{\gamma p+\f{B^2}{\mu_0}}-\f{\Big|\rho g\xi_z+\f{B^2}{\mu_0}\rmi k_y\xi_y\Big|^2}{\gamma p+\f{B^2}{\mu_0}}+\f{B^2}{\mu_0}k_y^2|\bxi|^2-g\f{\rmd\rho}{\rmd z}|\xi_z|^2\right)\rmd z,
\end{split}
\end{equation}
where $z=a$ and $z=b$ are the lower and upper boundaries of the
atmosphere, and $\delta\Pi$ is the Eulerian perturbation of total
pressure.  (Self-gravitation may be neglected.)

You may assume that the atmosphere is unstable if and only if the
integral on the right-hand side can be made negative by a trial
displacement $\bxi$ satisfying the boundary conditions, which are such
that $[\xi_z^*\,\delta\Pi]_a^b=0$.  You may also assume that the
horizontal wavenumbers are unconstrained.  Explain why the integral
can be minimized with respect to $\xi_x$ by letting $\xi_x\to0$ and
$k_x\to\infty$ in such a way that $\delta\Pi=0$.

Hence show that the atmosphere is unstable to disturbances with
$k_y=0$ if and only if
\begin{equation}
  -\f{\rmd\ln\rho}{\rmd z}<\f{\rho g}{\gamma p+\f{B^2}{\mu_0}}
\end{equation}
at some point.

Assuming that this condition is not satisfied anywhere, show further
that the atmosphere is unstable to disturbances with $k_y\ne0$ if and
only if
\begin{equation}
  -\f{\rmd\ln\rho}{\rmd z}<\f{\rho g}{\gamma p}
\end{equation}
at some point.

How does these stability criteria compare with the hydrodynamic
stability criterion $N^2<0$?

\subsection{Waves in a rotating fluid}
\label{e:inertial}

Write down the equations of ideal gas dynamics in cylindrical polar
coordinates $(r,\phi,z)$, assuming axisymmetry.  Consider a steady,
axisymmetric basic state in uniform rotation, with density
$\rho(r,z)$, pressure $p(r,z)$ and velocity $\bmu=r\Omega\,\bme_\phi$.
Determine the linearized equations governing axisymmetric
perturbations of the form
\begin{equation}
  \real\left[\delta\rho(r,z)\,\rme^{-\rmi\omega t}\right],
\end{equation}
etc.  If the basic state is homentropic and self-gravity may be
neglected, show that the linearized equations reduce to
\begin{equation}
  -\rmi\omega\,\delta u_r-2\Omega\,\delta u_\phi=-\frac{\p W}{\p r},
\end{equation}
\begin{equation}
  -\rmi\omega\,\delta u_\phi+2\Omega\,\delta u_r=0,
\end{equation}
\begin{equation}
  -\rmi\omega\,\delta u_z=-\frac{\p W}{\p z},
\end{equation}
\begin{equation}
  -\rmi\omega W+\frac{v_\text{s}^2}{\rho}\left[\frac{1}{r}\frac{\p}{\p r}(r\rho\,\delta u_r)+\frac{\p}{\p z}(\rho\,\delta u_z)\right]=0,
\end{equation}
where $W=\delta p/\rho$.

Eliminate $\delta\bmu$ to obtain a second-order partial differential
equation for $W$.  Is the equation of elliptic or hyperbolic type?
What are the relevant solutions of this equation if the fluid has
uniform density and fills a cylindrical container $\{r<a,\,0<z<H\}$
with rigid boundaries?

\section{Electromagnetic units}
\label{s:gaussian}

These lecture notes use rationalized units for electromagnetism, such
that Maxwell's equations take the form
\begin{equation}
  \f{\p\bmB}{\p t}=-\curl\bmE,\qquad
  \div\bmB=0,\qquad
  \curl\bmB=\mu_0\left(\bmJ+\epsilon_0\f{\p\bmE}{\p t}\right),\qquad
  \div\bmE=\f{\rho_\rme}{\epsilon_0}.
\end{equation}
These involve the vacuum permeability and permittivity $\mu_0$ and
$\epsilon_0$, related to the speed of light $c$ by
$c=(\mu_0\epsilon_0)^{-1/2}$, but do not involve factors of $4\pi$ or
$c$.

In astrophysics it is common to use Gaussian units for
electromagnetism, such that Maxwell's equations take the form
\begin{equation}
  \f{\p\bmB}{\p t}=-c\curl\bmE,\qquad
  \div\bmB=0,\qquad
  \curl\bmB=\f{1}{c}\left(4\pi\bmJ+\f{\p\bmE}{\p t}\right),\qquad
  \div\bmE=4\pi\rho_\rme.
\end{equation}
In the limit relevant for Newtonian MHD, the $\p\bmE/\p t$ term is
neglected.  Different factors then appear in several related
equations.  The magnetic energy density in Gaussian units is
\begin{equation}
  \f{B^2}{8\pi}\qquad\hbox{rather than}\quad\f{B^2}{2\mu_0},
\end{equation}
the electromagnetic energy flux density (Poynting vector) is
\begin{equation}
  \f{c}{4\pi}\bmE\btimes\bmB\qquad\hbox{rather than}\quad\f{\bmE\btimes\bmB}{\mu_0},
\end{equation}
the Maxwell stress is
\begin{equation}
  \f{1}{4\pi}\left(\bmB\bmB-\f{B^2}{2}\bfI\right)\qquad\hbox{rather than}\quad\f{1}{\mu_0}\left(\bmB\bmB-\f{B^2}{2}\bfI\right),
\end{equation}
and the Lorentz force is
\begin{equation}
  \f{1}{c}\bmJ\btimes\bmB=\f{1}{4\pi}(\curl\bmB)\btimes\bmB\qquad\hbox{rather than}\quad\bmJ\btimes\bmB=\f{1}{\mu_0}(\curl\bmB)\btimes\bmB.
\end{equation}
The perfectly conducting fluid approximation of ideal MHD corresponds
to
\begin{equation}
  \bmE=-\f{1}{c}\bmu\btimes\bmB\qquad\hbox{rather than}\quad\bmE=-\bmu\btimes\bmB.
\end{equation}

The fields $\bmE$, $\bmB$ and $\bmJ$ can be converted from
rationalized to Gaussian units by replacing
\begin{equation}
\begin{split}
  \bmE&\mapsto\left(\f{1}{4\pi\epsilon_0}\right)^{1/2}\bmE=c\left(\f{\mu_0}{4\pi}\right)^{1/2}\bmE,\\
  \bmB&\mapsto\left(\f{\mu_0}{4\pi}\right)^{1/2}\bmB,\\
  \bmJ&\mapsto(4\pi\epsilon_0)^{1/2}\bmJ=\f{1}{c}\left(\f{4\pi}{\mu_0}\right)^{1/2}\bmJ.
\end{split}
\end{equation}

For historical reasons, rationalized electromagnetic units are
associated with the MKS (metre--kilogram--second) system of mechanical
units, while Gaussian electromagnetic units are associated with the
CGS (centimetre--gram--second) system.  The most common system of
rationalized units is SI units, in which $\mu_0$ has the exact value
$4\pi\times10^{-7}$ (in units of $\mathrm{N}\,\mathrm{A}^{-2}$ or
$\mathrm{H}\,\mathrm{m}^{-1}$).  In principle, rationalized units can
be used within CGS, in which case $\mu_0$ has the value $4\pi$.

\section{Summary of notation}

$A$: poloidal Alfv\'en number

$\bfA_i$: matrix describing hyperbolic structure

$\bmA$: magnetic vector potential

$\bma$: particle acceleration; initial position vector

$B$: Bernoulli constant

$\bmB$: magnetic field

$\bmB_\rmp$: poloidal magnetic field

$C_{ij}$: cofactor of deformation tensor

$c$: speed of light; velocity dispersion

$c_\rmp$: specific heat capacity at constant pressure

$c_\rms$: isothermal sound speed

$c_\rmv$: specific heat capacity at constant volume

$\rmD/\rmD t$: Lagrangian time-derivative

$\bmE$: electric field

$e$: specific internal energy

$\bme$: basis (unit) vector

$F$: determinant of deformation tensor

$F_{ij}$: deformation tensor

$\bmF$: flux density of conserved quantity

$\bmF_\rmm$: Lorentz force per unit volume

$\mathcal{F}$: force operator

$f$: distribution function

$f_\rmM$: Maxwellian distribution function

$G$: Newton's constant

$G_{ij}$: inverse of deformation tensor

$g$: gravitational acceleration

$\bmg$: gravitational field

$H$: Heaviside step function; scale height

$H_\rmc$: cross helicity

$H_\rmk$: kinetic helicity

$H_\rmm$: magnetic helicity

$h$: specific enthalpy

$I$: trace of inertia tensor; moment of inertia

$I_{ij}$: inertia tensor

$\bfI$: unit tensor

$J$: Jacobian determinant

$J_0$, $J_1$: Bessel functions

$J_{ij}$: Jacobian matrix

$\bmJ$: electric current density

$K$: polytropic constant; kinetic energy

$K_{ij}$: kinetic energy tensor

$k$: Boltzmann's constant; wavenumber; mass loading; potential Love number

$\bmk$: wavevector

$L$: characteristic length-scale; Lagrangian

$\mathcal{L}$: Lagrangian density

$\ell$: angular momentum invariant; specific angular momentum

$M$: magnetic energy; mass

$\bfM$: Maxwell stress tensor

$\mathcal{M}$: Mach number

$m$: particle mass

$m_\rmH$: mass of hydrogen atom

$N$: buoyancy frequency

$n$: number of degrees of freedom; number density

$\bmn$: unit normal vector

$p$: pressure

$p_\mathrm{g}$: gas pressure

$p_\mathrm{m}$: magnetic pressure

$p_\mathrm{r}$: radiation pressure

$q$: electric charge; density of conserved quantity

$R$: shock radius

$R_\pm$: Riemann invariants

$r$: cylindrical radius; spherical radius

$r_0$: footpoint radius

$r_\rma$: Alfv\'en radius

$r_\rms$: sonic radius

$S$: bounding surface; action

$s$: specific entropy

$T$: temperature; characteristic time-scale; torque

$T_\mathrm{m}$: magnetic tension

$\bfT$: stress tensor

$\mathcal{T}$: trace of integrated stress tensor

$\mathcal{T}_{ij}$: integrated stress tensor

$t$: time

$U$: internal energy

$\bfU$: state vector

$u_\mathrm{sh}$: shock speed

$\bmu$: velocity field

$V$: volume (occupied by fluid)

$\hat V$: exterior volume

$V_{ij}$: second-rank potential energy tensor

$V_{ijkl}$: fourth-rank potential energy tensor

$v$: specific volume; wave speed

$v_\rms$: adiabatic sound speed

$\bmv$: particle velocity; relative velocity of frames

$\bmv_\rma$: Alfv\'en velocity

$\bmv_\rmg$: group velocity

$\bmv_\rmp$: phase velocity

$W$: gravitational energy; potential energy functional

$x$: Cartesian coordinate

$\bmx$: position vector

$Y$: spherical harmonic

$y$: Cartesian coordinate

$z$: Cartesian coordinate

$\beta$: plasma beta

$\Gamma_1$: first adiabatic exponent

$\gamma$: ratio of specific heats; adiabatic exponent

$\Delta$: Lagrangian perturbation; divergence of displacement

$\delta$: Eulerian perturbation; Dirac delta function

$\delta m$: material mass element

$\delta\bmS$: material surface element

$\delta\bmu$: velocity difference; velocity perturbation

$\delta V$: material volume element

$\delta\bmx$: material line element

$\delta\Phi$: material flux element

$\epsilon_{ijk}$: Levi--Civita tensor

$\varepsilon$: energy invariant

$\bmeta$: secondary displacement

$\theta$: polar angle; angle between wavevector and magnetic field

$\lambda$: mean free path; force-free field scalar; scaling parameter

$\mu$: mean molecular weight; scaling parameter

$\mu_0$: vacuum permeability

$\xi$: similarity variable

$\bxi$: (Lagrangian) displacement

$\Pi$: total pressure

$\rho$: mass density

$\rho_\rme$: charge density

$\sigma$: Stefan's constant; collisional cross-section

$\tau$: relaxation time

$\Phi$: gravitational potential

$\Phi_\rme$: electrostatic potential

$\Phi_\mathrm{ext}$: external gravitational potential

$\Phi_\mathrm{int}$: internal (self-) gravitational potential

$\phi$: azimuthal angle

$\varphi$: phase

$\chi$: scalar field in gauge transformation

$\chi_\rho$: inverse isothermal compressibility

$\Psi$: secondary gravitational potential

$\psi$: magnetic flux function

$\Omega$: angular velocity

$\omega$: wave frequency

$\bomega$: vorticity

\bibliographystyle{jpp}

\end{document}